\pdfoutput=1

\documentclass[12pt,preprint]{aastex}

\slugcomment{}

\shorttitle{The EBEX Instrument - Optics, Receiver, and Polarimetry}
\shortauthors{EBEX Collaboration}

\usepackage{subfiles}
\usepackage[titletoc]{appendix}
\usepackage[nolist]{acronym}
\usepackage[margin=0.75in]{geometry}                
\usepackage{graphicx}
\usepackage{subfigure}
\usepackage{amssymb}
\usepackage{xcolor}
\usepackage{fixltx2e}
\usepackage{amsmath}
\usepackage{hyperref}
\usepackage{float}
\usepackage{tikz}
\usepackage{subfiles}
\usepackage{multirow}
\usepackage{multicol}
\usepackage{calrsfs}
\usepackage{morefloats}

\usetikzlibrary{calc,fit,shadows}
\newcommand{\planck}{{\it Planck}}

\newcommand{\FIG}{Figure }
\newcommand{\TAB}{Table }

\begin{document}


\title{The EBEX Balloon Borne Experiment - Optics, Receiver, and Polarimetry }


\author{The EBEX Collaboration: Asad~M.~Aboobaker\altaffilmark{1},
Peter~Ade\altaffilmark{2}, Derek~Araujo\altaffilmark{3}, 
Fran\c{c}ois~Aubin\altaffilmark{4}, 
Carlo~Baccigalupi\altaffilmark{5,6}, Chaoyun~Bao\altaffilmark{4},
 Daniel~Chapman\altaffilmark{3}, 
Joy~Didier\altaffilmark{3}, Matt~Dobbs\altaffilmark{7,8}, 
Christopher~Geach\altaffilmark{4},
Will~Grainger\altaffilmark{9}, 
Shaul~Hanany\altaffilmark{4, *}, Kyle~Helson\altaffilmark{10}, Seth~Hillbrand\altaffilmark{3}, 
Johannes~Hubmayr\altaffilmark{11}, Andrew~Jaffe\altaffilmark{12}, 
Bradley~Johnson\altaffilmark{3}, Terry~Jones\altaffilmark{4}, 
Jeff~Klein\altaffilmark{4}, Andrei~Korotkov\altaffilmark{10}, Adrian~Lee\altaffilmark{13},
Lorne~Levinson\altaffilmark{14}, 
Michele~Limon\altaffilmark{3}, Kevin~MacDermid\altaffilmark{7},
Tomotake~Matsumura\altaffilmark{4, 15}, 
Amber~D.~Miller\altaffilmark{3}, Michael~Milligan\altaffilmark{4}, 
Kate~Raach\altaffilmark{4}, Britt~Reichborn-Kjennerud\altaffilmark{3},
Ilan~Sagiv\altaffilmark{14}, Giorgio~Savini\altaffilmark{16},
Locke~Spencer\altaffilmark{2,17}, 
Carole~Tucker\altaffilmark{2}, Gregory~S.~Tucker\altaffilmark{10},
Benjamin~Westbrook\altaffilmark{11}, 
Karl~Young\altaffilmark{4}, Kyle~Zilic\altaffilmark{4}}


\altaffiltext{1}{Jet Propulsion Laboratory, California Institute of Technology, Pasadena, CA 91109} 
\altaffiltext{2}{School of Physics and Astronomy, Cardiff University, Cardiff, CF24 3AA, United Kingdom} 
\altaffiltext{3}{Physics Department, Columbia University, New York, NY 10027} 
\altaffiltext{4}{University of Minnesota School of Physics and Astronomy, Minneapolis, MN 55455} 
\altaffiltext{5}{Astrophysics Sector, SISSA, Trieste, 34014, Italy} 
\altaffiltext{6}{INFN, Sezione di Trieste, Via Valerio 2, I-34127 Trieste, Italy} 
\altaffiltext{7}{McGill University, Montr´eal, Quebec, H3A 2T8, Canada} 
\altaffiltext{8}{Canadian Institute for Advanced Research, Toronto, ON, M5G1Z8, Canada} 
\altaffiltext{9}{Rutherford Appleton Lab, Harwell Oxford, OX11 0QX} 
\altaffiltext{10}{Brown University, Providence, RI 02912} 
\altaffiltext{11}{National Institute of Standards and Technology, Gaithersburg, MD 20899} 
\altaffiltext{12}{Department of Physics, Imperial College, London, SW7 2AZ, United Kingdom} 
\altaffiltext{13}{Department of Physics, University of California, Berkeley, Berkeley, CA 94720} 
\altaffiltext{14}{Weizmann Institute of Science, Rehovot 76100, Israel} 
\altaffiltext{15}{Kavli Institute for the Physics and Mathematics of the 
Universe (Kavli IPMU, WPI), Todai Institutes for Advanced Study, The University of Tokyo, 
Kashiwa City, Chiba 277-8583, Japan} 
\altaffiltext{16}{University College London, London WC1E 6BT, UK} 
\altaffiltext{17}{University of Lethbridge, Lethbridge, AB, T1K 3M4, Canada} 
\altaffiltext{*}{Corresponding Author: Shaul Hanany (hanany@umn.edu)}


\begin{abstract}



The E and B Experiment (EBEX) was a long-duration balloon-borne 
cosmic microwave background polarimeter that flew over Antarctica in 2013. 
We describe the experiment's optical system, receiver, and polarimetric approach, and 
report on their in-flight performance.  
EBEX had three frequency bands centered on 150, 250, and 410~GHz. 
To make efficient use of limited mass and space we designed 
a 115~cm$^{2}$sr high throughput optical system 
that had two ambient temperature mirrors and four anti-reflection coated polyethylene lenses 
per focal plane. All frequency bands shared the same optical train. 
Polarimetry was achieved with a continuously rotating achromatic 
half-wave plate (AHWP) that was levitated with a superconducting magnetic bearing
(SMB). Rotation stability was 0.45~\% over a period of 10~hours, and angular position accuracy 
was 0.01~degrees. This is the first use of a SMB in astrophysics. The measured 
modulation efficiency was above 90~\% for all bands. To our knowledge the 109~\% fractional 
bandwidth of the AHWP was the broadest implemented to date. 
The receiver that contained one lens and the AHWP at a temperature of 
4~K, the polarizing grid and other lenses at 1~K, and the two focal planes at 0.25~K
performed according to specifications giving focal plane temperature stability with 
fluctuation power spectrum that had $1/f$ knee at 2~mHz. 
EBEX was the first balloon-borne instrument to implement technologies characteristic 
of modern CMB polarimeters including high throughput optical systems, and large
arrays of transition edge sensor bolometric detectors with mutiplexed readouts. 
\end{abstract}


\maketitle



\section{Introduction}
\label{sec:introduction}


Measurements of the \ac{CMB} have provided a wealth of information about the physical mechanisms responsible for
the evolution of the Universe.
In recent years, experimental efforts have focused on measuring the \ac{CMB}'s polarization 
patterns: {\it E}-modes and {\it B}-modes~\citep{zaldarriaga97_physrev}. 
The level and specific shape of the angular power spectrum of CMB {\it E}-mode
polarization can be predicted given the measured intensity anisotropy. 
Lensing of {\it E}-modes by the large scale structure of the Universe produces
cosmological {\it B}-modes at small angular scales, while an inflationary phase at sufficiently 
high energy scales near the big bang is predicted to leave another detectable {\it B}-mode signature at large 
and intermediate angular scales~\citep{Baumann_2009}.  

The {\it E}-mode polarization of the \ac{CMB}
was first detected by the DASI experiment~\citep{DASI_Emodes}, and other
experiments soon followed suit~\citep{Scott_2010}.  
The combination of all measurements is in excellent agreement with
predictions. {\it B}-mode polarization from gravitational lensing 
of {\it E}-modes and from Galactic dust emission has also recently been 
detected~\citep{SPT_Bmodes, PolarBear_Bmodes, ACT_Bmodes, bicep2detection, bicep+planck}. 
Intense efforts are ongoing by ground- and balloon-based instruments to
improve the measurements, separate the Galactic from the cosmological signals, and 
identify the inflationary {\it B}-mode signature.


\ac{EBEX} was a balloon-borne CMB polarimeter striving to 
detect or constrain the levels of the inflationary gravitational wave and lensing {\it B}-mode power spectra.  
\ac{EBEX} was also designed to be a technology pathfinder for future \ac{CMB} space missions.
To improve instrument sensitivity, we implemented a kilo-pixel array of \ac{TES} bolometers and planned
for a long duration balloon flight. We included three spectral bands centered on 150, 250, and 410~GHz to 
give sensitivity to both the CMB and the galactic dust foreground. 
The combination of the 400~deg$^2$ intended survey size and an optical system with 0.1~deg resolution gave 
sensitivity to the range $30 < \ell < 1500$ of the angular power spectrum. 
Polarimetry was achieved with a continuously rotating achromatic \ac{HWP}.

Several new technologies have been implemented and tested for the first
time in the \ac{EBEX} instrument.
It was the first balloon-borne experiment to implement a kilo-pixel array of \ac{TES} bolometric detectors.
It was also the first to implement a digital frequency domain
multiplexing system to read out the \ac{TES} arrays; this digital 
system was later adopted by a number of ground-based experiments.
Finally, it was the first astrophysical instrument to implement and operate 
a \ac{SMB}, which was used to levitate the \ac{HWP}.


Design and construction of the experiment began in 2005.
A ten-hour engineering flight was launched from Ft.\ Sumner, NM on June 11, 
2009, and the long-duration science flight was launched from Mc\,Murdo
Station, Antarctica on December 29, 2012.
Because the majority of the 25-day long-duration flight was in January 2013, we refer to this flight as EBEX2013.


This paper is one of a series of papers describing the experiment and its in-flight performance.
This paper, called \ac{EP1}, discusses the telescope and the polarimetric receiver; \ac{EP2}~\citep{EBEXPaper2} describes the detectors 
and the readout system; and \ac{EP3}~\citep{EBEXPaper3}, describes the gondola, the attitude control system, and other support systems.
Several other publications give additional details about the \ac{EBEX} experiment. Some are 
from earlier stages of the program~\citep{Oxley_EBEX2004, Grainger_EBEX2008, Aubin_TESReadout2010, Milligan_Software, ReichbornKjennerud_EBEX2010, klein_HWP, Sagiv_MGrossman2012, Westbrook_Design_Evolution}, and others discuss some
subsystems in more detail~\citep{Dan_thesis, Britt_thesis, Sagiv_thesis, Aubin_thesis, MacDermid_thesis, MacDermid_SPIE2014, Westbrook_thesis, Zilic_thesis, chappy_thesis, chappy_ieee_paper, joy_thesis, joy_ieee_paper, Aubin_MGrossman2015}. 

The science goals of \ac{EBEX} and the choice of technical implementation placed constraints on the design 
and operation of the instrument. In Section~\ref{sec:telescopeandoptics} we describe the EBEX optical system, including the warm telescope, the cold
optics, and the frequency bands. The optical system was designed to provide a resolution of 5-10 arcminutes, sufficient to probe 
the lensing signal at $\ell \simeq 1000$. It also gave a flat and telecentric focal surface to accommodate the array of 
transition-edge sensors that were fabricated on silicon arrays. This optical system required lenses to enhance the 
throughput provided by the two warm mirrors. The lenses, the focal planes, and the cryogenic refrigerators that kept them 
at temperatures below ambient are described in Section~\ref{sec:receiver}. 
We implemented an \ac{AHWP} because it made efficient use of the throughput for the three frequency bands. We used 
it in continuous rotation to avoid low frequency noise. In Section~\ref{sec:polarimetry} we discuss the implementation of the 
\ac{AHWP}, the \ac{SMB}, the rotation mechanism, and the polarimetric calibration. 

\section{Telescope and Optics}
\label{sec:telescopeandoptics}

\subsection{Optical Design}
\label{sec:telescope}

\begin{figure}[ht!]
  \includegraphics[width=\textwidth]{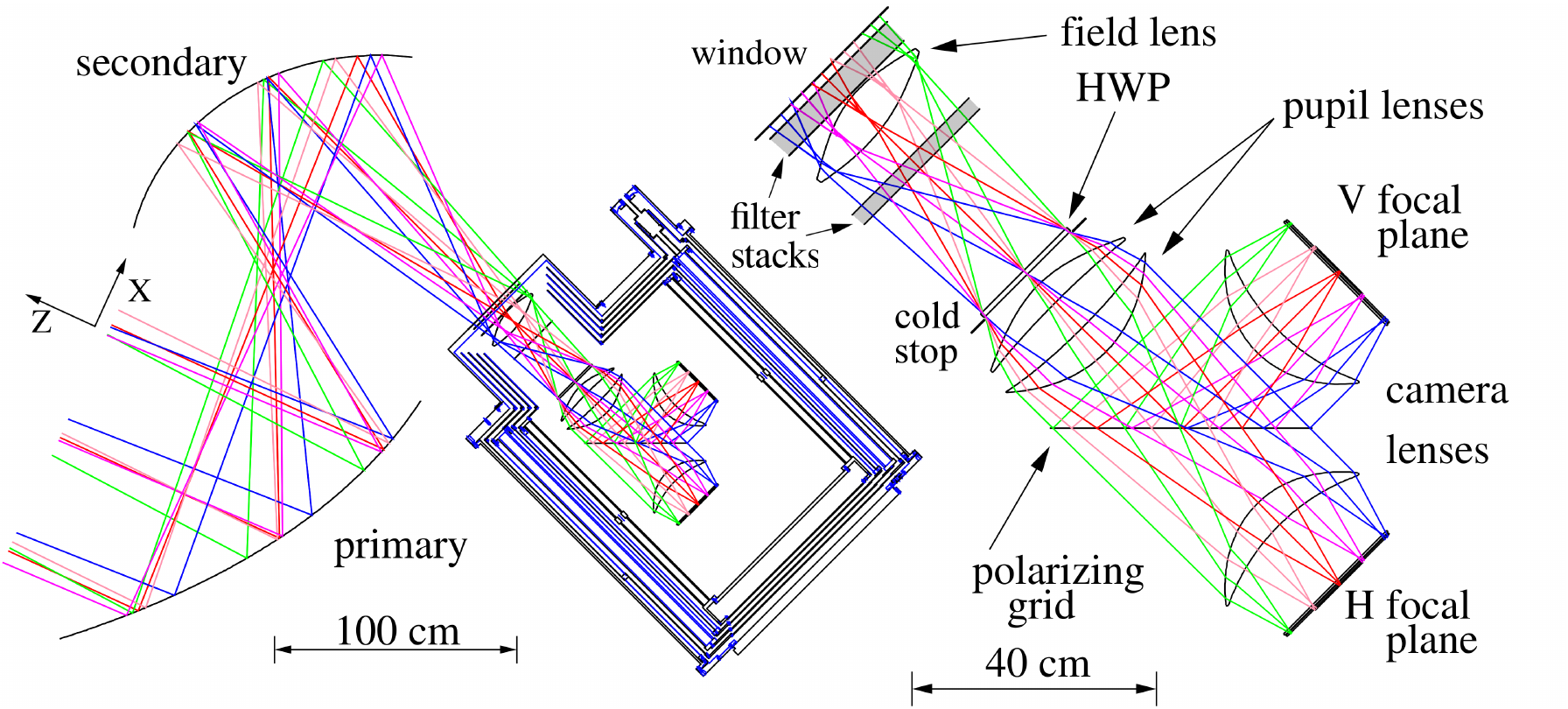}
  \caption[\ac{EBEX} optics ray trace]{Ray tracing of the \ac{EBEX}
    optical design consisting of two ambient temperature reflectors in 
    a Gregorian configuration and a cryogenic receiver (left). Inside the 
    the receiver (right), cryogenically cooled polyethylene lenses formed a cold stop
    and provided diffraction limited performance over a flat, telecentric, $6.6\degr$ field 
    of view. A continuously rotating achromatic
    half-wave plate placed near the aperture stop and a polarizing 
    grid provided the polarimetry capabilities.    	}
  \label{fig:raydiagram}
\end{figure}

The \ac{EBEX} optical system consisted of an off-axis Gregorian reflecting telescope
coupled to a cryogenic receiver containing refractive optics, a rotating \ac{AHWP} at a
cold aperture stop, and a polarizing grid that directed independent polarizations to each 
of two focal planes; see Figure~\ref{fig:raydiagram}. 

\begin{table}[ht!]
  \centering
  \begin{tabular}{|c|c|}
    \tableline
    effective focal length           & 198 cm \\
    aperture diameter    & 105 cm \\
    PR focal length                   & 80 cm \\
    $\angle$ between PR and SR axes  & $12.77\degr$ \\
    PR offset                      & 100 cm \\
    \tableline
    \tableline
    SR semi-major axis length, $a$           & 110.2 cm \\
    SR semi-minor axis length, $b$           & 98.21 cm \\
    SR conic constant, $K$                   & -0.2059 \\
    SR opening half-angle             & $52\degr$ \\
    \tableline
     \tableline
    PR maximum size  & $1.5\times1.8$~m  \\
    SR maximum size  & $1.2\times1.3$~m  \\
    \tableline
  \end{tabular}
  \caption{
    Five fundamental parameters define the geometry of an off-axis Gregorian telescope (upper panel).
    PR(SR) denotes the primary(secondary) reflector. 
    The middle panel lists derived parameters relevant to telescope fabrication.  
    The physical mirror dimensions (lower panel) are the full fabricated size of the mirrors and 
    are 1.4 times larger than the ray-tracing apertures. 
  \label{tab:op-params} }
\end{table}

The parabolic off-axis primary mirror collected incoming radiation and, via an elliptical secondary, 
formed the Gregorian focal surface 10~cm behind the vacuum window of the receiver. 
A field lens was placed coincident with this focal surface and was tilted $8.1\degr$ from the optical axis. The field lens 
formed an image of the primary at the cold aperture stop, which was immediately after the \ac{AHWP}.  
Past the cold stop a pair of pupil lenses collimated the ray bundle. A wire grid linear polarizer passed one state of polarization 
and reflected another, forming two optical branches.  Camera lenses on each branch formed the final flat, 
telecentric focal planes, denoted as H (horizontal) and V (vertical) for the branches which were transmitted and 
reflected by the grid, respectively. 
At the focal plane conical feeds backed by circular waveguides coupled the light from free space 
into the detector cavities. 
The reflectors and cryostat were mounted on the gondola's inner frame and were
surrounded by baffles to control sidelobe pickup.

\subsection{Ambient Temperature Telescope}
\label{sec:ambienttelescope}

\begin{figure}[ht!]
  \centering
  \includegraphics[width=.7\textwidth]{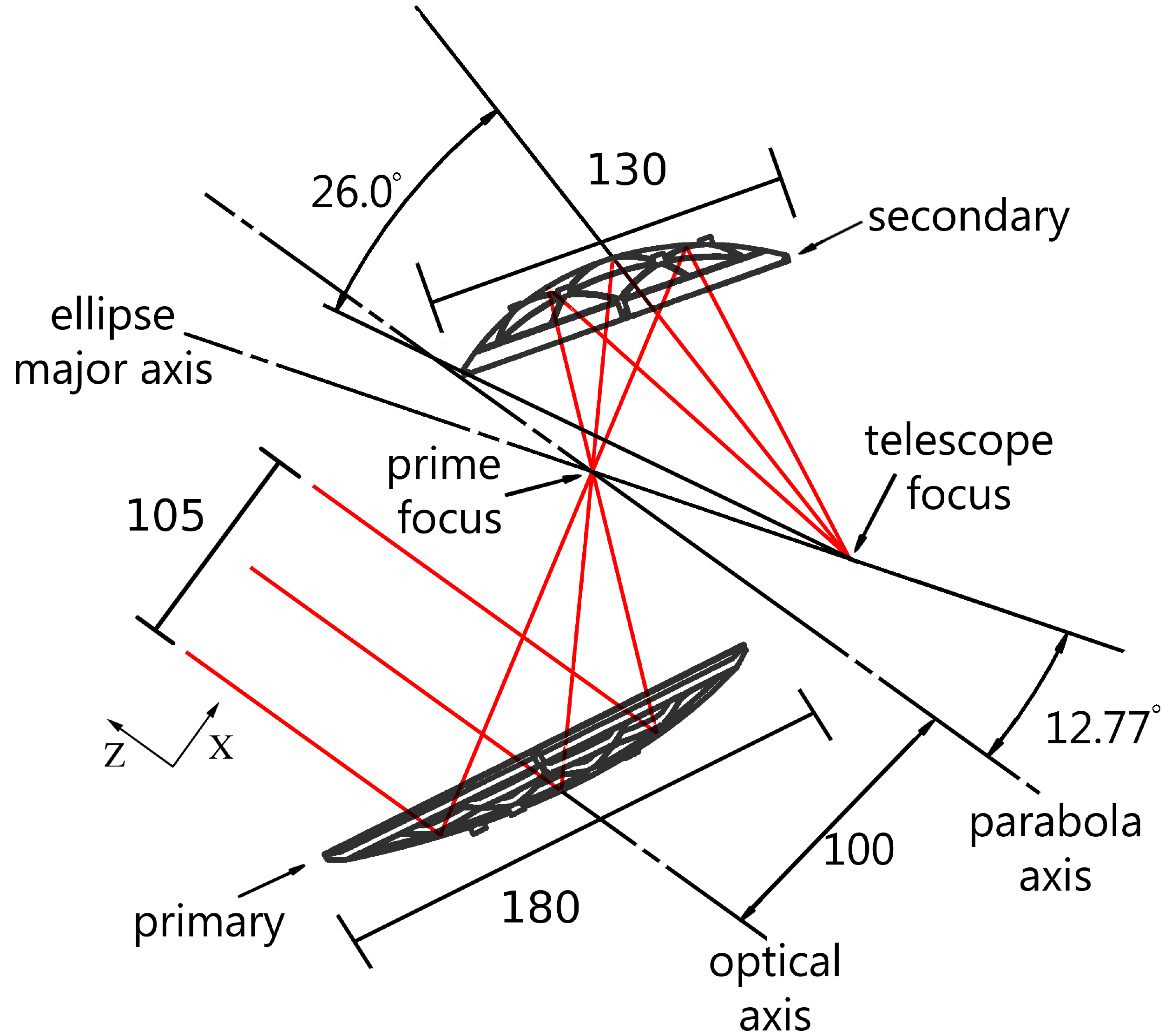}
  \caption[\ac{EBEX} telescope mirrors geometry]{Geometry of the \ac{EBEX}
    Gregorian-Dragone reflecting telescope. Lengths are in cm. }
  \label{fig:op-mirrors}
\end{figure}

The ambient temperature telescope was an off-axis 
Gregorian Mizuguchi-Dragone design~\citep{mizuguchi78,dragone82} with an entrance aperture of 
1.05~m, defined by the cold stop.
The f-number varied across the \ac{FOV} by up to 10\% with an average of $f/1.9$. 
The telescope geometry is shown in \FIG\ref{fig:op-mirrors}, and the parameters of the 
design are tabulated in \TAB\ref{tab:op-params}. To minimize sidelobe pickup due to spillover power,
both the primary and secondary mirrors were 1.4 times larger than the size defined by ray tracing
of a 1.05~m entrance aperture diameter. 
All ray tracing analysis using a 1.05~m aperture apodized by the Gaussian illumination 
from the feedhorns as appropriate for each frequency band; Section~\ref{sec:coldoptics}.  

The $1.5 \times 1.8$~m parabolic primary mirror weighed 42~kg and was previously used in Archeops~\citep{benoit02-archeops}. 
The $1.2 \times 1.3$~m and 22~kg secondary, a section of an ellipsoid of revolution, was fabricated 
for \ac{EBEX}\footnote{Machining by Remmele Engineering, New Brighton, Minnesota.}.
Each mirrors was machined from single billet of 6061-aluminum. The mirrors had a 5~mm thick reflecting surface backed by 
a hexagonal rib structure designed to provide stiffness during surface machining while minimizing weight. 
The rough machined mirrors were heat treated to the T6 condition 
before the last 250~$\mu$m were milled from the reflecting surface.   
The machined mirror surfaces had roughness of less than 2~\micron.  Additionally, small areas at the center 
and at the $\pm x$ and $\pm y$ edges of each reflector were polished to optical quality to enable laser alignment.
We measured the primary and secondary mirror surface contours using a tooling ball laser probe and found
RMS figure accuracies of 51~$\mu$m and 48~$\mu$m, respectively.  This figure accuracy 
was $1/13$ of the wavelength at the highest edge of the highest frequency band. 

\subsection{Receiver Cold Optics and Focal Planes}
\label{sec:coldoptics}

The receiver cold optics formed a reimaging camera which transferred
the image formed at the Gregorian focus to the focal planes. The f-number was 
approximately preserved while the camera hosted an internal aperture stop, 
enlarged the diffraction limited field of view, and formed 
two flat, telecentric focal planes. The receiver section of the light path 
also included electromagnetic filters, the \ac{AHWP}, and a polarizing wire grid beam splitter. 
\begin{figure}[ht!]
  \centering
  \includegraphics[width=.5\textwidth]{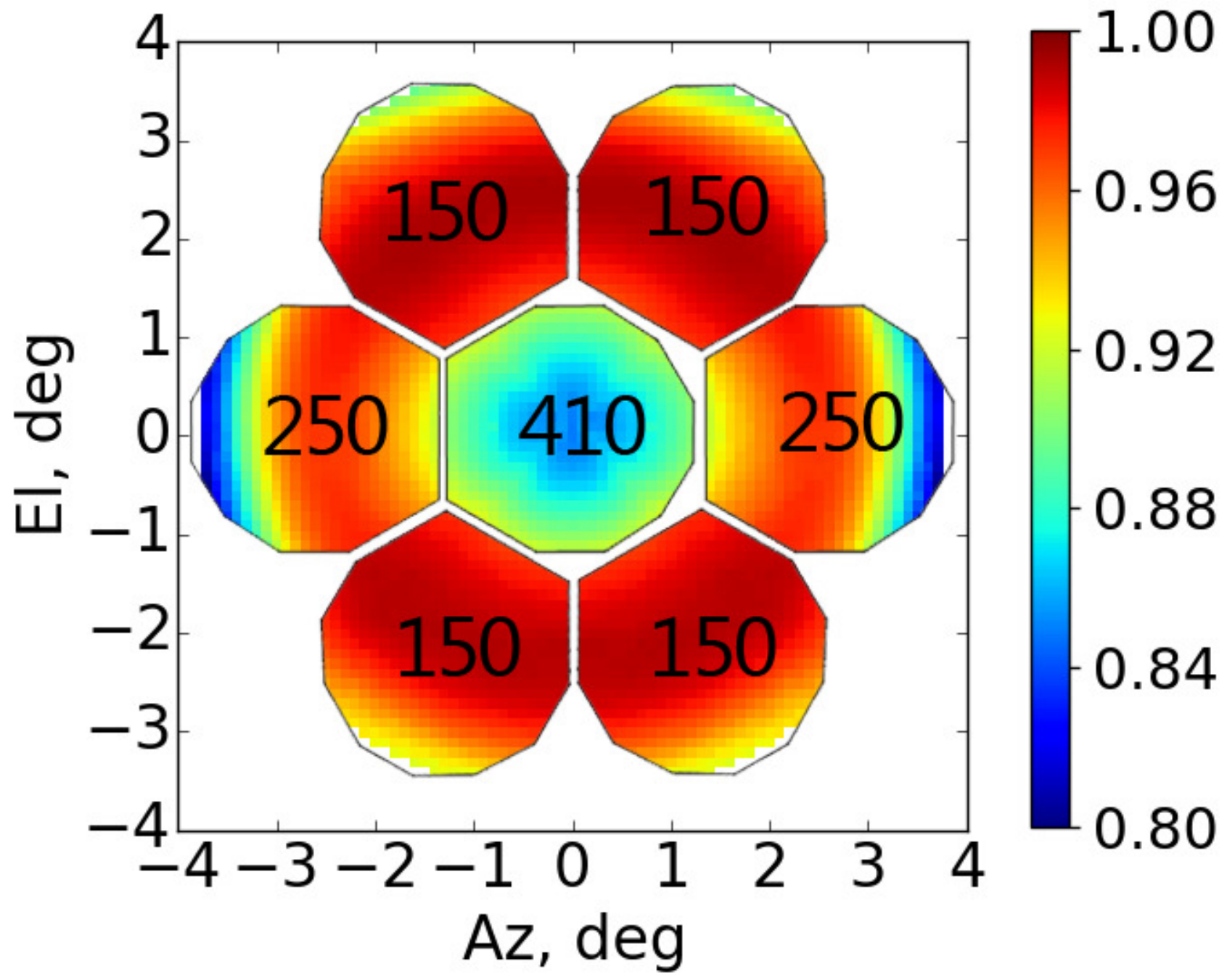}
  \caption{ Detector wafer outlines on the focal plane, overlaid by Strehl ratios for each frequency, which 
  is denoted as a number at the center of the wafer. We improved optical performance near the edge 
  of the field of view at the expense of performance at the focal plane center.  Strehl ratios are not shown 
  beyond $3.8\degr$ in radius as these fields are strongly vignetted, causing 
  ray tracing codes to fail. }
  \label{fig:strehl}
\end{figure}

Optical elements inside the receiver are heat-sunk to several distinct temperature stages. 
The vacuum window was at ambient 
temperature, the field lens and the \ac{AHWP} were at liquid helium temperature,  
the aperture stop, pupil lenses, and camera lenses were at 
approximately 1~K, and the focal planes operated near 0.25~K. 
\begin{figure}[ht!]
\begin{center}
\includegraphics[width=0.7\textwidth]{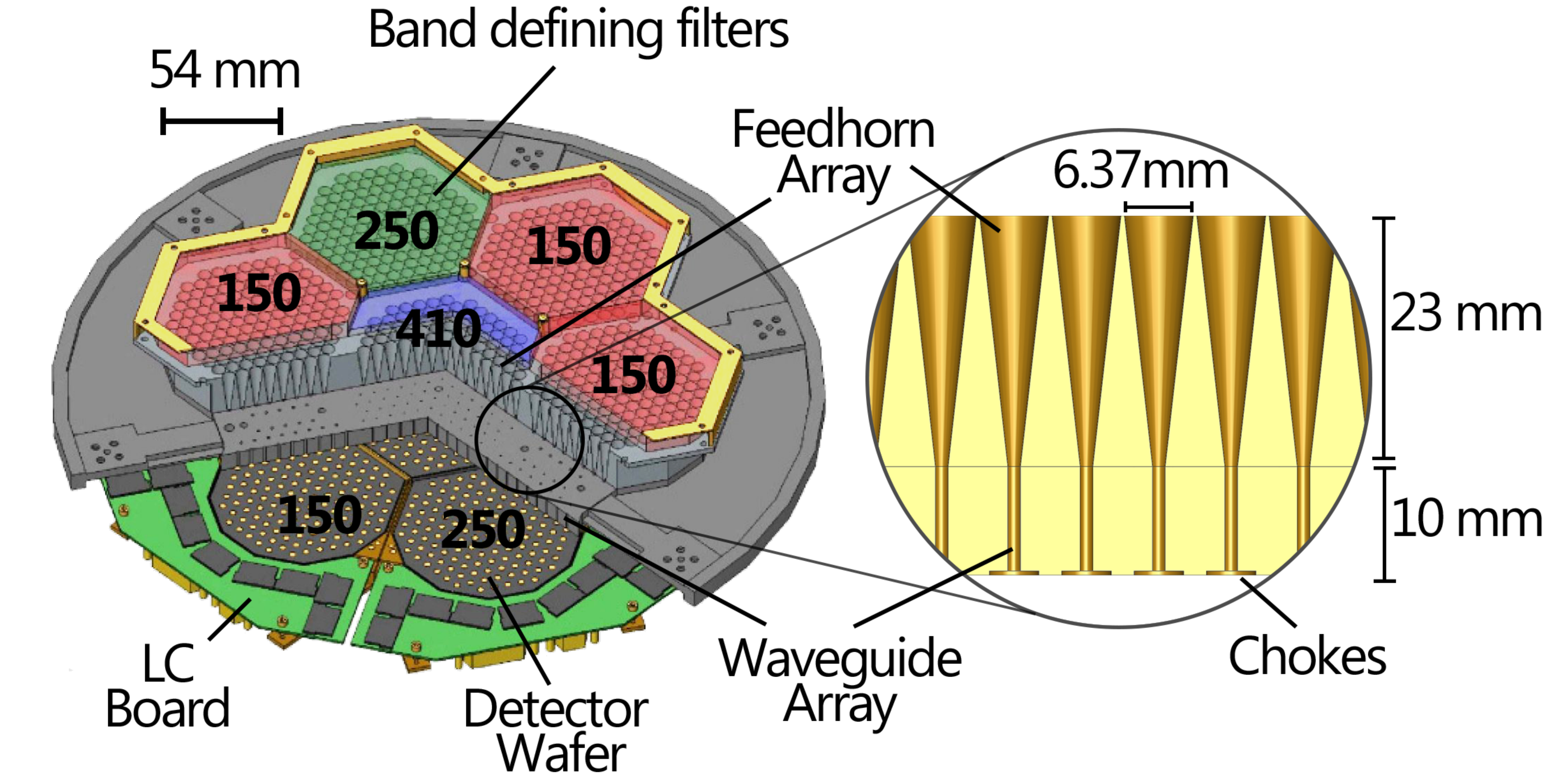} 
\caption{The \ac{EBEX} focal plane (left).  Filters and waveguides defined the observing bands, Section~\ref{sec:filtersandbands}.
         Conical feedhorns coupled the detectors to free space, Section~\ref{sec:coldoptics}.          
         Seven wafers held 141 bolometers each at 150, 250, or 410~GHz. 
         Below the wafers were \ac{LC} boards, the first step in the 
         readout chain.  Detector and readout details in \ac{EP2}. }
\label{fig:focal_plane}
\end{center}
\end{figure}

To ensure consistent material properties, we machined all the lenses
from a single block of ultra-high molecular weight \ac{PE}.  We measured a sample of this \ac{PE} at 
room temperature in a Fourier transform spectrometer and found the index of refraction to be $n = 1.503\pm 0.002$ with 
no detectable birefringence ($\delta n < 0.004$). 
Extrapolation to 4~K using the Lorentz-Lorenz equation with linear contraction between 
1.5 and 2~\% gives a predicted cold index between 1.53 and 1.54. 
We optimized the lens shapes using the ray tracing codes CodeV\footnote{Synopsys Optical Solutions} 
and ZEMAX\footnote{OpticStudio from ZEMAX}. 
The optimization constrained solutions to a flat and telecentric focal plane while maximizing the \ac{DLFOV}
so as to accommodate as large a number of detectors as possible. Due to an error, 
we used $n=1.52$ in the optimization. We discuss the consequences of this error in Section~\ref{sec:beams}. 
All designed lens surfaces were conic sections of revolution. With $n=1.52$ the designed lenses provided a \ac{DLFOV}
with Strehl ratio above 0.8 across the entire focal plane, as shown in Figure~\ref{fig:strehl}. The total throughput
of the optical system for each focal plane was 115 cm$^{2}$sr. 


Each of the two \ac{EBEX} focal planes consisted of a layer of band defining electromagnetic filters, a monolithic
array of feedhorns attached to a monolithic array of waveguides, 7 detector wafers, wafer holders 
and ``\ac{LC} Boards", and a back-cover which, together with the array of waveguides, completed a Faraday cage around the detectors; 
see Figure~\ref{fig:focal_plane}. The electromagnetic filters and waveguides defined three frequency bands
centered on 150, 250, and 410~GHz. The focal plane was arranged such that 4 wafers operated
at 150~GHz, 2 at 250~GHz, and 1 at 410~GHz. The \ac{LC} boards were part of the multiplexed frequency domain 
bias and readout of the detectors, which is discussed in more detail in \ac{EP2}. Each detector wafer had 128 wired 
detectors which were biased and read out with 8 pairs of wires. 

The array of smooth-walled feedhorns coupled the radiation from free space to an array of waveguides with chokes; 
see Figure~\ref{fig:focal_plane}. The horns were truncated cones with an entrance diameter of 6.37~mm and 
a length of 23~mm for all frequency bands. The exit diameter was 1.32, 0.81, and 0.48~mm for the 150, 250, and 
410~GHz bands, respectively, matching the waveguide diameters immediately below the feedhorns. The waveguide
plate was 10~mm thick. We spaced the horns and detectors at 1.74~f$\lambda$ for 150~GHz, which was 6.626~mm. 
The horn length maximized the 410~GHz band gain with a moderate reduction in the gain of the other two 
bands~\citep{king50}.  The predicted gain at 150 and 250~GHz was 84\% and 93\% of the maximum, respectively.
We machined the feedhorn array and waveguide array each from a single piece of aluminum. 
Before machining they were aligned with dowel pins and bolted together.
The horns were machined with a custom cutter which marked the waveguide centers. Then the horn array was removed, the 
waveguides were drilled, and both pieces were plated with a 0.127~\micron~gold layer.  The horn array was then 
reattached to the waveguides using the dowel pins to ensure alignment.
We optimized the optics for the 150~GHz band by aligning the 150~GHz feedhorns to the 
focus of the telescope optics. This defocused the 250 and 410 GHz feedhorns.  The predicted loss in coupling 
efficiency at 410~GHz, where the defocusing was the largest, was only 3\%.

\subsection{Telescope Alignment} 
\label{sec:telalignment}


The focal planes were the reference for the alignment of the entire optical system. They 
had no adjustable degrees of freedom for motion, and thus all optical elements were aligned to them, specifically to focal plane H. 
We made no attempt to align the two focal planes relative to each other and relied on their common, rigid mechanical construction. 
We also made no attempt to ensure that pairs of detectors at the two focal planes simultaneously observed 
the exact same sky location.  The combination of a rotating \ac{AHWP} and wire grid made each focal plane detector 
an independent polarimeter sensitive to the incident $I, Q,$ and $U$ Stokes parameters. 

The lenses and polarizing grid were mounted on custom-made adjustable supports
that gave a dynamic range of 3~mm and $1\degr$ in lateral and rotational positions.  They were 
aligned to the focal planes using a portable \ac{CMM}.\footnote{Microscribe MX} 
The alignment took into account the few~mm of differential thermal contraction of the receiver elements including 
the receiver shells, the Vespel legs that mount the optical elements to the 4~K cold plate, and the \ac{PE} from which 
the lenses were made~\citep{Zilic_thesis}. 

We transferred the alignment of the internal optics to a reference frame defined by three tooling balls
mounted on the cryostat shell near the vacuum window. This reference frame was used to align 
the secondary mirror to the receiver, and subsequently the primary mirror was aligned to the secondary. 
The mirrors were mounted to the inner frame of the gondola using custom-built hexapods with 
a 25~mm lateral and $3\degr$ rotational dynamic range.  The primary and secondary mirrors each had 
three tooling balls in known locations relative to the reflector surface. After the initial mounting of the 
secondary mirror, we used an inside micrometer to measure the nine relative distances 
between the receiver and secondary tooling balls and compared them to the distances required by the optical design. 
The repeatability of the inside micrometer's measurements was 0.075~mm.  We used a computer program to calculate 
the hexapod leg lengths that would bring the mirror to its required position.  
The hexapod legs were adjusted manually, locked in position, and the distances were re-measured to verify proper alignment.  
The procedure was repeated for the alignment of the primary mirror relative to the secondary. Post-alignment, the nine distance measurements 
between the secondary and the receiver (the primary and the secondary) were between 25 and 175 (25 and 350)~$\mu$m from the 
design values, with an RMS of 110 (180)~$\mu$m. 

The final nine distances measurement were used to recreate the most probable relative positioning of the optics. That geometry 
was programmed into CodeV, and the beam sizes were compared to the nominal alignment.  The 250 and 410~GHz beams were expected to 
grow by 10\% and there was no change expected for the 150~GHz beams. 

\subsection{Anti-Reflection Coatings} 
\label{sec:lensarc}

\begin{figure}[ht!]
  \centering
  \includegraphics[width=.6\textwidth]{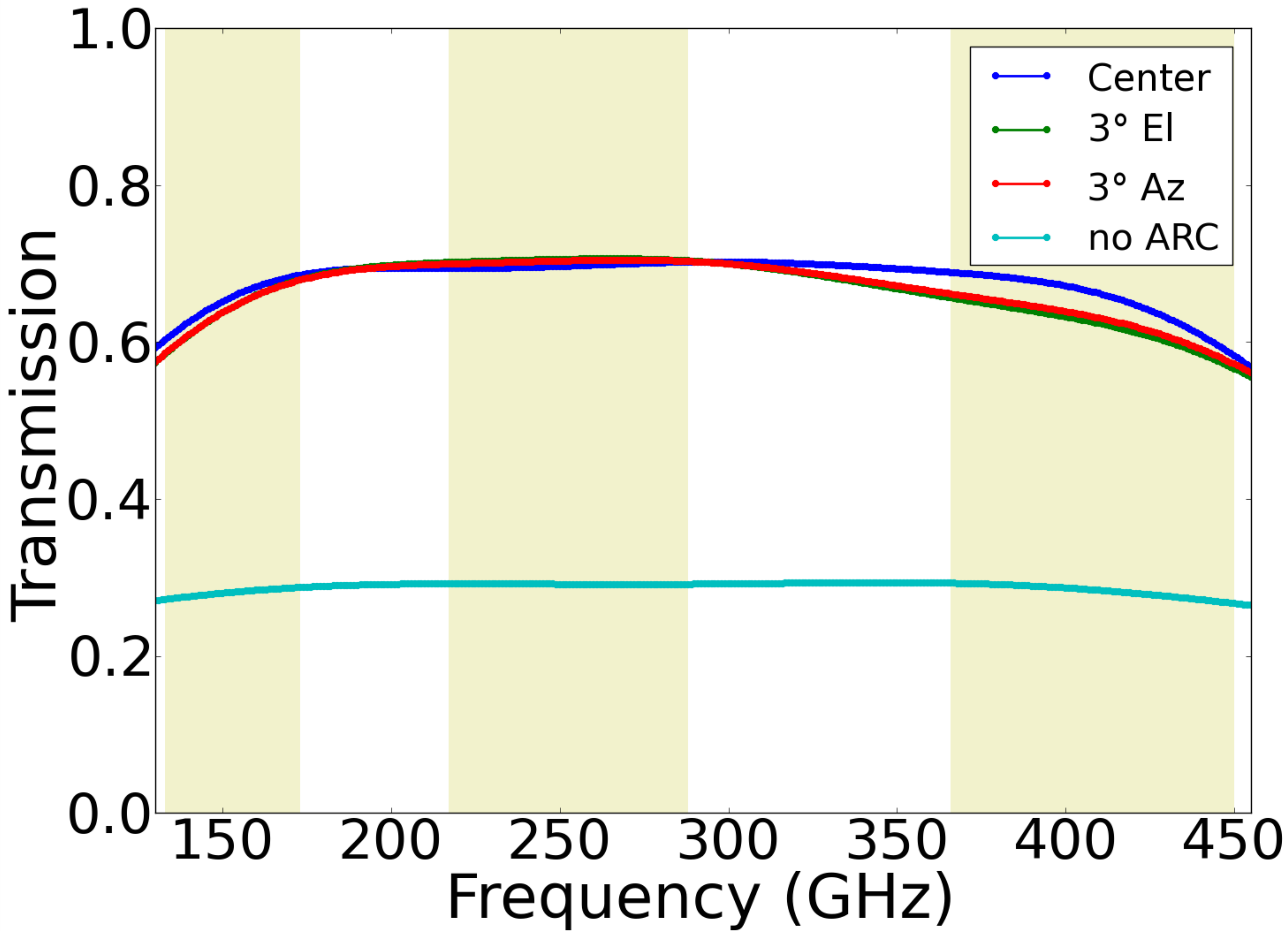}
  \caption{Total transmission as a function of frequency (red, green, blue) including the \ac{ARC} applied on 
  the vacuum window, lenses, filters, and 
  AHWP, compared to the transmission without \ac{ARC} (cyan). We used CodeV to calculate the $II$ Mueller
  matrix element at three locations on the focal plane: 
  the center (blue), center top (green), and center right (red). Absorption is not included, as reliable information about 
  the absorption at cryogenic temperatures was not available.  The vertical bars (khaki) show the \ac{EBEX} frequency bands. 
  \label{fig:arc} }
\end{figure}

The \ac{PE} lenses and vacuum window were coated with microporous Teflon\footnote{Porex, U.K.} to 
provide a broad-band \ac{ARC}. The Teflon had an index of refraction $n=1.23$.  
A 200~$\mu$m thick layer was bonded to the sky-facing side of each lens and a
400~$\mu$m thick layer was bonded to the focal plane side. For the vacuum window 
the thicknesses were 180~\micron~and 460~\micron, respectively. Figure~\ref{fig:arc} 
gives the calculated end-to-end transmission as a function of frequency including all 
optical elements.  

We chose the \ac{ARC} thicknesses of the porous Teflon by ray-tracing a range of available thicknesses and choosing 
the combination that produced the lowest instrumental polarization. 
For each coating option we calculated the cumulative Mueller matrix of the optical system
traced from the sky to, but not including, the \ac{AHWP}.
This Mueller matrix was averaged across the entrance pupil and over five frequencies within each band.
We choose the \ac{ARC} that minimized instrumental polarization $\sqrt{IQ^2 + IU^2}$ 
at the azimuth extreme of the FOV, where
$IQ$ and $IU$ are Mueller matrix components.
The design provided maximum instrumental polarization at
150, 250, and 410~GHz of 1.2\%, 1.8\%, and 0.5\%, respectively. The instrumental 
polarization was dominated by the tilt and curvature of the field lens. 

\begin{figure}[h]
  \centering
  \includegraphics[width=.8\textwidth]{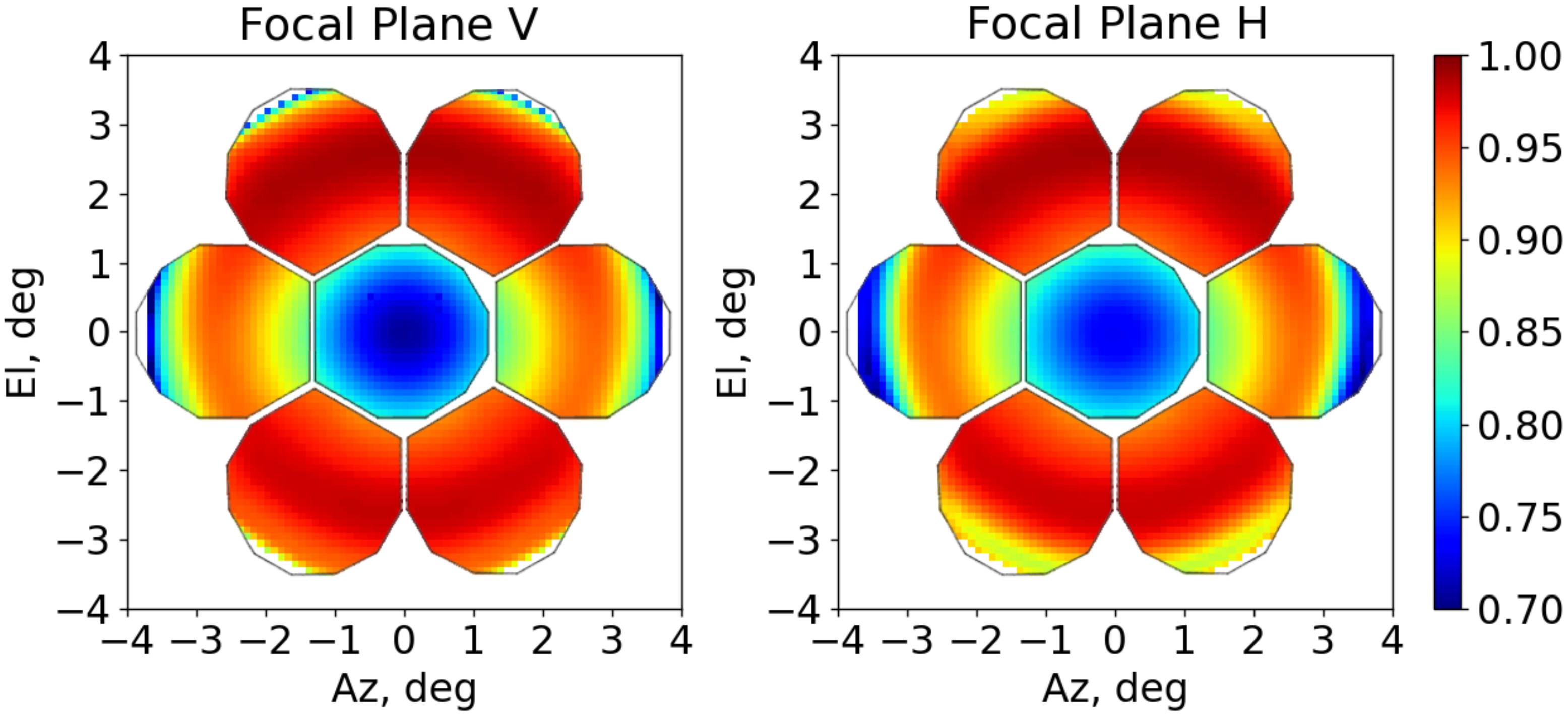}
  \caption{Strehl ratios at the focal plane as calculated using the measured 
  shape of the lenses after applying the \ac{ARC}. 
  The change in Strehl differed between the focal 
  planes because each focal plane had a different camera lens. }
  \label{fig:strehl_change}
\end{figure}

The Teflon \ac{ARC} was heat-bonded to the \ac{PE} lenses. 
During this process the lenses distorted relative to their machined, designed shape.  
We measured the final lens shapes including the \ac{ARC} and ray-traced the final optical system.  
Figure~\ref{fig:strehl_change} shows the effect on the Strehl ratios; compare to Figure~\ref{fig:strehl}. 
The average change in Strehl ratio was a decrease of 0.02, 0.03, 0.11 for the 150, 250, and 410~GHz 
bands, respectively.

\begin{figure}[htp]
  \begin{center}
     \includegraphics[width=0.85\textwidth]{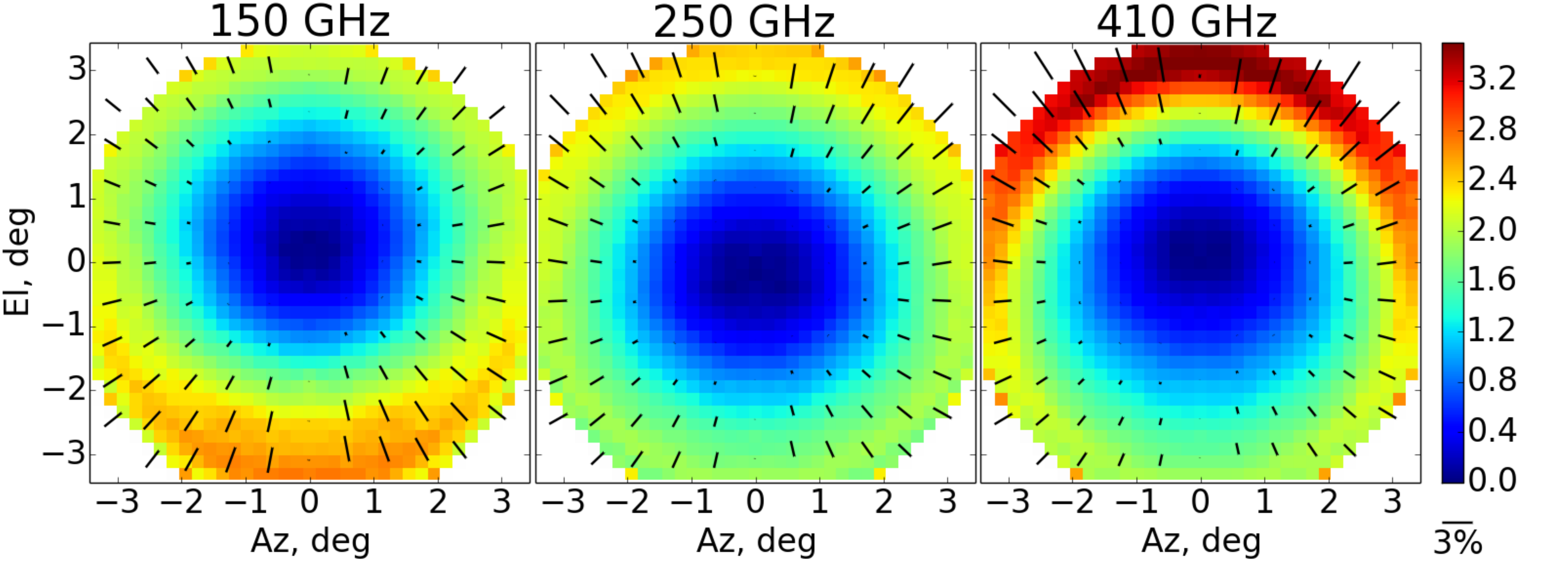}
  \end{center}
  \caption{Magnitude and orientation of calculated \ac{IP} for the \ac{EBEX} optics up to and including 
  the field lens.  Orientation is indicated by the polarization vectors (black bars). Both the color scale and 
  the length of the polarization vectors give the \ac{IP} magnitude.
  \label{fig:focal_plane_IP} }
\end{figure}

Figure~\ref{fig:focal_plane_IP} gives the calculated instrumental polarization of the final 
optical system including optical elements up to and including the field lens. The field 
lens is the dominant contributor to the instrumental polarization and this is represented
by the radial orientation of the polarization vectors and the increase in magnitude with radius. 


\subsection{Electromagnetic Filters and Frequency Bands}
\label{sec:filtersandbands}

\begin{figure}[ht]
\begin{center}
\includegraphics[width=0.6\textwidth]{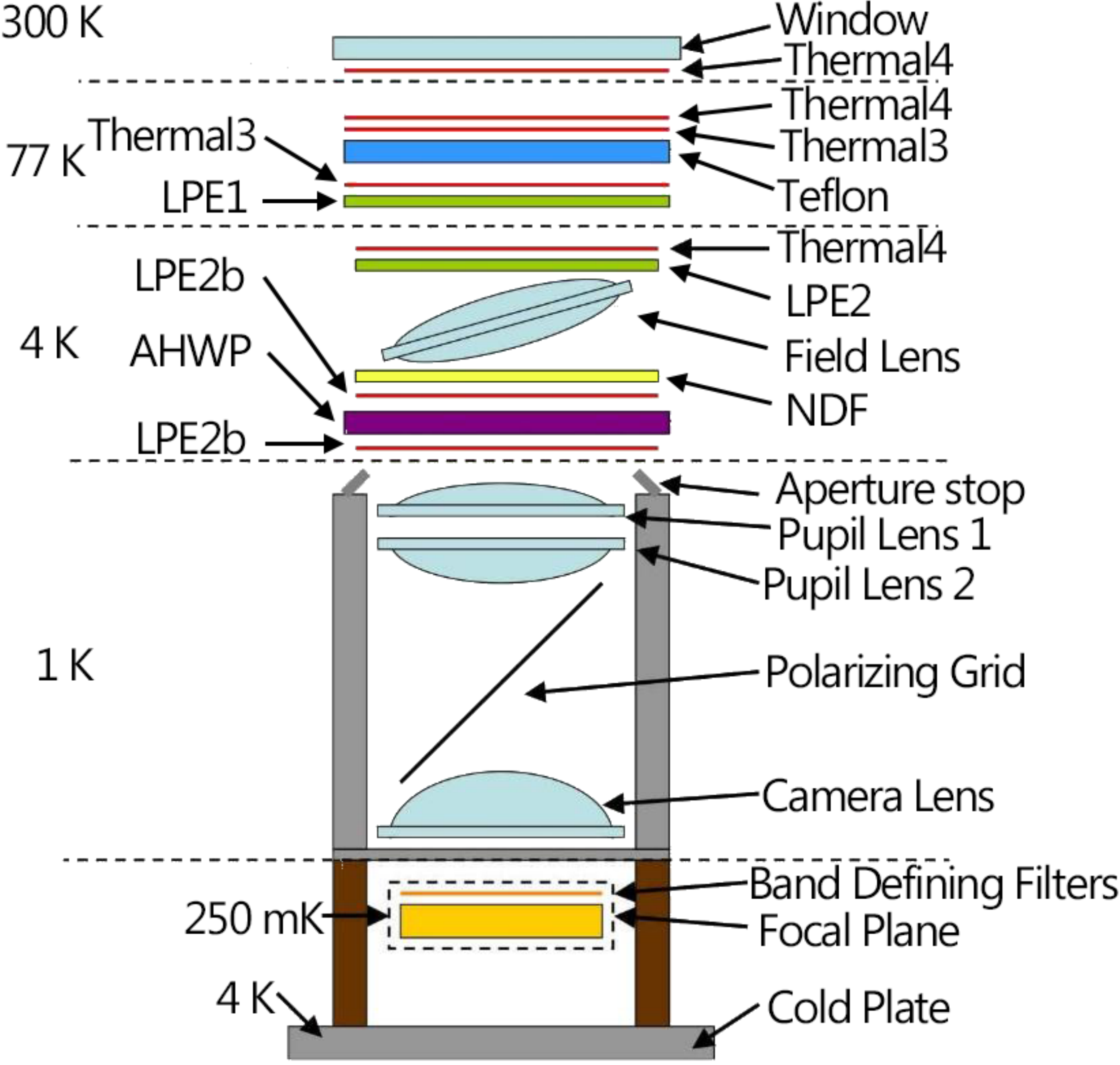} 
\caption{Ordering of filters along the optical path and their thermal stages.  }
\label{fig:filters}
\end{center}
\end{figure}

A set of reflective and absorptive low pass filters, as shown in Figure~\ref{fig:filters} and Table~\ref{tab:band_filters}, 
together with the horn-array waveguides were dominant in determining the transmission properties of the instrument. 
We used metal mesh low-pass filters~\citep{Ade:metal_mesh}, an absorptive Teflon filter, and a neutral density filter 
that was used only for ground operations. 

\begin{figure}[ht!]
\begin{center}
\includegraphics[width=0.6\textwidth]{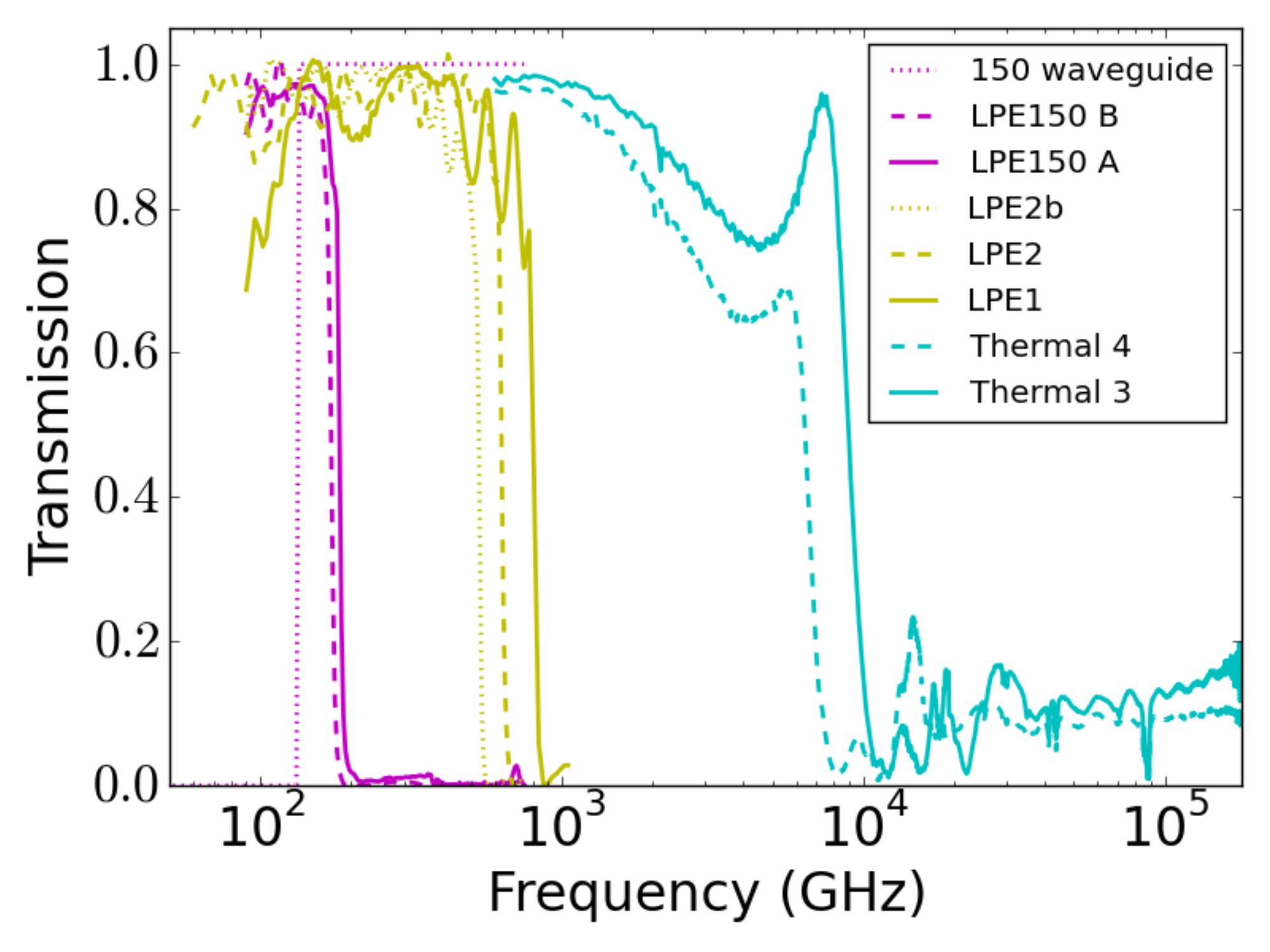} 
\caption{Measured transmission curves for the low-pass common filters, and, as an example, for the additional band-specific 150 GHz band filters. 
The high-pass response shown for the 1.32~mm diameter waveguide of the 150~GHz band is based on calculations. }
\label{fig:filter_trans}
\end{center}
\end{figure}

We used two types of metal mesh low-pass filters, a `thermal' and `\ac{LPE}'.
The thermal filters were 10~\micron~thick polypropylene 
with a copper mesh layer sized to reflect frequencies in the infrared band.  
The \ac{LPE} filters were made up of multiple layers of single mesh layers embedded in polypropylene 
with microporous Teflon\footnote{Porex, U.K.} antireflection coating.
Three types of \ac{LPE} filters were placed sky-side of the polarizing grid to reject thermal radiation and 
two \ac{LPE} filters per frequency band were placed on each focal plane to define the high frequency side of each of the bands. 
The 3~dB cutoff for each of these filters is given in Table~\ref{tab:band_filters} and the filter transmission curves 
for the 150~GHz filters are shown in Figure~\ref{fig:filter_trans}. 
The low-frequency edge of each of the three \ac{EBEX} frequency bands was set by circular waveguides positioned between
the horns and the detector wafers; see Figure~\ref{fig:focal_plane}. The waveguides had diameters of 1.32, 0.81, and 0.48~mm 
for the 150, 250, and 410 GHz bands, respectively, giving the turn-on frequencies listed in Table~\ref{tab:band_filters}.  
The location and cutoff of the high-frequency 
edge of the bands due to the \ac{LPE} filters made the TE$_{11}$ mode dominant. 
Convolving the transmission of the \ac{LPE} filters with the transmission function for the TM$_{01}$ mode, which is 
the next most dominant mode, we find that it contributed 
1.4\%, 4.2\% and  0.07\% for the 150, 250, and 410~GHz, respectively, relative to the TE$_{11}$ mode. 
Contributions from higher modes was negligible. 

\begin{table}[ht!]
\begin{center}


\begin{tabular}{c c} 
\begin{tabular}[t]{| c | c |}
\multicolumn{2}{c}{Common Filters}\\
\tableline
Name      & 3 dB [GHz] \\
\tableline
Thermal 3 & 8930 \\
\hline
Thermal 4 & 6400 \\
\hline
LPE1 & 803      \\
\hline
LPE2 & 631  \\
\hline
LPE2b & 531   \\
\tableline
\end{tabular} & 
\begin{tabular}[t]{| c | c | c |}
 \multicolumn{3}{c}{Band Filters}\\
\tableline
Band [GHz] & Name & 3 dB [GHz] \\
\tableline
     & LPE150 A & 183 \\
\cline{2-3}
150  & LPE150 B & 172 \\
\cline{2-3}
     & 150 waveguide & 133 \\
\hline
     & LPE250 A & 337 \\
\cline{2-3}
250  & LPE250 B & 285  \\
\cline{2-3}
     & 250 waveguide & 217 \\
\hline
     & LPE410 A & 558   \\
\cline{2-3}
410  & LPE410 B & 445   \\
\cline{2-3}
     & 410 waveguide & 364   \\
\hline
\end{tabular} \\
\end{tabular}

\caption{List of the filters common to the entire optical path and those specific for each frequency band. The `3~dB' column 
gives the 3~dB cutoff and turn-on of the low- and high-pass filters, respectively. } 
\label{tab:band_filters}
\end{center}
\end{table}

The absorptive IR filter was a 12.7~mm thick slab of Teflon 
heat sunk to the liquid nitrogen stage. We chose Teflon for its low index of refraction and strong infrared but relatively 
low mm-wave absorption. However, because of its low thermal conductance the central region of the filter reached 
temperatures of 110~K, making emission 
from this filter a significant fraction of the total calculated optical load on the detectors; more details are provided in EP2. 

\begin{figure}[ht!]
\begin{center}
\includegraphics[width=0.55\textwidth]{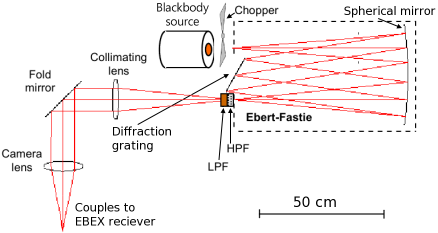} 
\caption{Diagram of the spectral response measurement using the Ebert-Fastie spectrometer.  The diffraction grating was rotated to send 
         a specific band of frequencies through the exit aperture. At the exit aperture, high- and low-pass filters (HPF, LPF) 
         selected a single diffraction order coming 
         from the grating.  The Ebert-Fastie was mounted on top of the EBEX receiver (not shown) and two lenses coupled 
         the output to the EBEX optics, illuminating a single detector on the focal plane.}
\label{fig:ef_diagram}
\end{center}
\end{figure}

For ground work we added a \ac{NDF} in the focal-plane side of the field lens.  
Without the \ac{NDF} the atmospheric load on the ground would saturate the detectors in all bands.
The \ac{NDF} was made from a slab of Eccosorb MF110\footnote{Emerson and Cuming Microwave Products, Inc.} 
machined down to produce seven hexagonal segments of different thicknesses.  Since the NDF was near a field stop, these 
seven segments corresponded to the seven wafers on the focal plane.  The central section, corresponding to the 410~GHz wafer, 
was 6.6~mm thick while the 250~GHz and 150~GHz segments were 10.8~mm and 18.3~mm thick, respectively.  
The \ac{NDF} was coated with a 0.5~mm layer of Teflon. The predicted transmission of the \ac{NDF} was 1.4\%, 1.3\% and 1.6\% 
at the 150, 250, and 410~GHz bands, respectively. 

\begin{figure}[ht!]
\begin{center}
\includegraphics[width=0.6\textwidth]{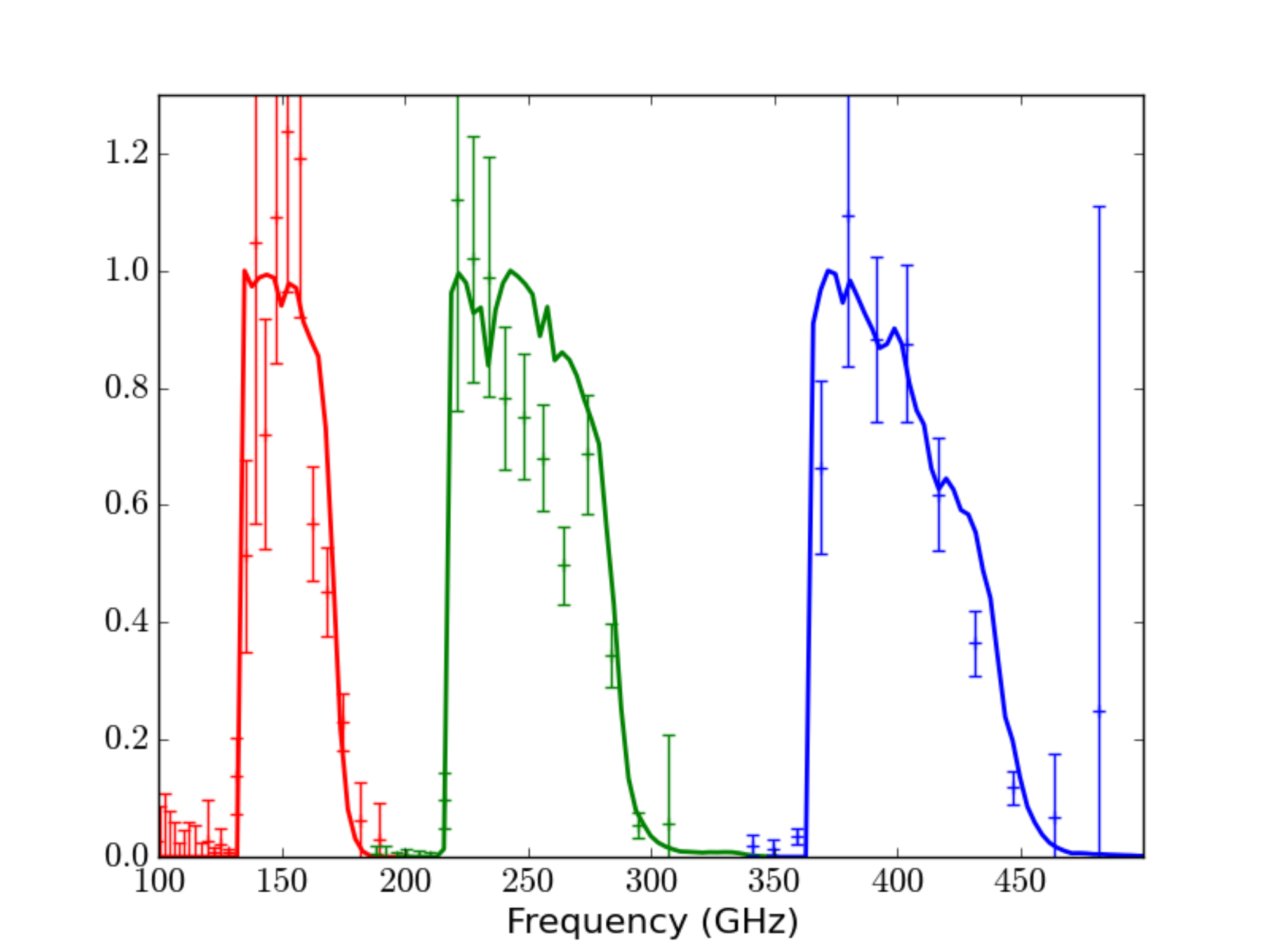} 
\caption{Measured (points) and predicted (solid lines) bands at each \ac{EBEX} frequency. The native measurements were made 
with higher frequency resolution and were binned to reduce clutter. The predictions are normalized to 1.0.}
\label{fig:band_measured}
\end{center}
\end{figure}

We measured the end-to-end frequency response of the instrument using an Ebert-Fastie 
spectrometer~\citep{fastie_1952a,fastie_1952b,Dan_thesis,Zilic_thesis} 
that had a 1200~K black body and a chopper as an input source. The output radiation was coupled to the 
receiver only -- without the warm telescope -- specifically to the  
throughput of individual focal plane detectors using lenses and a fold mirror; see Figure~\ref{fig:ef_diagram}. 
A translation stage was used to couple the spectrometer to between 10 and 12 detectors from each frequency band. 
The 12~mm exit aperture of the spectrometer gave an output frequency bandwidths between 1.3 
and 3.0~GHz, 2.4 and 5.1~GHz, and 4.1 and 7.0~GHz for the 150, 250, and 410~GHz bands, respectively. 
We measured the frequency response of the detectors with a resolution that was approximately half 
the width of the output bandwidth. For each frequency band we find an average response as a function of frequency 
by averaging individual spectra weighted by signal-to-noise. We then least squares minimize the average 
and the predicted responses with only an overall scaling as the free parameter; during that step 
the highest predicted response is normalized to 1. The measured bands are shown in 
Figure~\ref{fig:band_measured} and are interpreted as the end-to-end band shapes because within these 
narrow bands, the ambient temperature mirrors are achromatic. 


\subsection{Beams}
\label{sec:beams}

Viewed in the time reversed sense, the focal plane feedhorns launch antenna patterns that propagate through the 
optical system into a far field pattern. In addition to the horns, the polarizing grid and the aperture stop are central 
in determining the far field shape and size of the beam. The polarizing grid breaks the rotational symmetry of the 
horn, and therefore the beams are inherently elliptical. As the aperture stop is an image of the primary mirror, the Fourier
transform of its illumination gives the far field pattern. 

A rigorous prediction of the far field antenna pattern requires physical optics calculation including all 
elements in the optical path. We do not have the capability to carry out this calculation. 
We make approximate predictions by simulating the antenna pattern of the horn, calculating 
the shape of its illumination on the stop, and carrying out the Fourier transform to find the far field beams. 
To simplify the calculation of the 
Fourier transform we use a circularly symmetric Gaussian pattern that is the best fit to the elliptical pattern. 
The predictions for the illumination on the stop and for the far field beams 
are given in Table~\ref{tab:beams}. The table shows that at 150~GHz a significant fraction 
of the beam is intercepted by the cold stop. This is the main reason we kept the stop at temperature 
below that of the sky. 
\begin{table}[ht!]
\begin{center}
\begin{tabular}{| c | c | c || c | c | c | c |}
\tableline
 Band   &  \multicolumn{2}{| c ||}{Predicted } & \multicolumn{4}{c |}{Measured (')} \\
             &   Taper    &    FWHM        &   \multicolumn{2}{c |}{One} &  \multicolumn{2}{c |}{Average} \\
 (GHz)    &  (dB) & (')  &  FWHM$_a$   & FWHM$_b$ &   FWHM$_a$   & FWHM$_b$     \\
\tableline
150 & -7.2     & 7.8      &  8.6  & 7.6 &  8.9  & 8.3   \\
250 & -19.4    & 5.8     &   8.2  & 6.4 &  7.6  & 7.4 \\
410 & -50.1   & 5.0     &  8.6  & 6.5 &  11.9  & 10.1  \\
\tableline
\end{tabular}
\caption{ Predicted and measured beam \ac{FWHM} for 
the long ($_{a}$) and short ($_{b}$) axes. The `one' column gives the parameters for the 
beams measured for one detector in each frequency band, as shown in Figure~\ref{fig:beam_maps}. The  
`average' is for a signal-to-noise weighted mean of all the detectors measured per frequency band. 
The parameters measured are for the long and short FWHM
of a 2-dimensional Gaussian fit. 
\label{tab:beams} }
\end{center}
\end{table}

\begin{figure}[htp]
  \begin{center}
    \subfigure{\label{fig:150_beammap}
      \includegraphics[width=0.3\textwidth]{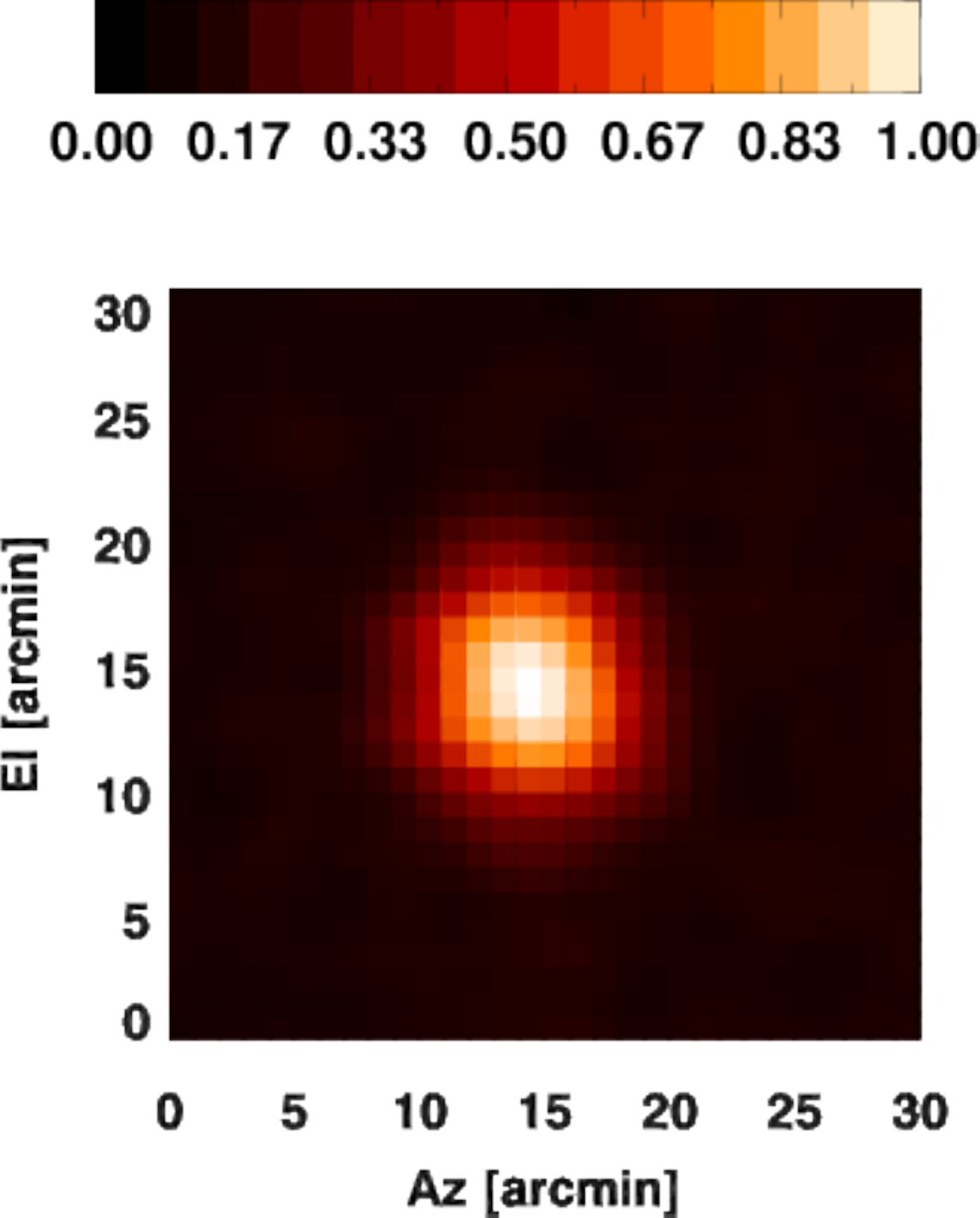} }
    \subfigure{\label{fig:250_beammap}
      \includegraphics[width=0.3\textwidth]{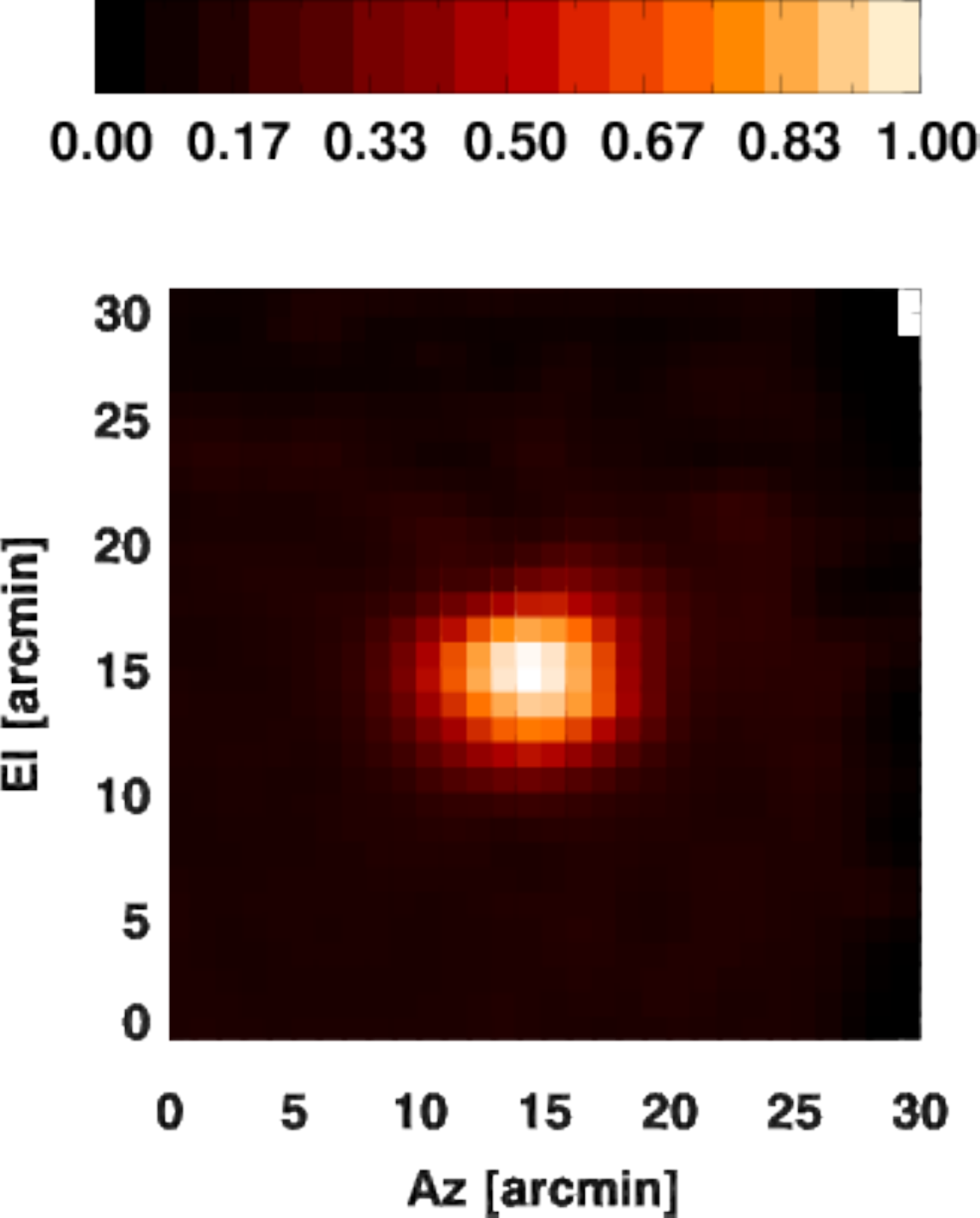} }
    \subfigure{\label{fig:410_beammap}
      \includegraphics[width=0.3\textwidth]{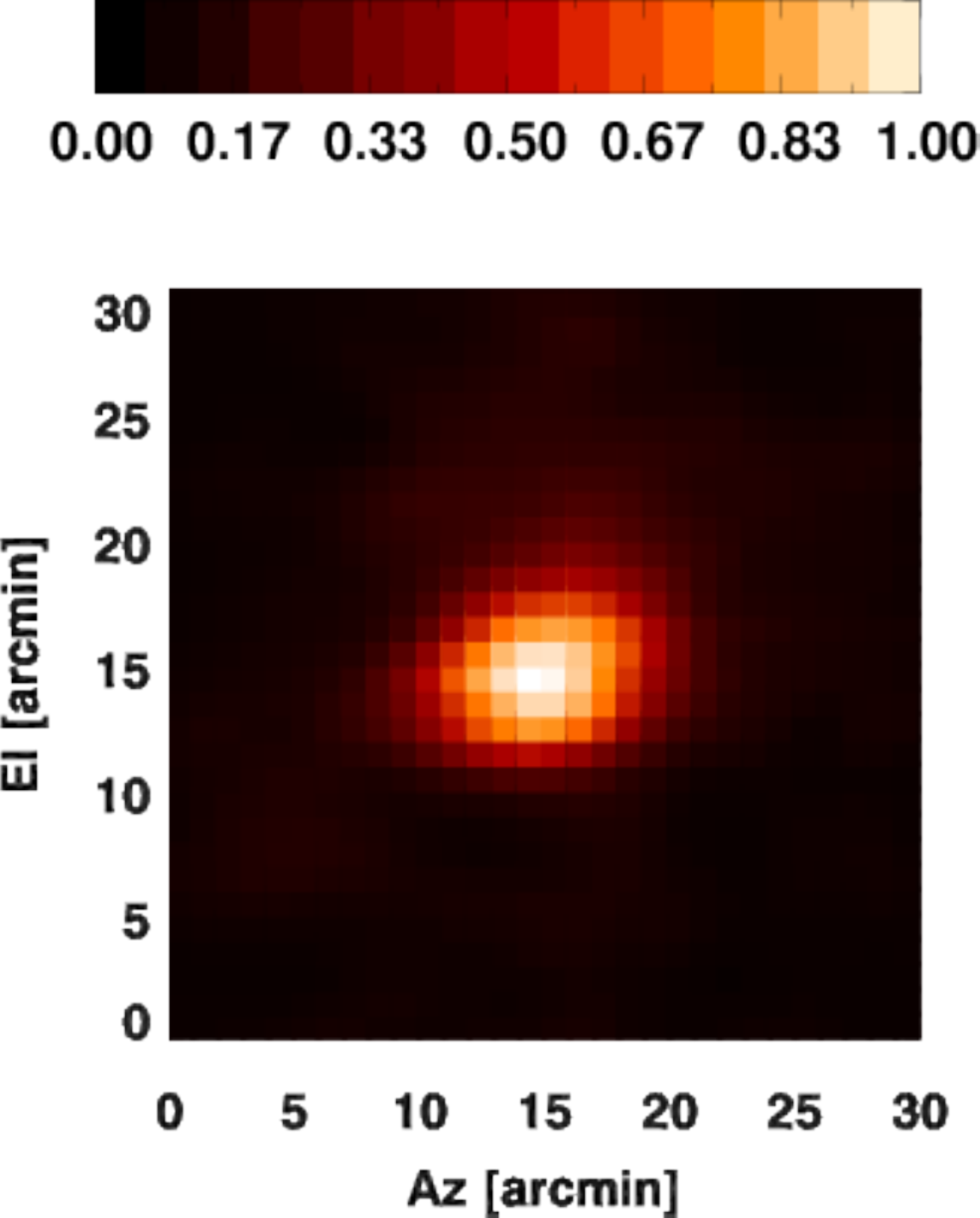} }
  \end{center}
  \caption{Example ground calibration beam map from one detector for each of the 150 (left), 250 (middle), and 410~GHz (right) 
  frequency bands.  
  \label{fig:beam_maps}  }
\end{figure}

\begin{figure}[htp]
  \begin{center}
    \subfigure{\label{fig:150_aggregate_map}
      \includegraphics[width=0.3\textwidth]{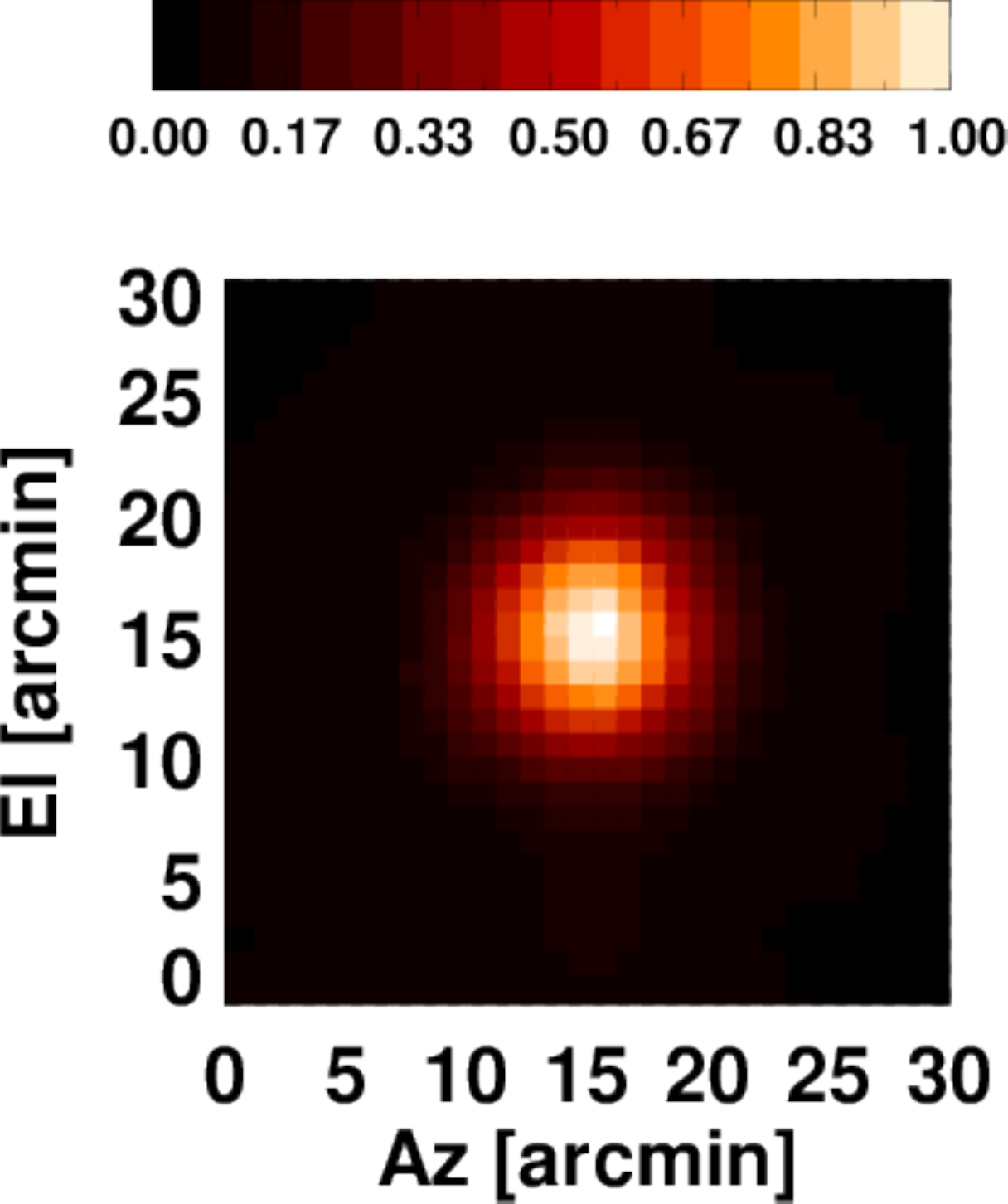} }
    \subfigure{\label{fig:250_aggregate_map}
      \includegraphics[width=0.3\textwidth]{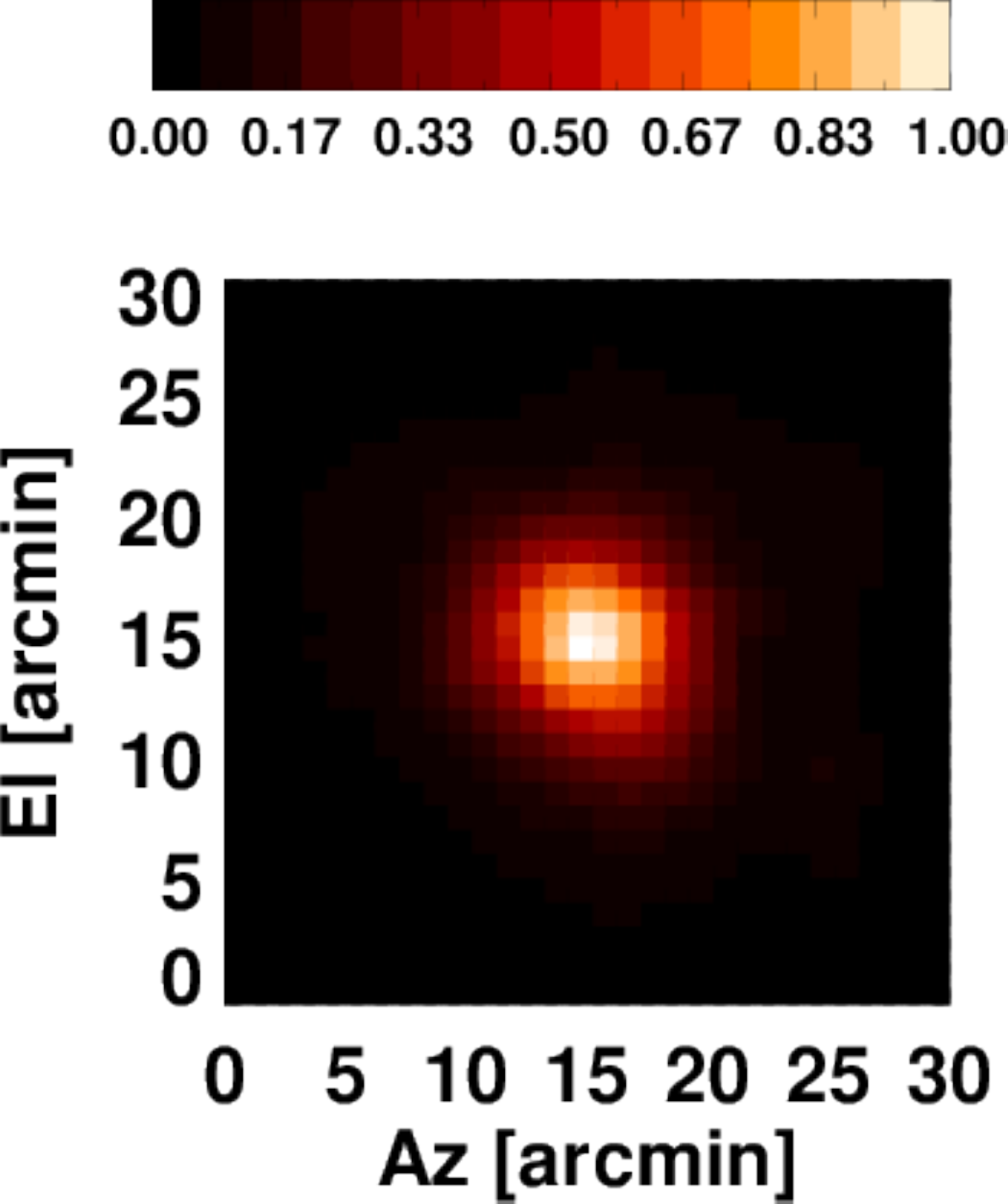} }
    \subfigure{\label{fig:410_aggregate_map}
      \includegraphics[width=0.3\textwidth]{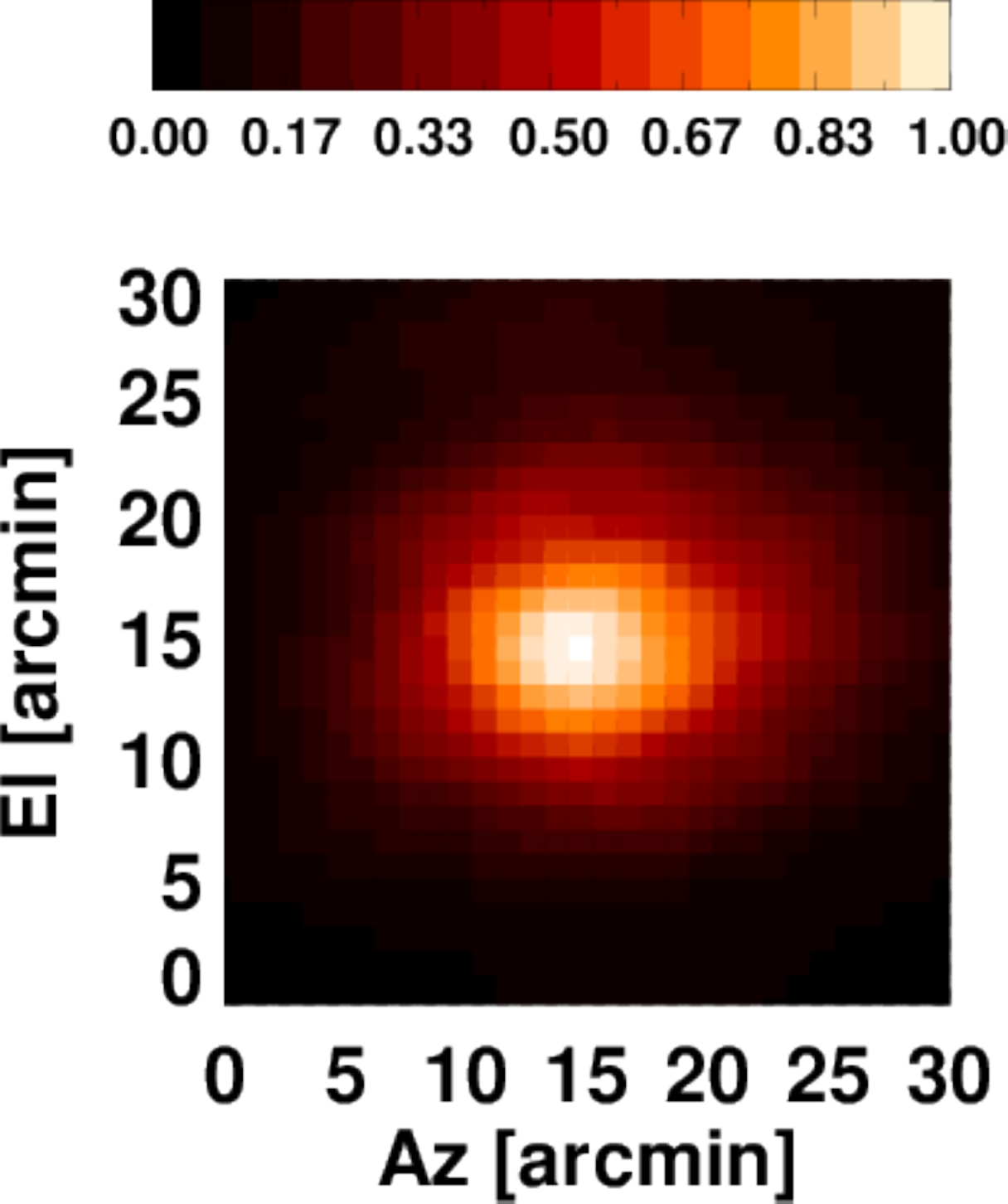} }
  \end{center}
  \caption{Signal-to-noise weighted ground calibration beam maps made from beam maps of all detectors of 
  each frequency band (left to right: 150, 250, and 410~GHz). 
  \label{fig:beam_aggregate_maps}  }
\end{figure}

We measured the beam size and shape of the \ac{EBEX} optical system on the ground and in-flight. 
On the ground we used a Gunn diode with a modulating power source and tunable between 125 and 
140~GHz. It was mounted on a water tower 
that was 50~m high and 104~m horizontal distance from the payload, giving a total distance of 115~m. 
The source had a wire grid polarizer at the exit aperture 
to ensure highly polarized emission. 
For the 150~GHz band measurements the source frequency was set to 140~GHz. We used a doubler 
and a tripler to set the source to 254 and 410~GHz for the two higher frequency bands, respectively.  
We raster-scanned the source while simultaneously 
rotating the \ac{AHWP} to make detector time streams that had three modulations: the raster scan period, 
the rotation of the \ac{AHWP}, and 
the on/off modulation of the source. Using the double demodulation analysis technique described in 
Appendix~\ref{sec:polcaleqns} we made intensity maps of the source with a subset of the detectors. 
We fitted the measured antenna response of 
each bolometer to a 2-dimensional Gaussian and extracted the two FWHMs; Figure~\ref{fig:beam_maps}
shows a beam map for one of the detectors at each frequency band.  
To give an indication of the average beam per frequency band, Figure~\ref{fig:beam_aggregate_maps} 
shows a signal-to-noise ratio weighted map made from all the beam maps of all detectors at each frequency band.  
Table~\ref{tab:beams} gives the measured sizes of the beams. We find that the beams at the higher frequency 
bands are larger than the design. We ascribe this difference to the error in index (see Section~\ref{sec:coldoptics}) 
and to a slight misalignment of the telescope, which would affect the higher frequencies more than the 150~GHz band. 

The in-flight beam size is inferred from maps made of passes of the galactic source RCW38. 
Because of a malfunction with the azimuth motor, described in \ac{EP3}, many detectors have only few 
passes in the vicinity of the source and we can not reliably reconstruct their beam shapes. 
We therefore construct one effective temperature beam map per frequency band using 
all detectors for which we have valid absolute calibration. 
(The absolute calibration is discussed by~\citet{Aubin_MGrossman2015}.)
The 150, 250, and 410~GHz maps use data from 331, 231, and 80 detectors, respectively. The data maps 
are compared to \planck\ temperature reference maps that are made 
with various smoothing scales, as described below. We deduce the beam size from the reference map 
that best fits the data. 

For the 250 and 410~GHz bands we generated the \planck\ reference temperature maps by summing the \planck\ component 
maps, which have been scaled to and integrated over the measured EBEX bands~\citep{PlanckPaper10}.  
For the 150~GHz band we used the \planck\ 143~GHz map. At this frequency the \planck\ component 
map reconstructs RCW38 poorly, and the 143~GHz data is sufficiently close to the \ac{EBEX} 150~GHz band. 
We made a bank of reference maps by smoothing the \planck\ temperature map with a two-dimensional symmetric Gaussian 
to a range of scales from 8 to 40~arcmin. The pixelization was 1.7~arcmin. 
Using the \acused{EBEX2013}\ac{EBEX2013} pointing information we created detector-specific simulated time-ordered 
data by scanning the bank of reference maps. The time ordered data was subject to the same filtering 
and processing as the corresponding \ac{EBEX2013} flight detector data. 
We compared the measured data and each of the simulated maps for a square region that is 
$1.2\deg \times 1.2\deg$ around RCW38. 

\begin{figure}[htb]
  \begin{center}
      \includegraphics[width=0.3\textwidth]{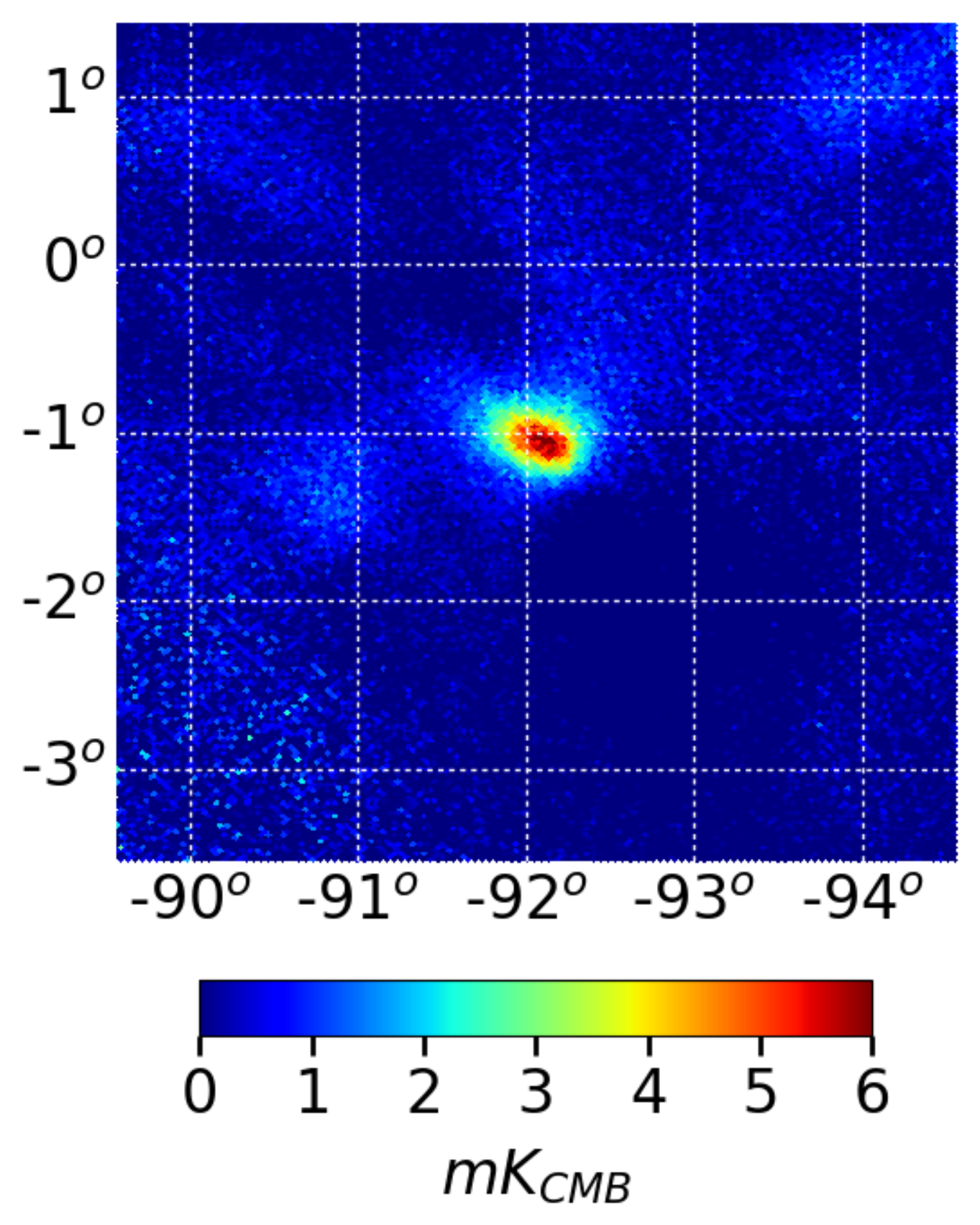}
      \includegraphics[width=0.3\textwidth]{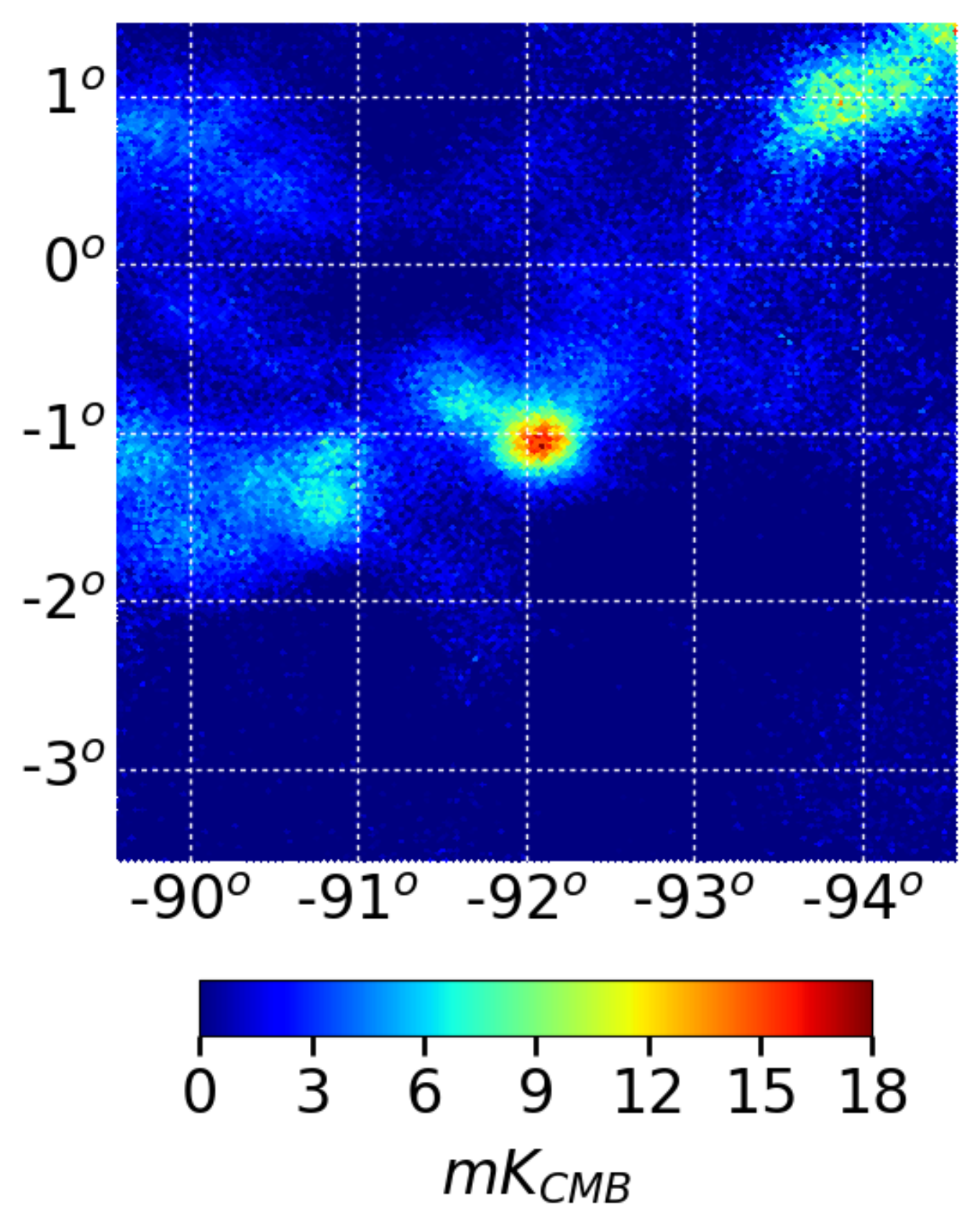}
      \includegraphics[width=0.3\textwidth]{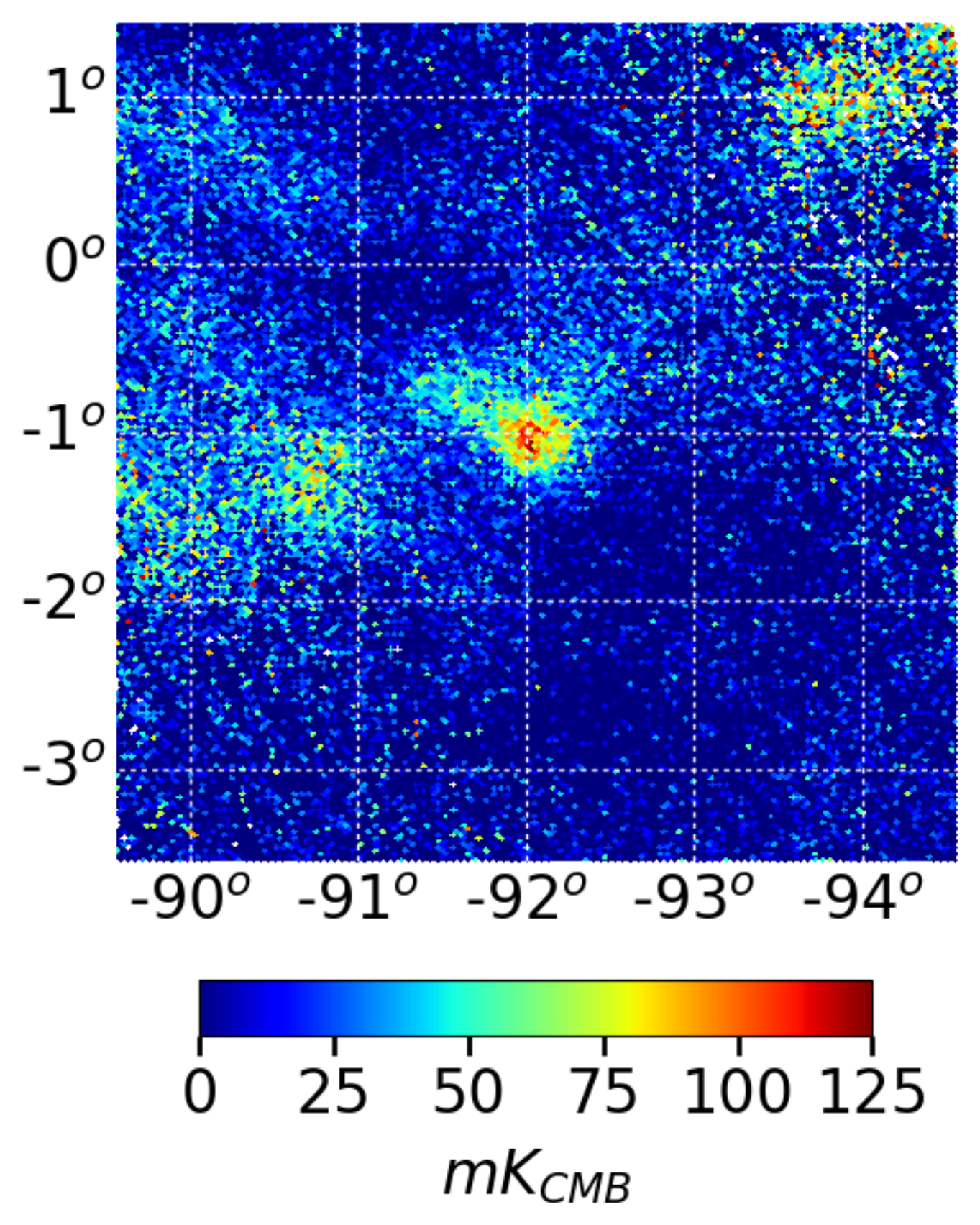}
  \end{center}
  \caption{Maps of RCW38 made with 331, 231, and 80 detectors at 150, 250, and 410~GHz, respectively (left to right).
  \label{fig:flight_beams} }
\end{figure}

We found in-flight beam sizes of 32, 30, and 30~arcmin for the 150, 250, and 410~GHz bands, respectively; 
see Figure~\ref{fig:flight_beams}. These were larger than beams measured on the ground. There 
are two possible reasons for this difference: (a) detector pointing uncertainty; and 
(b) non-optimal alignment of the telescope. 
Making combined maps from many detectors requires knowledge of their relative positions
on the focal plane with an accuracy much smaller than the inherent beam size, namely, 1-3'. As discussed earlier, we could not use
flight data to determine these relative positions for many of the detectors. Instead, we relied on the mechanical design of the 
focal plane. For a handful of the detectors at 250~GHz for which there were data, there was evidence that this assumption 
gave offset errors on the order of the beam size, leading to smearing and hence to larger combined beam maps. 
The ground beam measurements were conducted before shipping 
the payload to Antarctica. The short flight season in Antarctica did not allow time to repeat beam pattern measurements 
after re-assembling the payload, and it is possible that telescope alignment was not optimal. Further evidence for this 
hypothesis is provided by the apparent elliptical shape of RCW38 at the EBEX 150~GHz band; see Figure~\ref{fig:flight_beams}. 
The Planck map of RCW38 at 143~GHz shows a more circularly symmetric shape. 


\section{Receiver}
\label{sec:receiver}


\label{sec:receiverstructure}

The \ac{EBEX} receiver was designed to contain the cold optical elements, as described in Section~\ref{sec:coldoptics}, and to 
provide sufficient cryogens for 10 days of flight. In addition to 
providing appropriate heat sink temperatures between 0.25 and 4~K for optical elements, detectors, and \ac{SQUID} 
preamplifiers, the design had to implement several features that were unique to the balloon 
application and to our detector and readout scheme. They include accommodating 
a double vacuum-window mechanism, minimizing sensitivity to detector radio-frequency pick-up, 
and reducing as well as characterizing magnetic pickup in detector readout components.
These features will be discussed in subsequent subsections. 

\begin{figure}[ht]
  \centering   
  \includegraphics[width=0.50\textwidth]{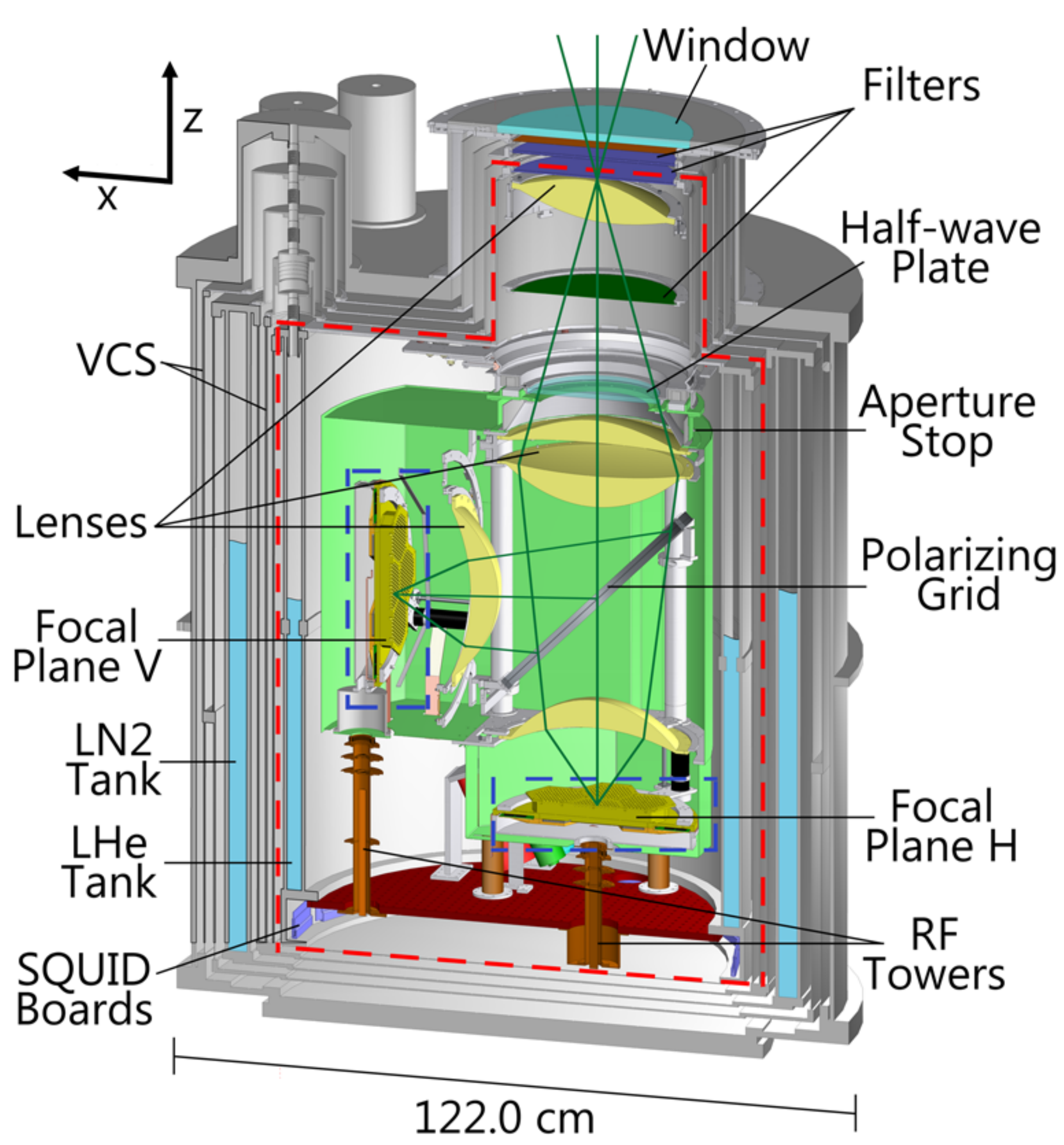}
  \caption{Electromagnetic radiation
  (solid green lines) entered the \ac{EBEX} receiver through a vacuum window, traversed filters, a lens, and the half-wave 
  plate before reaching 
  the aperture stop of the optical system. Two lenses collimated the beam. A polarizing grid transmitted one polarization state
  to focal plane H and reflected the other to focal plane V. Wires from the focal plane were channeled through 
  Faraday caged `RF Towers' 
  to \ac{SQUID} boards. The focal planes operated near 0.25~K (area enclosed in blue dash),  most of 
  the internal optics was maintained 
  near 1~K and was enclosed by absorber-lined metallic shield (shown as green shield), and all components inside of the 
  red dash line were cooled to liquid helium temperature.Two sub-Kelvin refrigerators are not shown.   
  \label{fig:cryo_cutaways} }
\end{figure}

An overall view of the receiver\footnote{Fabricated by Precision Cryogenics, Inc.} 
and its main elements is given in Figure~\ref{fig:cryo_cutaways}. Its bare, dry weight was 642~kg and it was 809~kg  
when full with cryogens and including all flight hardware.

\subsection{Cryogenics}
\label{sec:cryogenics}

The receiver had 8 thermal stages. Five of these stages were provided by the mechanical construction of the 
cryostat, which included an outer vacuum-shell at ambient temperature, a nitrogen vapor-cooled layer that was typically at 
180~K, a 130~liter liquid nitrogen (LN) reservoir, a helium vapor-cooled layer that was typically at 25~K, and a 130 liter 
liquid helium (LHe) reservoir; see Figure~\ref{fig:cryo_cutaways}. Three other temperature stages were 
provided by sub-K refrigerators that will be described below. 

During flight we maintained both LHe and LN near atmospheric pressure using valves\footnote{Tavco, Inc.} 
that maintained a  pressure of 15.7$\pm$1 psi. The reservoirs were kept near atmospheric pressure to stabilize 
the cryogen temperatures at design values and 
to reduce the boil-off rate of the cryogens. We also had commandable, motorized gate valves\footnote{Varian, Inc.} 
to vent the tanks to 
ambient pressure, if necessary, specifically in case of flight termination before cryogens expired. 
These were only used when the flight ended, long after the cryogens expired. Table~\ref{tab:cryo_loading} gives the pre-flight
calculated heat loads on the LN and LHe reservoirs. With 130 L for each of the cryogens these give a hold time of 11.8 and 10.8 
days for LN and LHe, respectively. LHe cryogen ran out after 10.8 days; we did not monitor the hold time of LN2.
\begin{table}[ht!]
  \center
  \begin{tabular}{|c|c|c|}
    \hline
    \bf{Element} & \bf{Load on LN (mW)} & \bf{Load on LHe (mW)}\\
    \hline
    Support Structure & 2260 & 56\\
    \hline
    Radiation & 12500 & 138 \\
    \hline
    Wiring & 880 & 33\\
    \hline
    AHWP Operation (Average) & N/A & 15 \\
    \hline
    Refrigerator Operation (Average) & N/A & 116\\
    \hline
    \hline
    Total & 15640 & 358 \\
    \hline
  \end{tabular}
  \caption{Pre-EBEX2013 calculated heat load contributions to the LN and LHe stages during flight conditions. 
  Values labeled average are the time-average values over the duration of the Antarctic flight based on individual duty cycles.}
  \label{tab:cryo_loading}
\end{table}
 
Cooling to sub-kelvin temperatures was achieved with two closed-cycle, pressurized helium 
adsorption refrigerators\footnote{Chase Research Cryogenics, Inc.}. 
A 30~STP liters, two-stage $^4$He refrigerator cooled the aperture stop and downstream optical elements with the exception 
of the focal planes. When the \ac{AHWP} was rotating (stationary) the base temperature was 1.2 (1.0)~K, 
the hold time 48 (87) ~hours, and the calculated load was 320 (70)~$\mu$W. 
The source of power dissipation by the AHWP mechanism is discussed in Section~\ref{sec:smboperation}. 
The second most significant source of heat load on the 1~K stage was conduction through the polyimide 
legs\footnote{Vespel SP-1, Dupont} with which the optics box was mounted to the cold plate. A 
cryogenic stepper motor was used as a mechanical heat switch between cold plate and the optics 
box during initial cooling of the equipment. This reduced the cool-down period from 12 to 4 days.

Two heat sink temperatures, 320 and 240~mK, were provided by a three stage adsorption
refrigerator with a $^4$He pre-cooling stage and two $^3$He refrigerators. The focal planes
including the detectors and their associated LC-boards (see \ac{EP2}) 
were operated near 250~mK.  Both the 320~mK and 1~K stages 
were used as heat sinks for wires leading from the focal planes to warmer temperature stages. The  
refrigerator had 2~STP liters of $^3$He and a total heat load of 0.5 $\mu$W on the coldest stage, and a hold time of 
84~hours. The heat load on the coldest stage was dominated by a 0.2~$\mu$W load due to detector wiring.

\begin{figure}[ht]
\centering
   \includegraphics[width=0.48\textwidth]{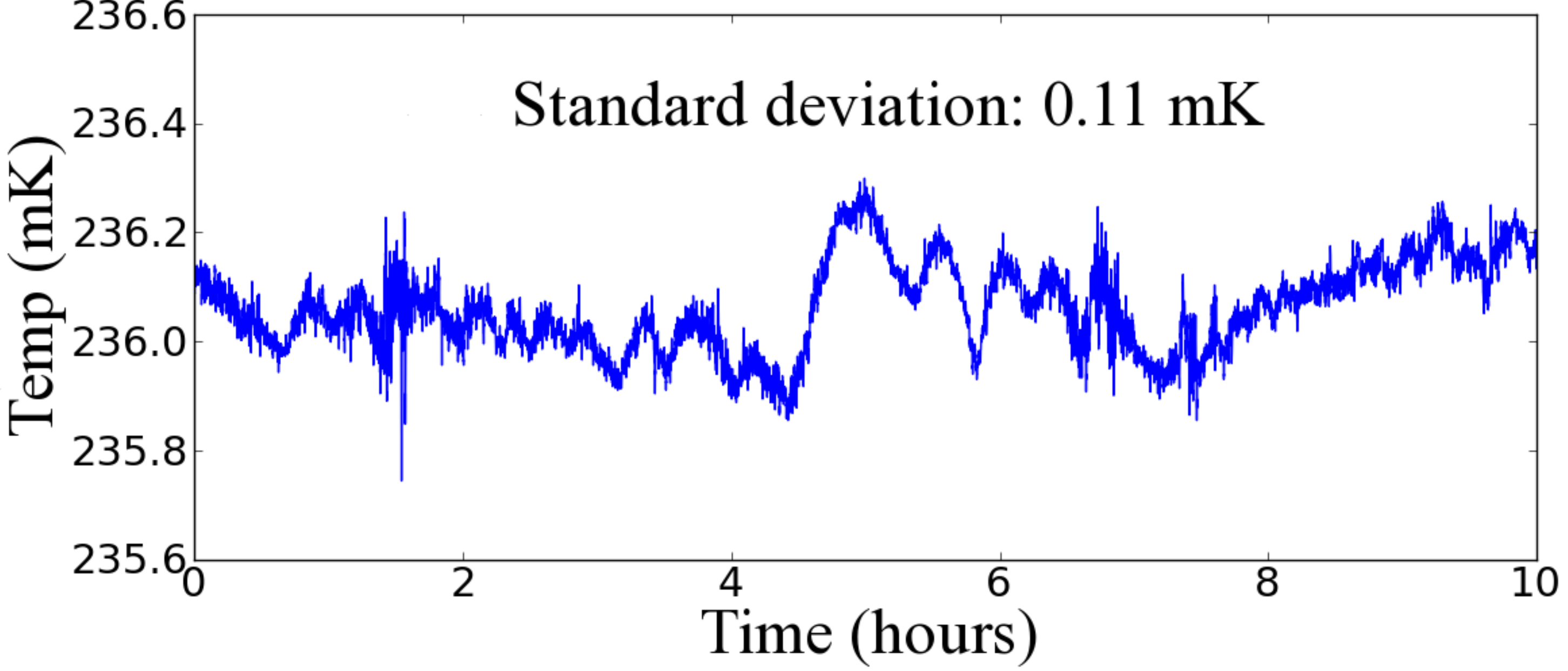}
   \includegraphics[width=0.48\textwidth]{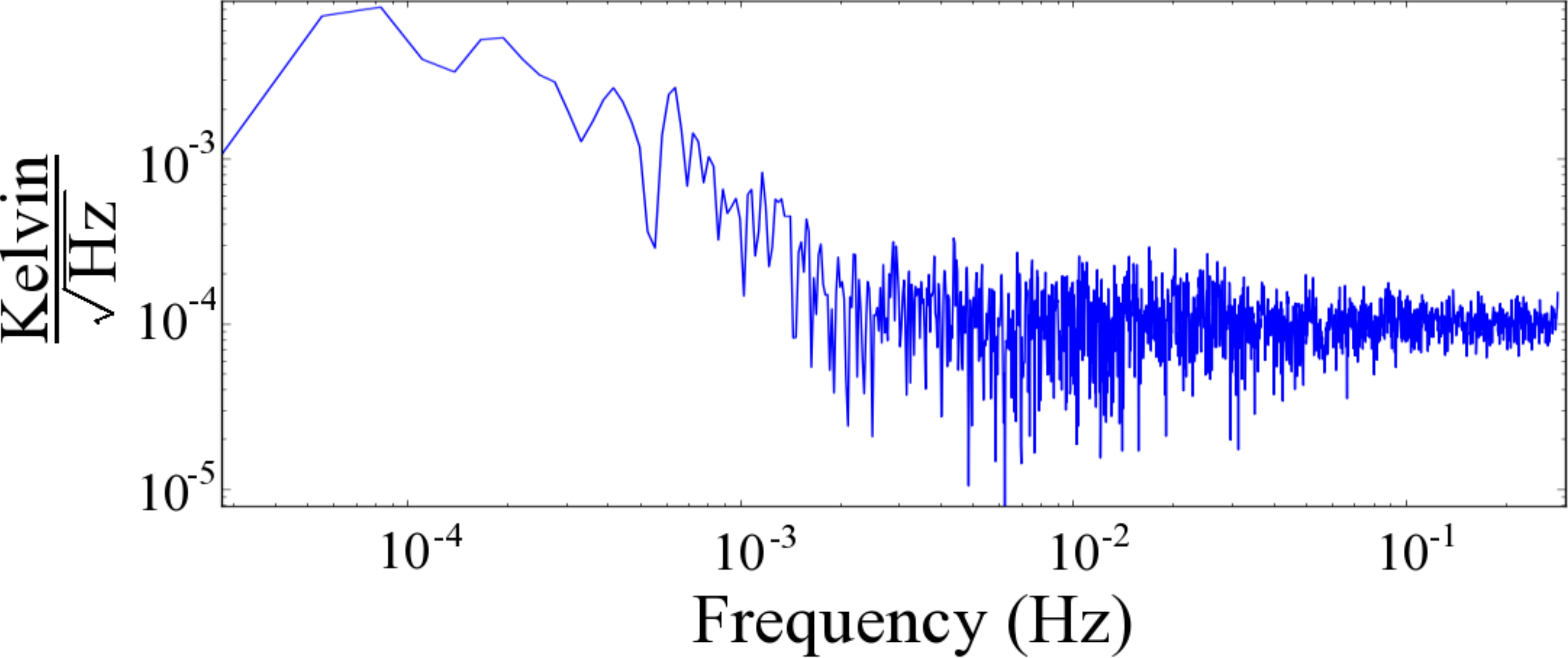}
   \caption{Left: temperature of the coldest stage of the three-stage adsorption refrigerator over 10~hours of the
                EBEX2013 flight. The 0.11~mK RMS temperature stability exceeded the requirement on gain fluctuations.
                Right: power spectrum of the temperature fluctuations for the same data section gives a $1/f$ knee 
                at 2~mHz and is identical to that measured on the detector wafers (not shown). 
                For frequencies higher than 10~mHz, the power spectrum is averaged at constant fractional bandwidth of 0.3\%.
                \label{fig:ultra_temp_stab} }
\end{figure}

The cryogenic system had temperature stability better than our requirement on gain fluctuations. Temperature fluctuations 
over a representative period of several hours are shown in Figure~\ref{fig:ultra_temp_stab}. The 0.1~mK RMS temperature 
fluctuations produced bolometer gain fluctuations of 0.05\%, which were negligible compared to other sources 
of calibration uncertainties. 

\subsection{Double Vacuum Window Mechanism}
\label{sec:doublewindow}

At the top of the 300~K shell, a 30~cm open aperture window separated the vacuum environment of the receiver from 
ambient pressure. Several materials were considered for this vacuum window, including sapphire, 
zotefoam\footnote{Zotefoams, PLC}, 
polypropylene, and polyethylene. We chose polyethylene (PE) because of the combination of its optical properties, 
durability, cost, ease of implementation, and the availability of broad-band anti-reflection coating.
\begin{figure}[ht!]
  \centering
  \includegraphics[width=0.49\textwidth]{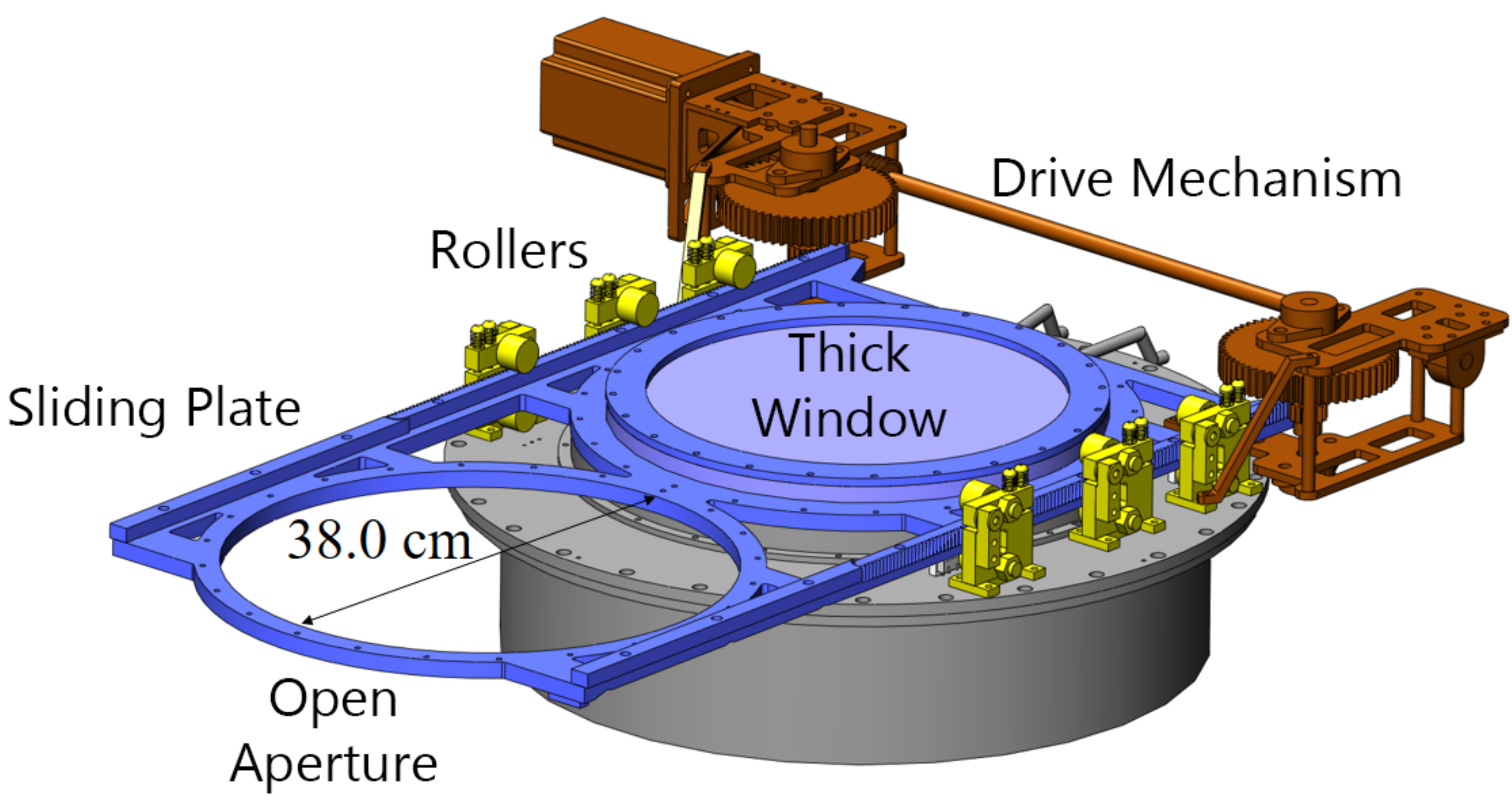}
  \includegraphics[width=0.49\textwidth]{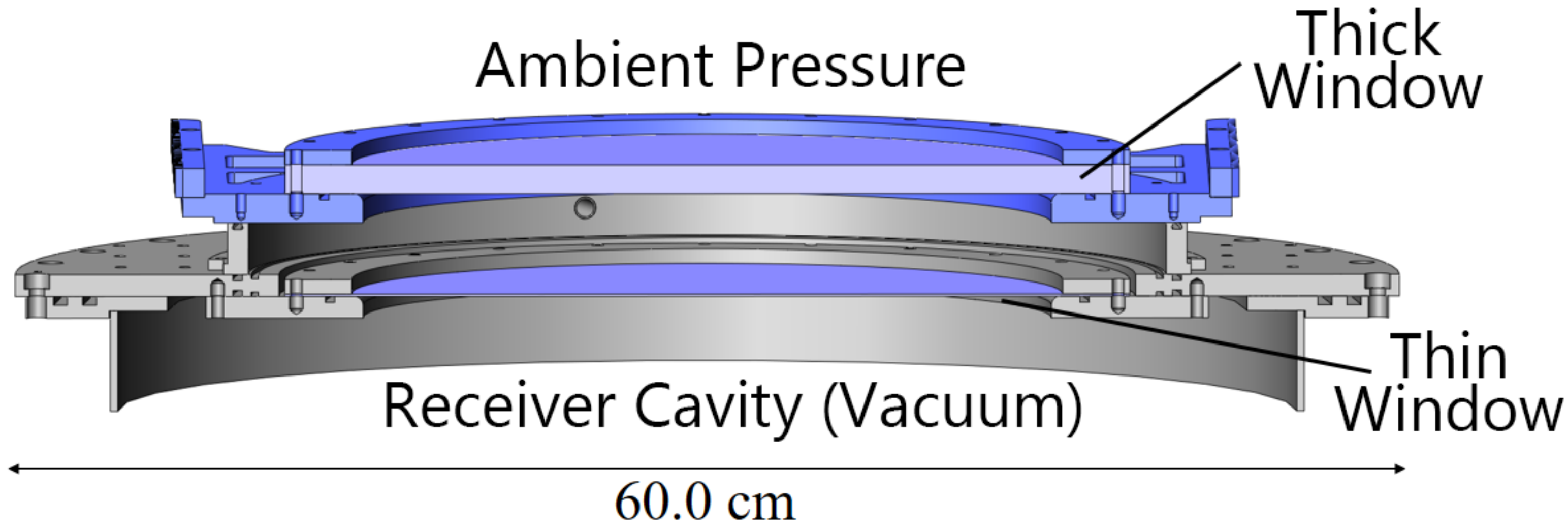}
  \caption{Components of the double vacuum window mechanism. The open aperture and thick window on the sliding plate are 
  moved along roller assemblies by a stepper motor drive mechanism. The thick window was placed above the thin 
  window in ground operations.}
  \label{fig:dwm}
\end{figure}

We used a 12.7~mm thick \ac{PE} when operating the receiver in the laboratory. Thinner material would bow 
inward and (1) deflect and damage the 
infra-red blocking filters that were mounted beneath the window (see Figure~\ref{fig:filters}), and (2) induce 
instrumental polarization due to differential reflection. Thermal emission from a 12.7~mm thick PE window operating 
near 300~K, which was a typical receiver shell temperature at float, would have given 30\% higher in-band load
at 150~GHz  compared to the CMB. To reduce this load we  implemented a double vacuum 
window mechanism (DVWM) that consisted of both the 12.7~mm thick and a thinner window. 
The mechanism is shown in Figure~\ref{fig:dwm}. 
During ground operations, the thick window was placed above the thin window with the cavity between the windows evacuated 
to put the pressure differential of the atmosphere on the thick window. When the payload reached altitude above which the 
ambient pressure was less than 10~torr, a ground operator 
commanded a motor that moved the thick window aside exposing an open aperture above the thin window. 
The thick window was moved back before flight termination to protect the thinner window from atmospheric pressure. 

\begin{figure}[ht!]
  \centering
  \includegraphics[width=0.5\textwidth]{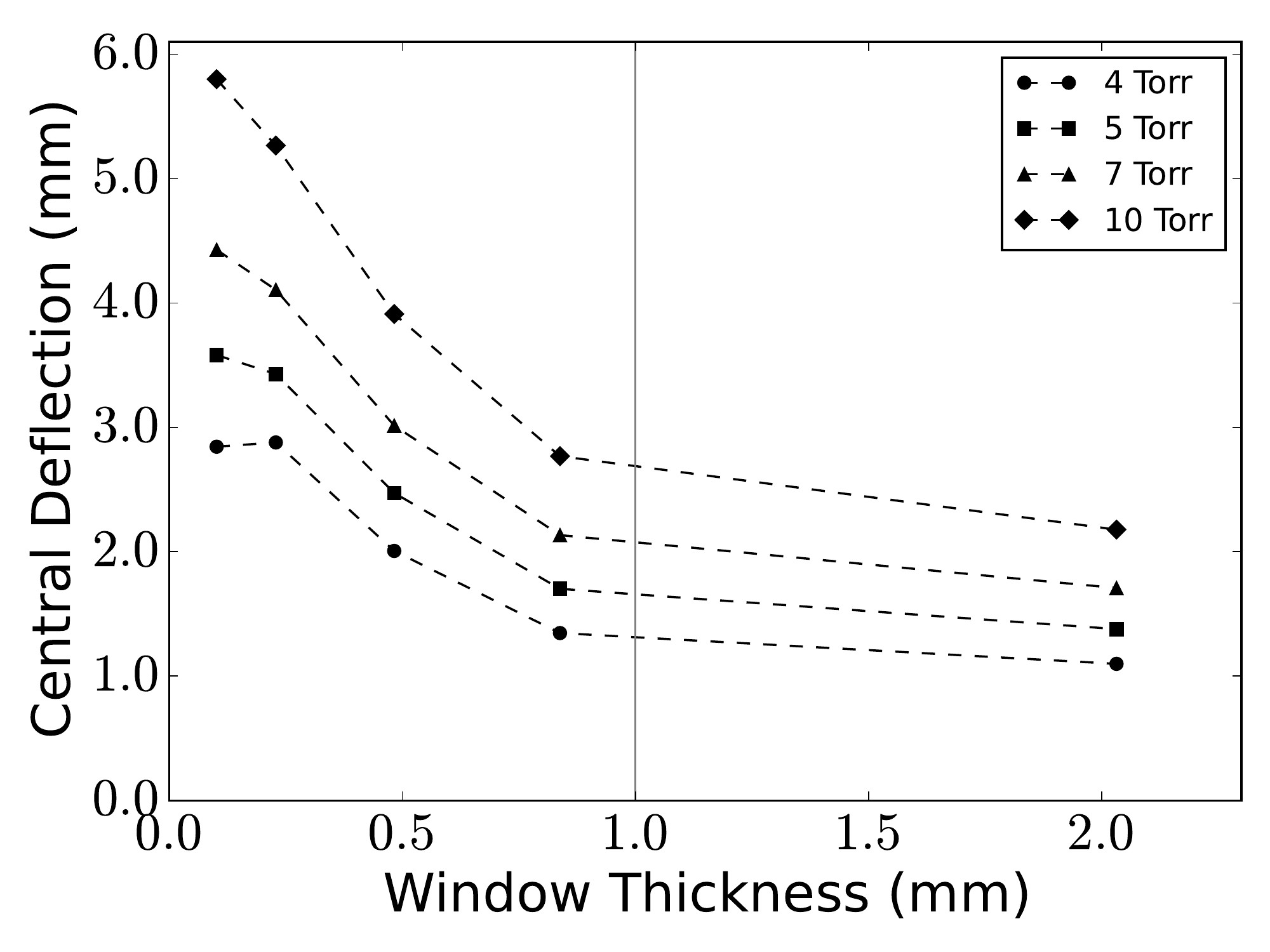}
  \caption{Central deflection of a 30~cm diameter polyethylene window supported at the edge as a function 
  of window thickness under various differential pressures.  For EBEX2013 we chose a thin window thickness
  of 1~mm (vertical solid)
  \label{fig:thin_wind_defl} }
\end{figure}

We conducted deflection tests on 30~cm PE windows to measure the maximum central deflection 
as a function of window thicknesses and differential pressures; see Figure~\ref{fig:thin_wind_defl}. 
These tests indicated that at the 
expected float pressures of up to 10~torr a 1~mm thick PE window would be adequate. A 10~torr pressure differential 
was considered a conservatively high estimate because it corresponds to an altitude of 28 km. 
Flight altitude ranged from 34 to 36.5~km. 
Thermal emission from the thin window was a factor of 10 less than in-band CMB power at 150~GHz. 

During the EBEX2013 flight we removed the thick window when the payload ascended through a pressure of 
8~torr. The thick window was returned to the optical path after the liquid cryogens were exhausted, 11~days after launch. 
In-flight readout of the DVWM position and post-flight inspection indicated nominal operation.
The DVWM is described elsewhere in more detail~\citep{zilic_dvwm}. 

\subsection{Mitigation of Radio Frequency Interference}
\label{sec:rf_environment}

\begin{figure}[ht]
  \centering
  \includegraphics[width=0.50\textwidth]{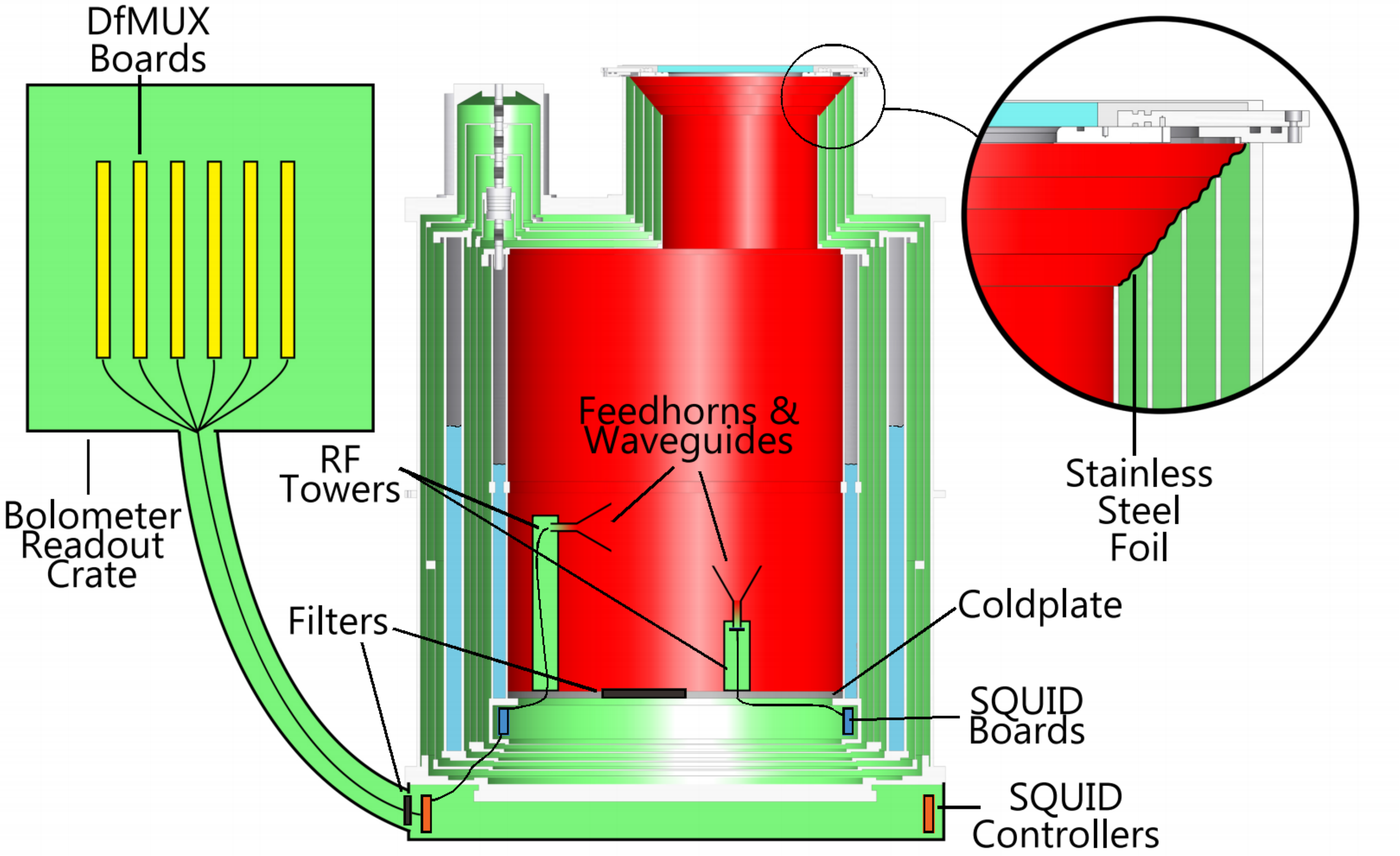}
  \caption{Mitigation against RF contamination relied on filtering and a Faraday cage for the \ac{SQUID}s, wires, and electronic boards. 
  We glued stainless steel foil between different cryogenic shells near the vacuum window (zoom in top right) to prevent RF radiation 
  from entering the inter-shell region of the cryostat. Waveguides at the back of the feedhorns in the focal plane provided high pass filtering. 
  The Faraday cage consisted of the walls of the cryostat, the \ac{SQUID} controller can, a dryer hose, and the walls of the readout crates.  
  Filters, shown on the cold plate and near the \ac{SQUID} controllers, were installed on individual wires. 
  \label{fig:rf_shielding} }
\end{figure}

The transition edge sensor bolometers, the \ac{SQUID} pre-amplifiers, and the bias and readout wiring were sensitive to 
interference by radio-frequency (RFI) electro-magnetic waves. CSBF's radio and video transmitters operating at 
frequencies between 0.9 and 2.5~GHz were a source of RFI. To mitigate RFI, we constructed 
an RF shielded environment that encompassed all the RFI sensitive equipment. The shielded environment 
is shaded green in Figure~\ref{fig:rf_shielding}. 

The majority of the RFI cavity consisted of a Faraday cage 
provided by metallic walls. It included the receiver's vacuum jacket, a can at the bottom of the cryostat that contained the 
\ac{SQUID} controllers, a metallic dryer hose that shielded cables between the 
receiver and the readout crates, and the walls of the readout crates. Connections between
Faraday cage walls had RFI gaskets\footnote{Parker Hannifin, rectangular strip EMI gasket} or relied on closely spaced
screws to give a waveguide cut-off below 6 GHz. 

The 30~cm diameter vacuum window was many RF-wavelengths across allowing RF radiation to 
enter the cryostat cavity (red shaded region in Figure~\ref{fig:rf_shielding}). 
The bolometer wafers were protected because they were enclosed in a Faraday cage and because 
the waveguides between the feedhorns and the bolometer cavities acted as high-pass filters. Bolometer wiring 
passed from the focal planes via `RF Towers', which will be described below, to SQUID amplifiers mounted below 
the cold plate, and from there through the various cryogenic shells to the 300~K vacuum jacket; see Figure~\ref{fig:rf_shielding}. 
It was therefore important to ensure that the inter-shell region was RFI free. 
We prevented RF radiation from entering the 
cryostat's inter-shell region by gluing a 25.4 $\mu$m thick stainless steel foil between the cryogenic shells near the 
vacuum window. The glue was electrically conductive.  The additional thermal load imposed
by this foil was 1\% the total load on the LHe stage and smaller for the warmer stages. The foil's RF
attenuation between 0.9 and 2.5~GHz varied from -16 to -26~dB,~respectively. 

\begin{figure}[ht]
  \centering
  \subfigure[RF tower model with cutaway.]{
    \includegraphics[width=0.4\textwidth]{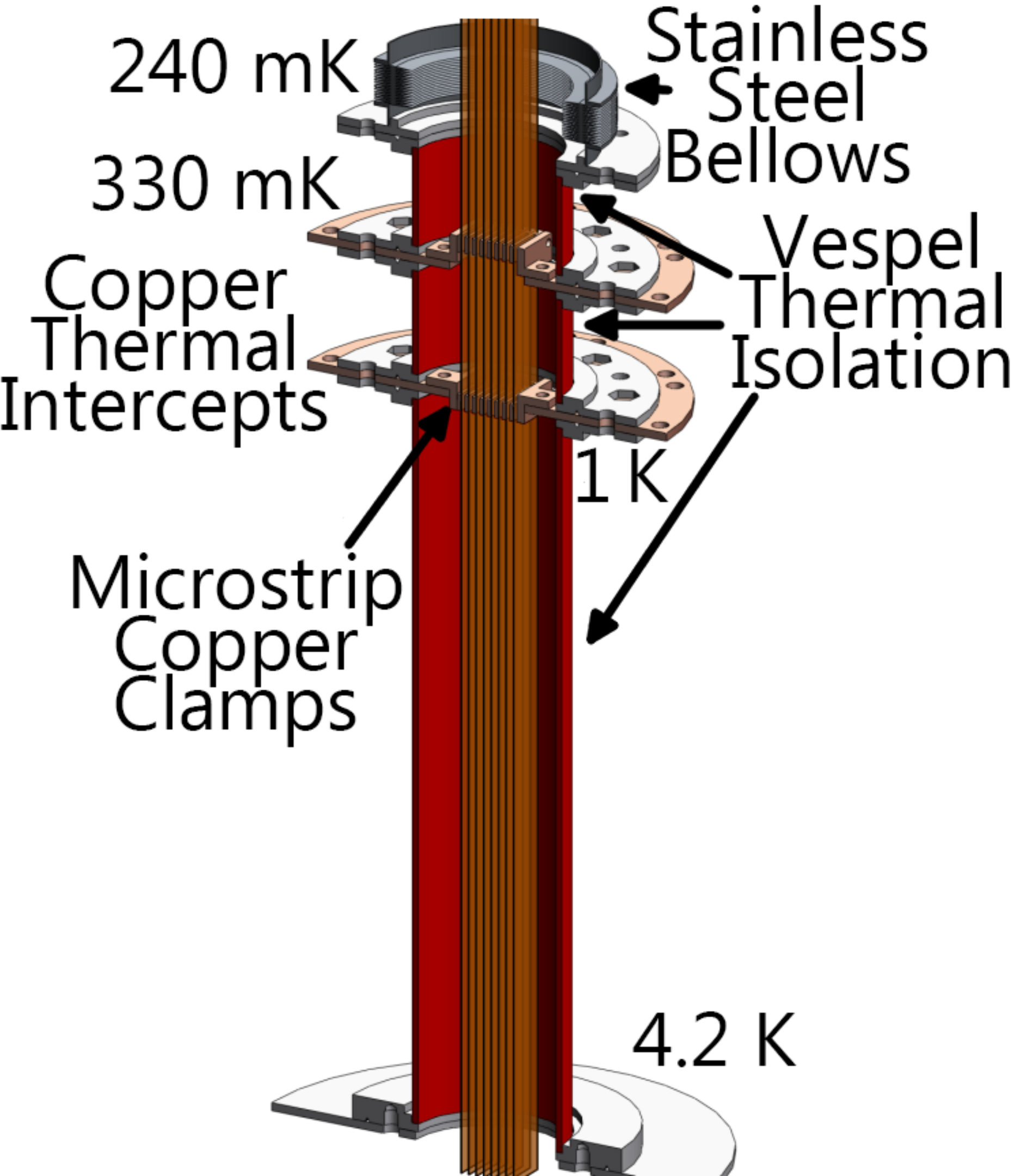}
    \label{fig:rft_interior}}
  \subfigure[RF tower assembly in situ.]{
    \includegraphics[width=0.30\textwidth]{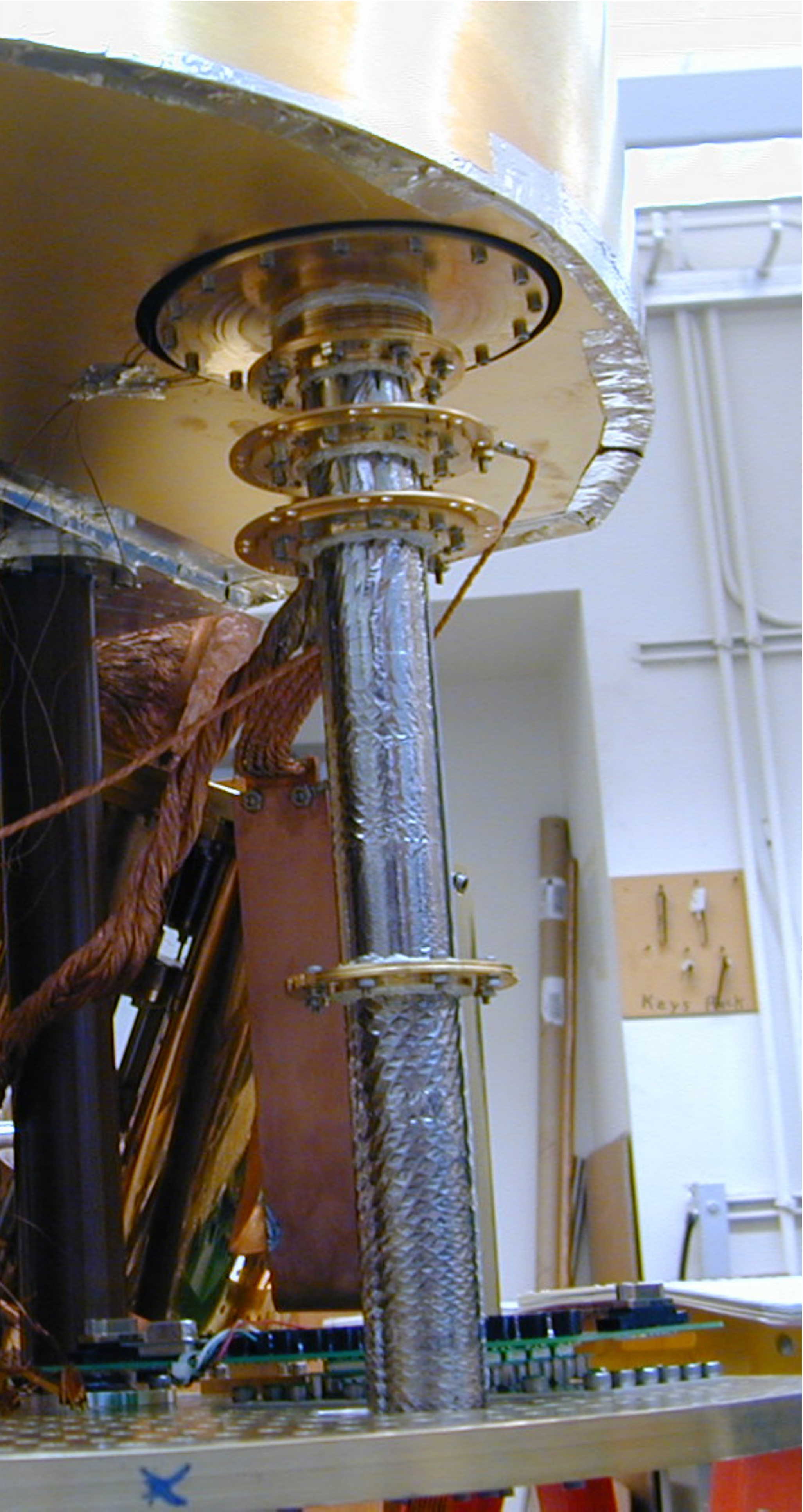}
    \label{fig:rft_real}}
     \caption{RF tower model with cutaway to show interior microstrip assemblies and in situ picture. 
     The RF tower shown in situ was for the V focal plane and has additional length to reduce thermal conductive loading.}
  \label{fig:rft}
\end{figure}

Between the focal planes and the cold plate of the instrument were two structures, one per focal plane, called `RF towers'. 
The RF towers provided RF-clean environment for the wires leading from the focal plane to the \ac{SQUID} amplifiers while 
minimizing the thermal conductance between the 0.24~K and the 4.2~K thermal stages. 
Each RF tower was a cylinder comprised of sections of Vespel\footnote{Vespel SP-22, by Dupont} 
tubes interspersed with heat-sink terminals, and
terminated near the focal planes with a  a stainless steel bellows. Bolometer wiring passed
inside the cylinder. The bellows attached to the Faraday cage
surrounding each of the focal planes and facilitated the mechanical connection between a focal plane and its tower despite 
possible misalignments. The four heat-sink terminals gave thermal connection to the 0.25~K, 0.33~K, 1~K, and 4.2~K stages. 
An RF-clean environment was maintained inside the RF tower by using a 99.9\% purity $5$ micron thick 
niobium foil  to completely wrap the Vespel tube through which the wires passed;  see Figure~\ref{fig:rft}. 
The niobium was purchased in sheets and was wrapped 
and spot-welded along a vertical seam spanning the length of the RF towers. The welds were spaced every 5~mm for integrity 
of the seam and rejection of RF frequencies less than 10~GHz.
The foil was electrically connected to the metallic ends of each of the heat-sink points using electrically 
conductive silver-filled adhesive. We used a niobium foil due to its low thermal conductivity below its 
9.5~K superconducting temperature. For our RF-Tower geometry the heat loads on the 0.25~K, 0.33~K, and 1~K 
temperature stages were 0.06, 2.9, and 41.9 $\mu$W, constituting 9, 18, and 21 percent, respectively, of 
the total load on these stages.

We used commercial capacitive filters\footnote{Spectrum Control} for housekeeping wires -- such as for 
temperature sensors and refrigerator control -- 
that crossed the cold plate, which was the boundary between the RFI-contaminated and RFI-clean environments. 
At room temperature, the filters provided a capacitive coupling between the signal lines of wiring 
and the system ground with a 3 dB cutoff frequency of 640 kHz and greater than 50 dB attenuation above 1 GHz.
We measured the frequency response of the filters at liquid nitrogen and found that the frequency for the 3~dB point
increased by a factor of approximately 2. Otherwise the characteristic shape of the response remained the same as 
at room temperature. To compensate for the frequency increase of the 3~dB point we connected 
two filters in series. We assumed that the measured change is a consequence of thermal contraction of 
the embedded capacitors and that subsequent contraction to liquid helium temperature was negligible. 
Similar filters were also applied on wires between the SQUID controllers the readout crates; they were used as an added 
precaution.

\subsection{Magnetic Shielding}
\label{sec:magnetic_shielding}


Both the \ac{TES} bolometers and the \ac{SQUID} amplifiers are sensitive to varying magnetic fields. 
We discuss each in view of the following two sources of time varying ambient magnetic field: (1) changes in 
orientation relative to Earth's magnetic field vector as a function 
of the azimuthal motion of the gondola, and (2) changes due to the rotation of the magnet upon
which the \ac{HWP} is mounted because of spatial inhomogeneity in its magnetic field. 
The rate of variation of both of these sources is at most few tens of Hz.

\subsubsection{SQUIDs} 

\begin{figure}[ht!]
  \centering
  \includegraphics[width=0.45\textwidth]{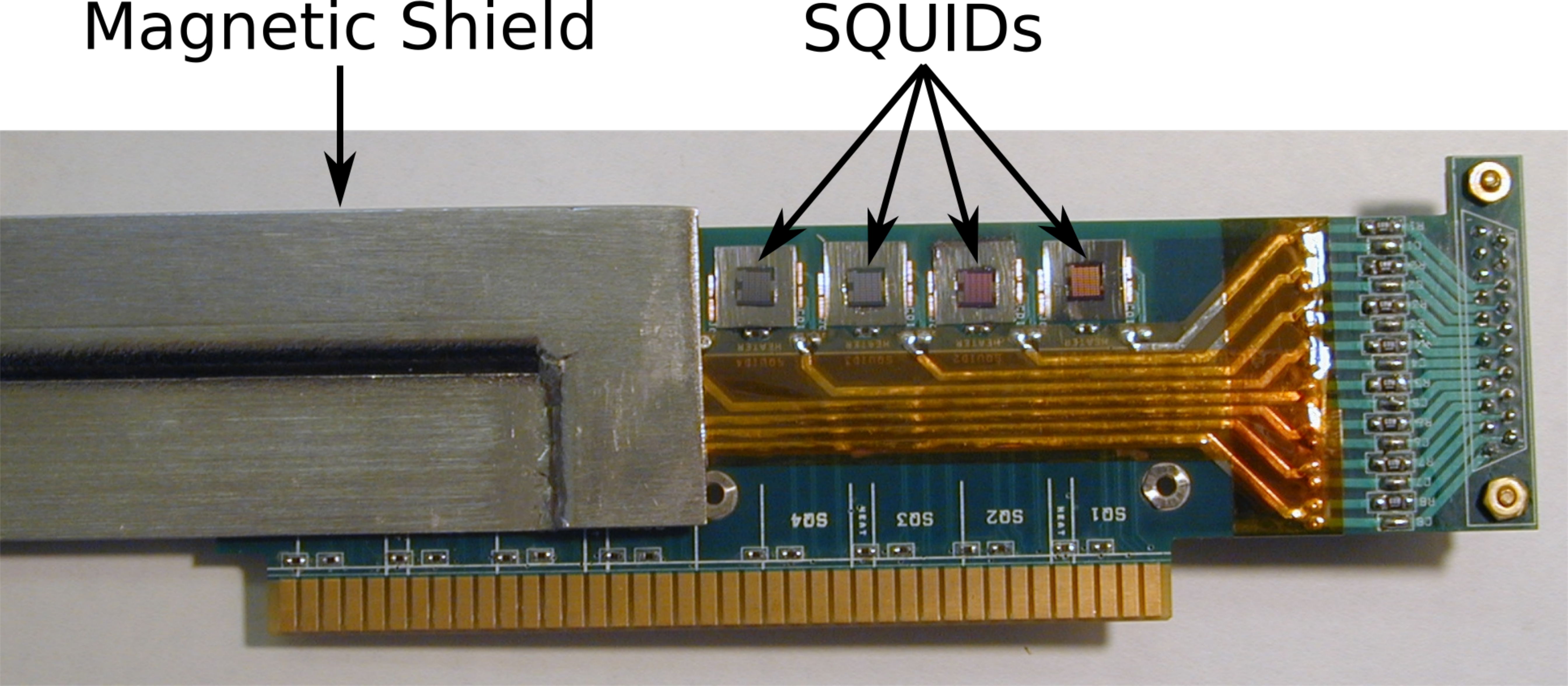}
  \caption{ A \ac{SQUID} board with 8 \ac{SQUID}s (four visible) and a magnetic shield.
   \label{fig:squid_layout} }
\end{figure}

Each \ac{SQUID} was fabricated with an underlying layer of niobium~\citep{Huber2001} and each 
board with 8 \ac{SQUID}s was inserted into a magnetic shield\footnote{The MuShield Company Inc.}, 
as shown in Figure~\ref{fig:squid_layout}. 


The few tens of Hz magnetic field variation in the sources is much smaller than the standard 0.1-1~MHz 
readout frequency of the \ac{SQUID}s.  We therefore do not expect
these sources to contribute spurious signals. 
We searched for azimuth- and \ac{HWP} rotation-synchronous signals in 
resistors and in `dark' \ac{SQUID}s; these are \ac{SQUID}s that were not connected to bolometers, but that 
otherwise shared the same readout path as regular detectors.
We binned 6 hours of flight time-ordered data in both gondola azimuth and \ac{HWP} rotation angle. 
Neither the raw data nor the binned data show any signature of synchronous signals above the noise. 

In addition to the MHz readout frequency that was used for all detectors, we had a readout channel 
that monitored the \ac{SQUID} amplifiers at very low frequencies, at and near DC. 
We did detect variations in Earth's magnetic field flux passing through the \ac{SQUID}s in that
readout channel, called `SQUID DC'. This detection is now used to quantify the attenuation of the 
\ac{SQUID}s magnetic shielding.  

\begin{figure}[ht!]
  \centering
  \includegraphics[width=0.45\textwidth]{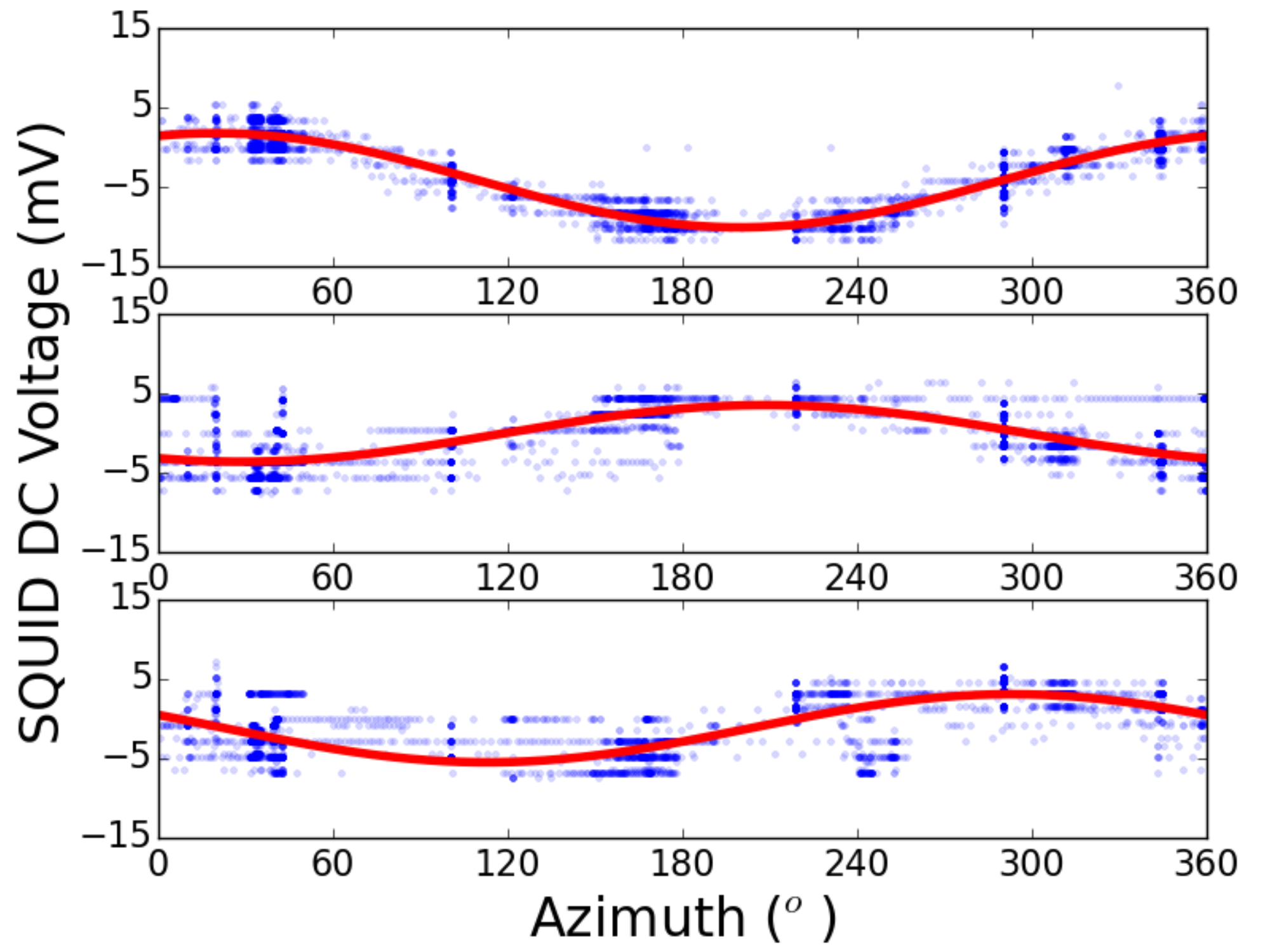}
  \includegraphics[width=0.45\textwidth]{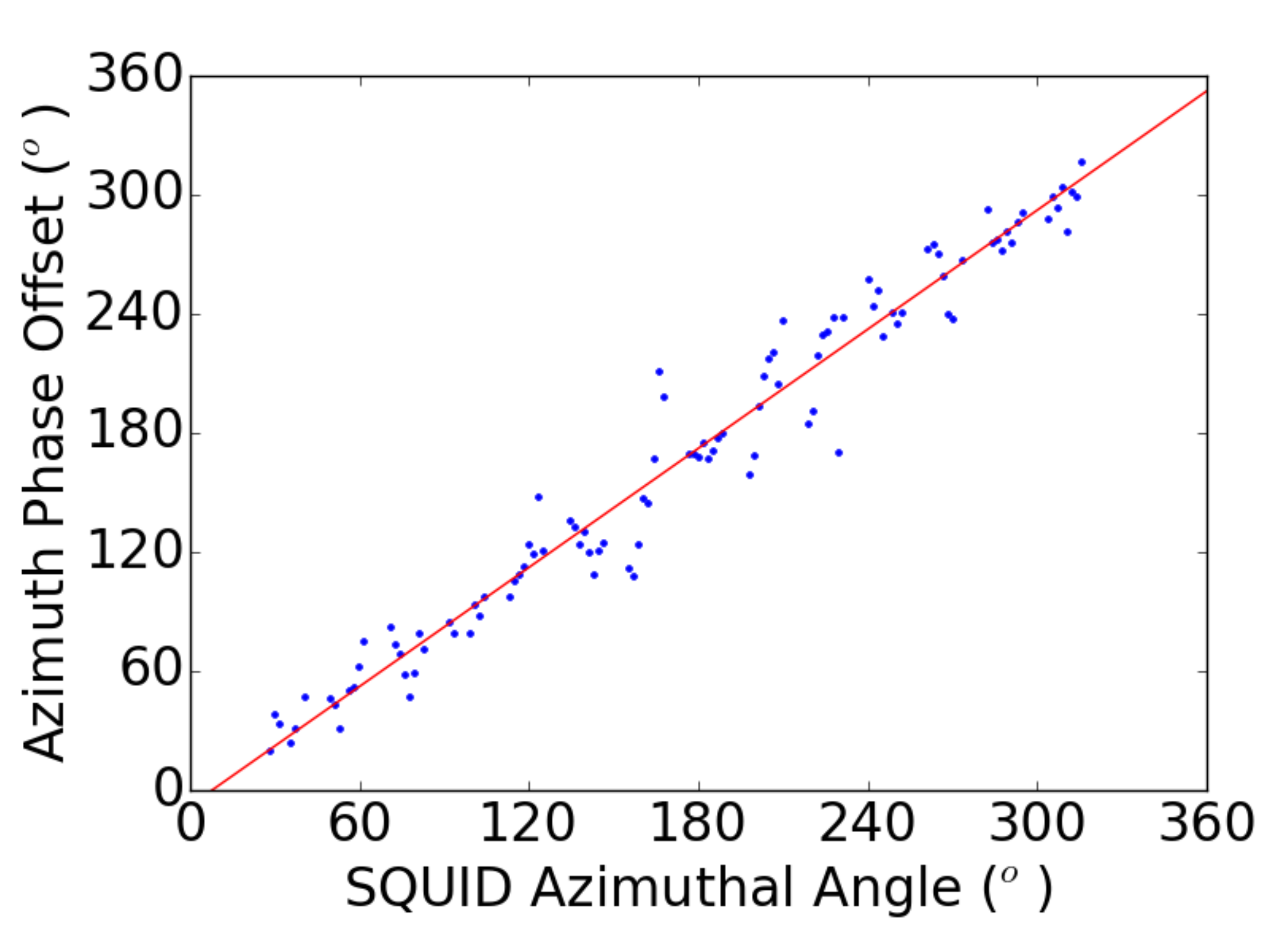}
  \caption{Left: \ac{SQUID} output DC voltage measured as a function of the payload azimuth (blue dots) and a 
  cosine fit (red solid) for \ac{SQUID}s mounted at azimuthal angle 28$^o$ (upper panel), 203$^o$ (middle panel) and 282$^o$ 
  (lower panel) relative to the orientation of Earth's magnetic field. More intense data point indicates more data in 
  the specific azimuth and voltage bin. 
  Right: Measured phases of the sinusoidal fits shown on the left for 105 SQUIDs (dots). Zero angle is 
  set to align with Earth's magnetic field at the time the data were recorded. The best fit offset of a line with unity slope (red)  
  is -8$^o$; the variance about the fit is 16$^o$.}
\label{fig:az_squid}
\end{figure}

Figure~\ref{fig:az_squid} shows data from the SQUID DC channels of three \ac{SQUID}s mounted on three different boards. 
There is a clear sinusoidal modulation at a frequency of one azimuthal rotation.
The boards are mounted at different azimuthal angles relative to each other, which is the source of 
the phase offset between the three data sets. We set the zero angle in the right panel of Figure~\ref{fig:az_squid} 
such that a \ac{SQUID} board mounted at that angle would be aligned with orientation of Earth's magnetic field at the time 
the data was taken and thus show a phase angle of zero. We measured an offset of -8$^o$ with variance of 16$^o$ 
compared to the best fit. The data support the interpretation of modulation in the signal due to Earth's magnetic field. 

During the 105 minutes of data shown, the horizontal component of Earth's magnetic field 
is $\Phi = 0.103$~G~\citep{Thebault2015}. 
The attenuation due to the magnetic shielding, defined as the 
magnetic flux measured with the shielding divided by the flux that would have been measured without, is  
\begin{equation}
\epsilon_B = V_{out} \left( \frac{1}{\Phi_{SQUID}} \right) \left( \frac{1}{dV/d \Phi} \right) G,
\label{eq:squidShielding}
\end{equation}
where $V_{out}$ is the measured \ac{SQUID} channel voltage, $\Phi_{SQUID}$ is the magnetic flux through the \ac{SQUID}, 
$dV/d \Phi$ is the \ac{SQUID} response function, which is known for each of our \ac{SQUID}s, and 
$G$ is a known gain factor that depends on the details of the electronics. 
For 105 \ac{SQUID}s we find a mean attenuation of $1.7\times 10^{-4}$ with a dispersion of $0.6\times 10^{-4}$.

\subsubsection{TES Bolometers} 

We assessed that no protection was necessary for the detectors and none was provided. The combination
of \ac{HWP} 1.235~Hz rotation speed and design azimuthal scan period of 50~s placed the polarization signals 
near 5~Hz, far from the tens of mHz frequency expected from modulation due to Earth's magnetic field. 

The rotation of the \ac{HWP} itself was very stable (see Section~\ref{sec:smboperation}) putting any modulation 
of the bolometer response due to magnetic field inhomogeneity of the rotor exactly 
at this frequency and its harmonics. Such signals are degenerate with other 
rotation-synchronous signals, which are removed during our data analysis process. 
The characteristics of the EBEX2013 time domain data are discussed in more detail in 
EP2, \citet{joy_thesis}, and \citet{kate_thesis}.

\section{Polarimetry}
\label{sec:polarimetry}


\ac{EBEX} used a combination of a continuously rotating \ac{AHWP} and a 
stationary wire grid for its polarimetry. Above a temperature of 30 K thermal emission
from the \ac{HWP} exceeds the power from the CMB. We therefore mounted the \ac{AHWP} to the 4~K temperature stage. 
The \ac{AHWP} was placed approximately 1~cm toward the sky side of the aperture stop of the optical system; see 
Figure~\ref{fig:raydiagram}. 
The aperture stop was heat sunk to the 1~K temperature stage; see Figure~\ref{fig:cutaway}. 
Tracing the optical path from the sky inward, the \ac{AHWP} is behind the mirrors, the vacuum window, 
and the field lens. Thus, the instrumental polarization induced by these elements is modulated by the 
the \ac{AHWP} and contributes to our observed polarization signal.

\subsection{Half Wave Plate and Grid}
\label{sec:hwpandgrid}

The \ac{AHWP} was made of a stack of five 24~cm diameter a-cut sapphire disks following a 
Pancharatnam design~\citep{pancharatnam55}. The aperture stop diameter was 19~cm. 
Each of the sapphire plates was approximately 1.66~mm thick, making it a standard \ac{HWP} 
for a frequency near 300~GHz. This was near the middle of the broad band required
from the \ac{AHWP}. The thickness of each of the plates is given in Table~\ref{tab:hwp}. 
X-ray diffraction analysis on a smaller, 15~cm diameter a-cut \ac{HWP} from the same 
vendor\footnote{Rubicon Technologies, Inc.} confirmed the crystal orientation and showed negligible level of 
defects. The stack is glued by interleaving 12.5~$\mu$m thick polyethylene sheets and hot-pressing the 
entire stack in an oven. The 22~cm diameter \ac{ARC} consisted of six glued layers per side.  
Starting from the outermost sapphire plate and listing
outward the layers were: Stycast1266\footnote{Emerson and Cummings, Inc.}, 
TMM6\footnote{Rogers Corporation},  Stycast1266, TMM3, 12.5~$\mu$m thick polyethylene, and microporous Teflon. 
The Stycast thickness was approximately 0.025~mm and the initial thickness 
of the TMM was 0.38~mm. After gluing with Stycast, each TMM layer was ground 
to its final thickness. 
We found that this ARC has survived 
several cryogenic cycles, although small cracks developed at the outside rim and slowly grew with
each cryogenic cycle. Inspection after the \ac{EBEX2013} flight showed that these cracks did not penetrate
the 19 cm optical diameter. 
\begin{table}[ht]
\begin{center}
\begin{tabular}{|c|c|}
\hline
Mean plate thickness (mm)   & (1.655, 1.657, 1.647, 1.657, 1.636) \\  \hline
Plate thickness SD(mm)      & (0.016, 0.015, 0.013, 0.013, 0.017) \\ \hline 
Fitted plate thickness (mm)  & (1.665, 1.677, 1.648, 1.675, 1.64)  \\  \hline
Fitted relative orientation (deg) & (0, 26.5, 94.8, 28.1, -2.6)  \\  \hline
\end{tabular}
\caption{Parameters of the \ac{EBEX} five-stack sapphire \ac{AHWP}. For each plate, 
the thickness was measured at room temperature in 80 locations; the second and third rows give the mean and standard 
deviation (SD). The thermal contraction of sapphire between room and cryogenic temperatures is less than 0.1\%. 
After the 5-stack was constructed we fit spectroscopy measurements to a model in
which the thicknesses and relative orientations of the plates are allowed to vary; see text. 
Here we give the best fit values (bottom two lines).
\label{tab:hwp} }
\end{center}
\end{table}


The ordinary and extraordinary axes of each plate were determined by placing the plates between two co-aligned wire 
grid polarizers and finding minimum transmission at the zero path difference of a Fourier transform spectrometer. The five plates
were then glued, a temporary ARC applied, and the transmission of the stack measured as a function of frequency for 17 
stack orientations when placed between two polarizers that were (1) co-aligned; (2) at 90$^{o}$ relative to each other; (3) at 45$^{o}$ relative 
to each other. 
The entire data was best fit 
for individual plate thicknesses and rotation angles.  For these fits we assumed the following room temperature indices of 
refraction $n$ and absorption coefficients $\alpha$ for the ordinary ($o$) and extraordinary axes ($e$)~\citep{pisano2006}
\begin{eqnarray*}
 n_{o} & = & 3.053 + 1.4\cdot 10^{-4} \nu + 2\cdot 10^{-7} \nu{^2} + 3\cdot 10^{-8} \nu{^3} \\
\alpha_{o} & = & 5.2 \cdot 10^{-3} \nu + 5.5 \cdot 10^{-4}  \nu^{2} + 8 \cdot 10^{-6}  \nu^{3} \\ 
n_{e} & =  & 3.387+ 4 \cdot 10^{-4} \nu^{2} \\
\alpha_{e}  & = & 5.2 \cdot 10^{-4} \nu^{2.2} , 
\end{eqnarray*}
where the frequency $\nu$ is in cm$^{-1}$. 
These measurements were reported by~\citet{matsumura2009} 
and the results are given in Table~\ref{tab:hwp}. Extrapolating the absorption values to 
80~K~\citep{moncelsi2014}, the absorption of the \ac{AHWP} stack is expected to be less than 0.7\% at 
all frequency bands. 

After the final ARC was applied we again measured the end-to-end transmission of the \ac{AHWP} as a 
function of frequency, this time for three relative orientations of the {\it two polarizers}, and for each of these 
orientations at every 5$^{o}$ for the \ac{AHWP}. The three polarizers' orientations were (1) polarizers' transmission 
axes co-aligned; (2) at 45$^{o}$ to each other; and (3) at 90$^{o}$ to each 
other (see also \citet{pisano2006,savini2009,moncelsi2014}). The measurements were done at room temperature. 
The effective ordinary axis for a broad band radiation 
with a low-pass cut-off at 600~GHz is marked on the side of the stack and is later used for alignment and inside
the \ac{EBEX} receiver. Using the transmission measurements we calculated the predicted Mueller matrix elements
as a function of frequency, and \ac{PME} per band assuming top hat band shapes. We find average \ac{PME}'s 
that are larger than 90\% for all three bands giving a total fractional bandwidth of 109~\%.  
Section~\ref{sec:polarizationcalibration} discusses the calibration of polarization rotation and measurements
of the \ac{PME} with the receiver. 

A 45~cm inner diameter wire grid is used to analyze the polarization information modulated by \ac{AHWP}. The plane 
of the grid is oriented at 45 degrees to the incident radiation such that radiation linearly polarized in one direction
is reflected to the \emph{V} focal plane, and radiation polarized in the orthogonal direction is transmitted to the 
\emph{H}; see Figure~\ref{fig:cryo_cutaways}. The wire grid is made by 
photo-lithographing 400~nm thick copper lines on 2.5 \micron\ thick mylar. The lines have a pitch of 20~\micron\ 
with 10~\micron\ wire spacing.  
The transmission of the grid for radiation parallel (perpendicular) to the transmission axis was measured
to be larger than 98.7\% (less than 0.14\%) in the \ac{EBEX} bands. 

The transmission axis of the grid is determined by using a microscope to find the orientation of the lithographed lines
relative to the grid holder. The error in this measurement is 0.25$^{o}$ degrees. 

\subsection{Continuous Rotation Mechanism} 
\label{sec:continuousrotation}

We considered using various types of mechanical bearing systems to achieve continuous rotation. They 
included small and large diameter stainless steel ball bearings, needle bearings with ceramics, and 
bearings made of special materials such as Teflon and Vespel. 
We had two requirements for a successful implementation: (1) low power dissipation, where low was 
defined as 10\% or less of the 360~mW total power load on the liquid helium stage, and (2) 
\ac{HWP} rotation life-time exceeding 
2~million rotations at cryogenic temperatures without significant wear-and-tear to sustain 
the entire long duration flight at rotation rates of up to 2~Hz including margin.  
Available information about friction and experimental testing indicated that 
none of the mechanical systems we considered satisfied both requirements
when considering the size and weight of the \ac{EBEX} rotor. Mechanical bearing failure due to wear-and-tear 
invariably resulted in a sharp increase in power dissipation. 
We thus chose a high temperature \ac{SMB} that provided a no-wear, microphonics-quiet, and relatively 
low-power dissipation rotation. 

\begin{figure}[ht!]
\begin{center}
\includegraphics[height=3in]{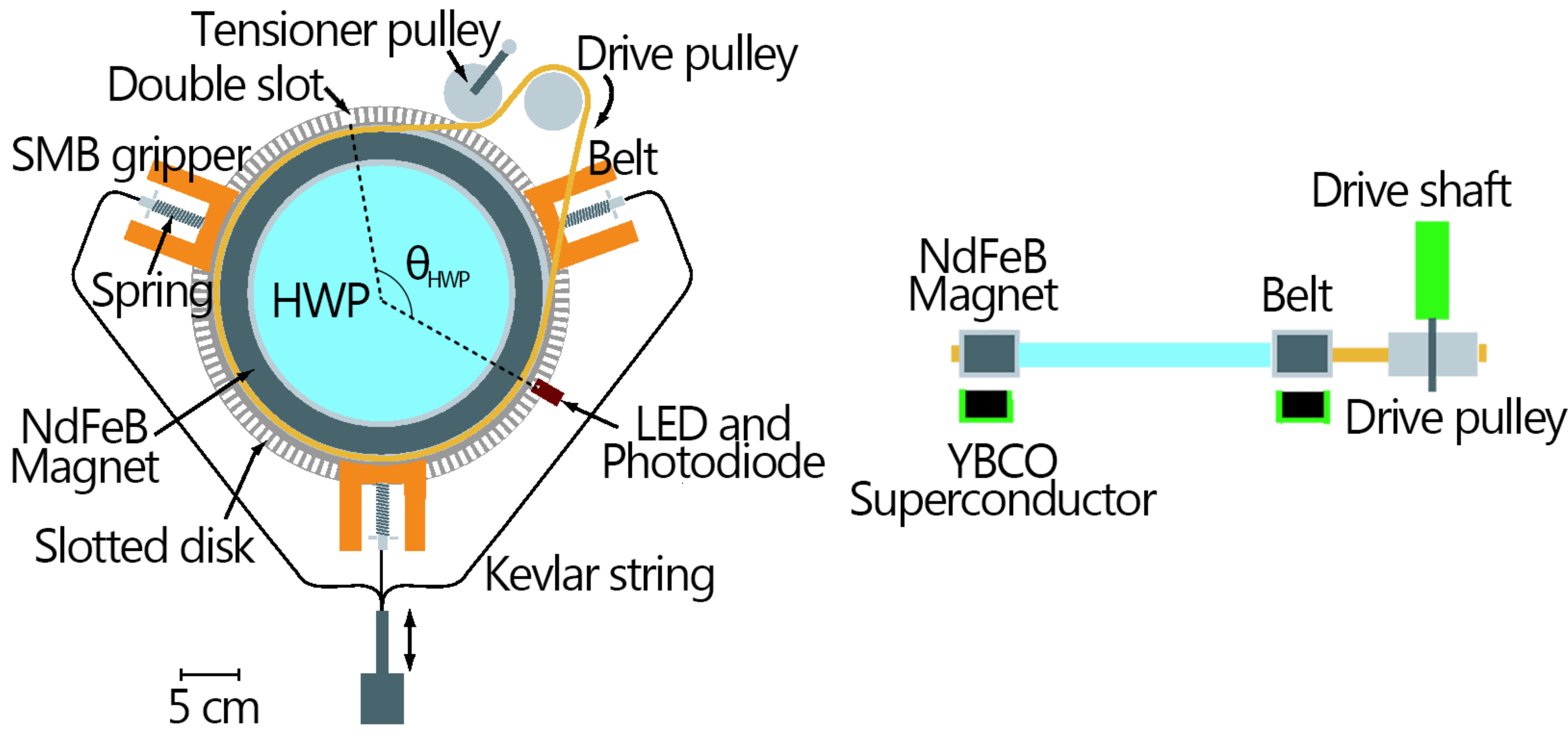}
\par\end{center}
\caption{
Left: Top (left) and side (right) views of the \ac{SMB} system. At temperatures below 95~K the rotor, made of 
NdFeB magnets and the \ac{HWP} levitate above the stator, made of YBCO superconductor tiles. 
A kevlar belt transfers the motion of a drive pulley to the rotor. A tensioner pulley keeps the kevlar tensioned.  
The drive pulley is coupled to a motor external to the cryostat. Three spring-loaded mechanical grippers support 
the rotor at temperatures above 95~K. They are connected with kevlar strings to a linear actuator. A combination 
of disk with slots, an LED and a photodiode are used to encode rotational position.  
\label{fig:HWP-setup}}
\end{figure}

\begin{figure}[tbph]
\begin{centering}
\includegraphics[height=2.75in]{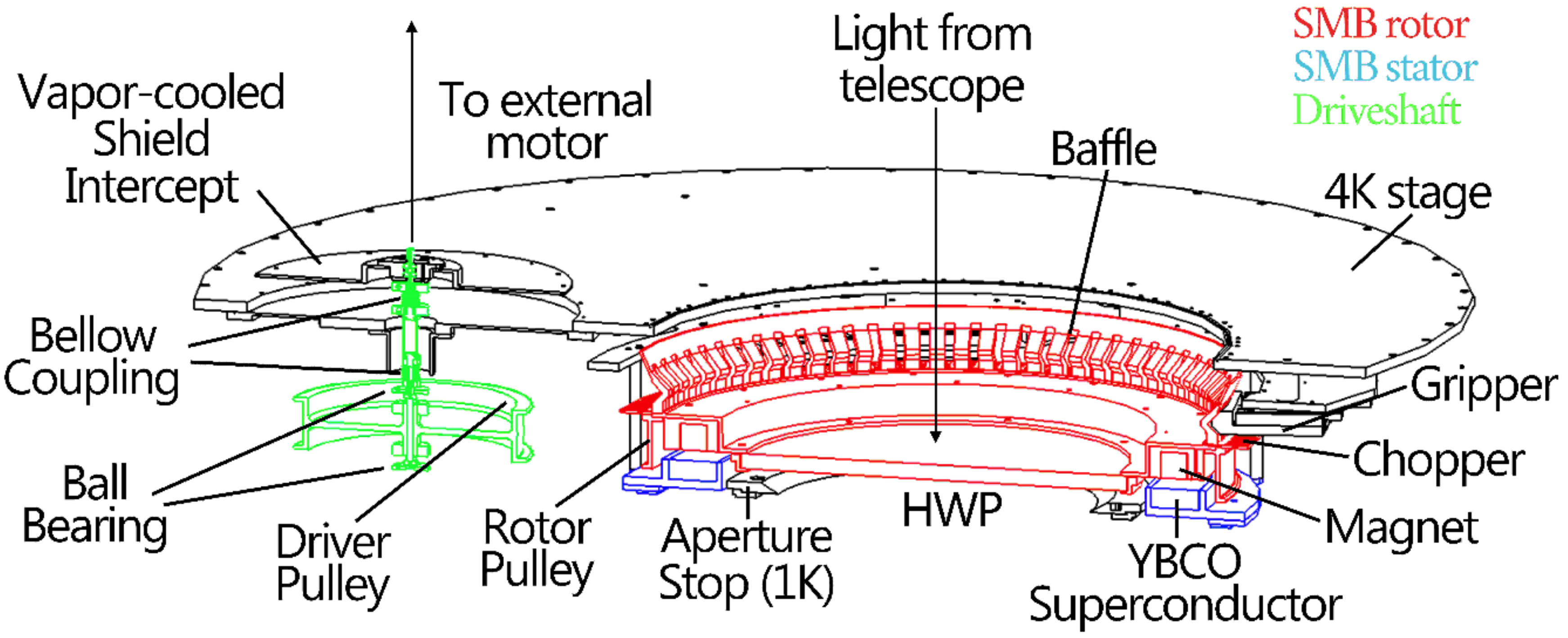}
\end{centering}
\caption{A cutaway of the implementation of the \ac{SMB} with color coding for the different 
functional elements.
The drive pulley and shaft are shown in green, rotating elements ---
the \ac{SMB} magnet, \ac{HWP}, holder, and baffle --- in red, and the \ac{SMB} stator
--- the YBCO superconductor tiles and holder --- in blue. 
\label{fig:cutaway} }
\end{figure}

\citet{hanany03} proposed using a \ac{SMB} for continuous rotation
of an \ac{HWP} and demonstrated a prototype system. Subsequent papers 
presented characterization of coefficient of friction, vibrational amplitudes, resonant frequencies,
and sources of power dissipation~\citep{hull1994, tomo_thesis}. An earlier paper discussed
the implementation of this system in the \ac{EBEX} 2010 North American
test flight~\citep{klein_HWP}. In this section we review the \ac{EBEX2013} implementation (see 
also~\citet{klein_thesis}). Figures~\ref{fig:HWP-setup} and~\ref{fig:cutaway} give 
a functional sketch and details of the technical implementation, respectively. 

\subsubsection{Superconducting Magnetic Bearing} 
\label{sec:smb}


The \ac{SMB} consisted of a stator made of tiles of YBCO and a rotor made of segments of NdFeB 
magnet\footnote{Adelwitz Technologiezentrum GmbH, Germany}. The stator is heat sunk to the liquid helium 
bath. It is made of 33 tiles that are glued into a ring with 271~mm and 331~mm inside
and outside diameters, respectively. The entire ring is glued into a G10 FR4 glass epoxy 
composite (G10) holder.  

The ring-shaped rotor was made of two layers, each of 8~arc sections that 
were stacked brick-like on top of each other. The inside and outside diameters were 284~mm 
and 316~mm, respectively, and the total height was 16~mm. All sections were glued into a G10 holder. 
The axial mean field at 5 mm distance from the surface at the mean radius was 2.1~$\pm$~0.1~kG. 
The \ac{AHWP} was mounted to the rotor with an aluminum holder. A wavy washer and an indium 
wire enhanced the thermal contact between 
the \ac{HWP} and the holder. The aluminum holder also had a slotted baffle that was painted with highly emissive 
material to increase the radiative coupling between the rotor and other receiver cold surfaces~\citep{bockgoop}. 
It was slotted to eliminate eddy currents induced by inhomogeneities in the magnetic field frozen into the 
superconductor stator. The overall mass of the rotor, including \ac{HWP}, magnet, and frame, was 5.6~kg and the 
moment of inertia was 0.11~kg~m$^2$.

\subsubsection{Warm Support and Drive Mechanism}
\label{sec:warmsupport}

At temperatures above 95~K the rotor was held 3.2~mm above the superconductor
by a warm support mechanism consisting of three aluminum linear motion
grippers. Each gripper was mounted on two parallel, free-motion, linear stages 
and was pushed into the rotor by a spring. At temperatures below the YBCO critical 
temperature we pulled the grippers to let the rotor levitate above the stator. For the pulling 
we used a linear actuator that was connected with kevlar strings to all the grippers; 
see Figure \ref{fig:HWP-setup}. The actuator maintained its 
last position when its power was turned off. 

The rotor was driven with a belt made of 2.5 cm wide Kevlar tape. The belt connected
the rotor to a pulley. The pulley was driven
by a thin, hollow, low thermal conductance shaft that extended to the outside of the cryostat and 
coupled to a DC brushless motor. 
We used a ferrofluidic vacuum feedthrough at the vacuum jacket of the cryostat\footnote{Ferrotec}.
Inside the cryostat the drive shaft was coupled to the helium vapor-cooled shield and to
the liquid helium stage using two extra-clearance, molybdenum disulfide
dry lubricated stainless steel ball bearings. A tensioner pulley that
was mounted with the same type of bearing was used to maintain belt
tension. Over the \ac{EBEX2013} flight these bearings withstood over 1.5 million rotations 
with no evidence of failure. 

To achieve steady state rotation we used two commandable settings for the 
DC motor, an initial low speed and a final higher speed setting. The initial 0.5~Hz low speed was 
a binary on/off state. When angular encoding showed the rotor rotating stably in its slow mode
we commanded the higher speed state which triggered an RC ramp circuit with a time constant 
of 1~minute that gradually increased the voltage to the motor. 

\subsection{Rotation Angle Encoding and Decoding}
\label{sec:angulardecoding}

We encoded the \ac{HWP} angular position using a chopper wheel
with 240 slots that interrupted a white LED shining onto a photodiode. 
One of the slots was double-width and marked an
arbitrary zero-position that was referenced to the ordinary axis of the \ac{AHWP}. 
The signals from the photodiode were sufficiency large for 
angle reconstruction when the LED consumed only 34~$\mu$W. 

\begin{figure}[htp]
\centering{}%
\begin{center}
\includegraphics[scale=0.7]{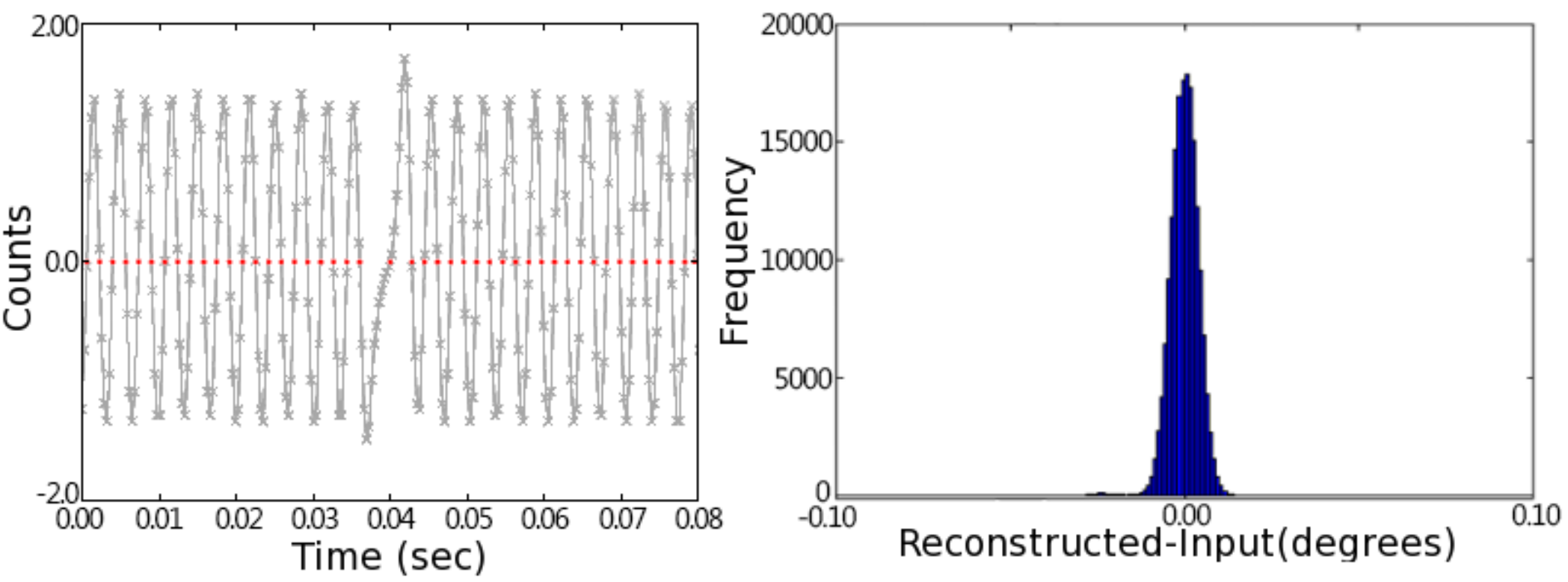}
\caption{Left: raw time domain angle encoding signal over one period of rotation after subtraction 
of an offset and a gradient. The wider pulse in the middle is the arbitrary zero angle mark. Red marks 
denote zero-level crossings. The temporal separation of the marks is assigned a fixed 1.5 deg angle 
separation. Right: histogram of the difference between the input angle and reconstructed angle in
a simulation of the \ac{HWP} angle reconstruction pipeline. The standard deviation is 0.01$^{\circ}$. 
\label{fig:HWP raw} }
\end{center}%
\end{figure}

DfMUX board of the same type used to gather bolometer data recorded and 
time-stamped the photodiode signals.  The
sampling rate was 3 kHz, 16 times higher than that of bolometer data. 
The time domain data of the angle readout was a sinusoidal wave to a good approximation; see Figure~\ref{fig:HWP raw}. 
We reconstructed angle by removing an offset and a gradient for sections that were approximately 1 hour long 
and then identifying times of zero-level crossings. Each zero-crossing was assigned an incremental angle of 
0.75~degrees. Each bolometer sample has its own time stamp and is associated with an angle by 
interpolating adjacent time-stamped angles. 
We constructed an end-to-end simulation of the angle reconstruction that included typical photodiode signal and noise, 
slot machining errors, and \ac{HWP} speed variations. A histogram of angle reconstruction errors gives a 
standard deviation of 0.01~degrees, making angle reconstruction a negligible contributor to the total 
uncertainty of the polarization angle calibration; see Figure~\ref{fig:HWP raw} and Section~\ref{sec:polarizationcalibration}. 

\subsubsection{EBEX2013 Operation and Performance}
\label{sec:smboperation}

For the \ac{EBEX2013} flight we released the warm support and set the \ac{AHWP} rotating at 
a frequency of 1.235~Hz well before payload launch. It continued to rotate without interruption through 
launch. The rotation was stopped and started several 
times during flight, but the warm support was re-engaged only after liquid helium was exhausted. 
Over the duration of the flight the rotor executed 584,000 rotations at the nominal
 rotation speed. The 
total number of rotations is slightly higher because we exclude a duration of about 4~hours that includes periods
during which the speed ramps up to or down from the nominal rotation speed. 

\begin{figure}[ht!]
\begin{center}
\includegraphics[height=2.5in]{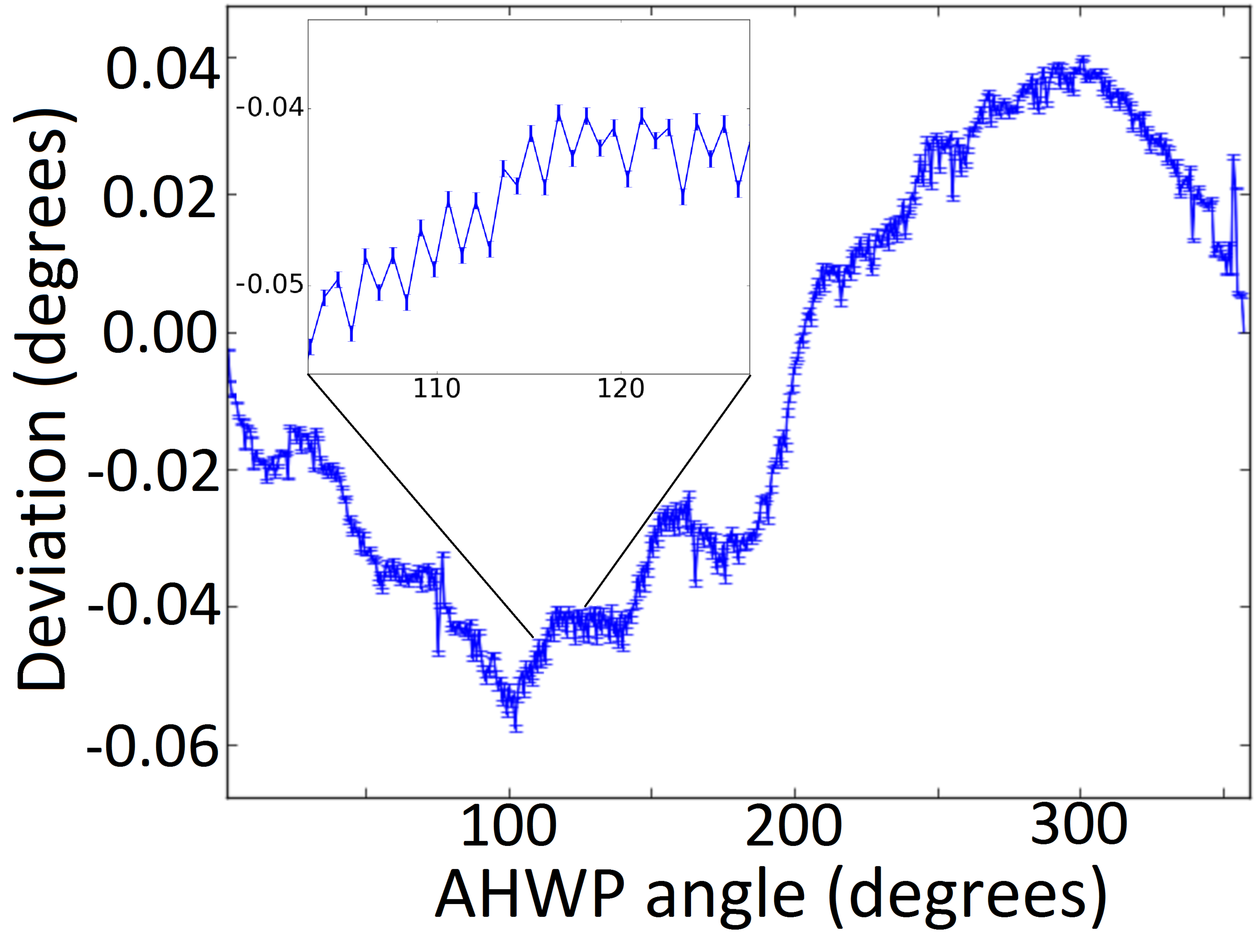}
\par\end{center}
\caption{Angle reconstruction over 4176 rotations, folded rotation by rotation, and averaged. A linear 
increase between 0 and 360~degrees has been removed. The deviations from zero are due to speed 
variations. The inset shows the effects of a non-uniform duty cycle in the chopper encoding the angle; see text.  
\label{fig:binnedangle} }
\end{figure}

\begin{figure}[ht!]
\begin{center}
\includegraphics[height=2.5in]{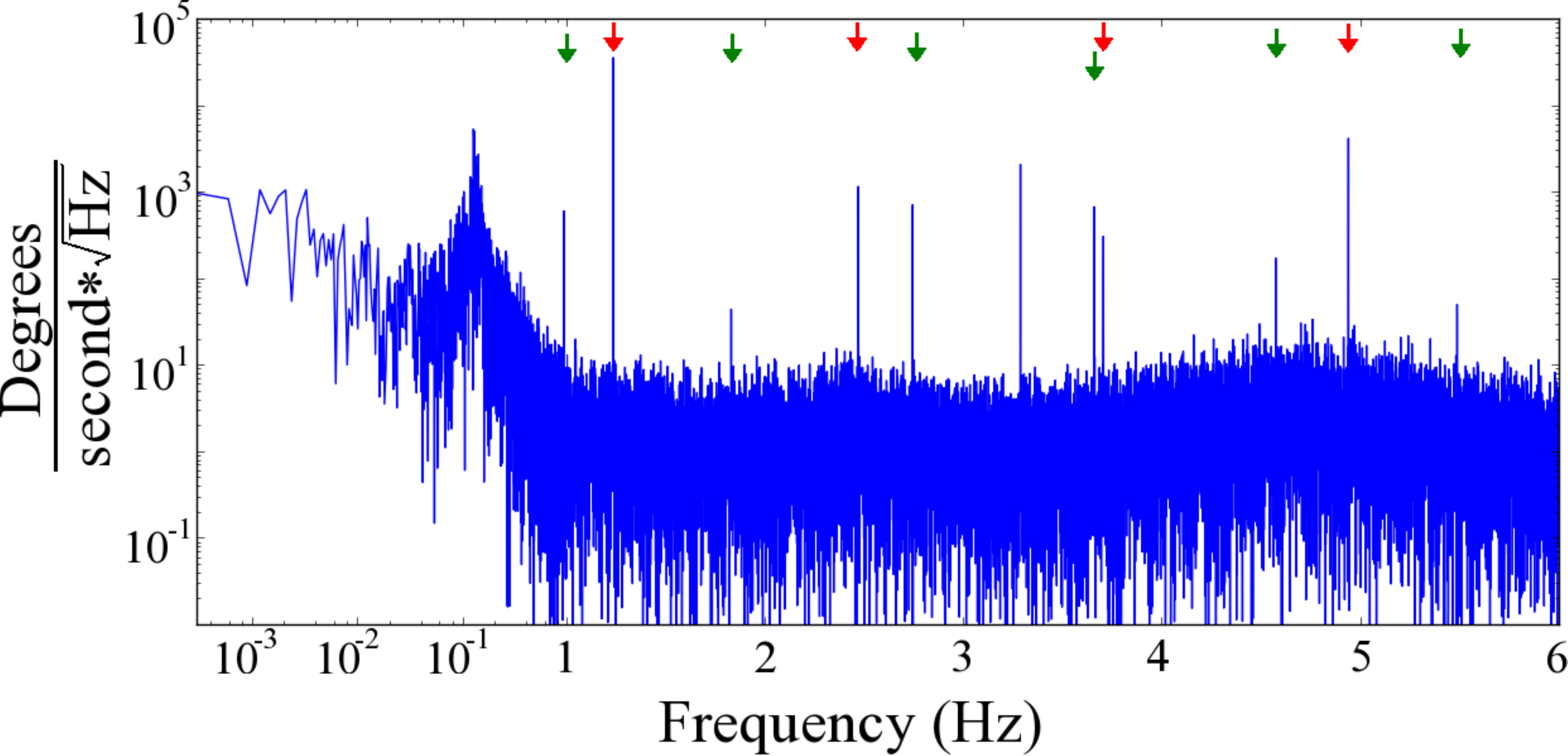}
\par\end{center}
\caption{Power spectrum of 60~minutes of \ac{SMB} rotation speed data.  The horizontal axis is logarithmic 
at frequencies below 1~Hz and linear above. High $Q$ peaks are present at the rotation speed and its harmonics (red
arrows) and at 0.91~Hz and its harmonics (green arrows). 
\label{fig:HWP-speed}}
\end{figure}

Figure~\ref{fig:binnedangle} shows angle reconstruction over a period of 56~min that was stacked rotation by rotation 
and averaged. A linear increase between 0 and 360~degrees was subtracted. The deviations displayed are 
the result of speed variations that were repeatable over many rotations. The most prominent deviation 
had a period of one rotation and was likely due to an overall dipole structure in the strength of the magnetic 
field between the rotor and stator. This period also gave rise to the largest amplitude peak in the power spectrum of the 
rotation rate, as shown in Figure~\ref{fig:HWP-speed}. The saw-tooth structure shown in the inset of Figure~\ref{fig:binnedangle} 
arises from the erroneous assumption during the angle reconstruction that each zero crossing 
corresponds to exactly 0.75~degrees. The construction of the chopper wheel and the alignment of the LED/photodiode gave
closed areas that were systematically slightly larger than the open ones; thus the assumption that 
each zero-crossing corresponds to 0.75~degrees gave a reconstructed angle per slot that was wrong by less than 0.005~deg. 
We did not attempt to correct for this error, which was negligible compared to other uncertainties. 

The rotation rate of the \ac{SMB} extracted from the reconstructed angle gives RMS speed variations of 
1.5\% over a period of 10~hours. This level decreases to 0.45\% when we first average 
the reconstructed angle over a full chopper slot period, namely over 1.5~degrees, thus eliminating 
the 0.005 degree systematic error in angle reconstruction discussed in the previous 
paragraph. Analysis of the data shown in Figure~\ref{fig:HWP-speed} reveals 
that 80\% of the RMS speed variations are due to broad-band 
readout and mechanical noise, rather than specific system resonances. 
The speed variation at frequencies near 0.1~Hz is much slower than the rotation rate of the rotor and 
we hypothesize that these variations are due to a torsional mode of the thin drive shaft. Figure~\ref{fig:HWP-speed} 
also shows high $Q$ peaks at harmonics of the rotational frequency and others at 0.91~Hz and 
harmonics thereof. We do not know the origin of the $0.91\cdot j\, (j=1,2,3,...)$~Hz peaks, 
but the high $Q$ suggests that these are not due to the tensioner pulley. 

Preflight measurements indicated that the power dissipation at 1.2~Hz was 15~mW and
that approximately half of that power came from Joule heating due to eddy currents and half due to friction in the
Kevlar belt and bearings of the pulleys~\citep{klein_thesis}. 
This estimate of total power dissipation proved accurate as the in-flight
total liquid helium hold time matched predictions within~2\%.


\subsection{Polarization Calibration}
\label{sec:polarizationcalibration}

The \ac{TOD} of a noiseless instrument with a combination of continuously rotating \ac{HWP} and 
wire grid analyzer is
\begin{equation}
  D(t) = \frac{1}{2} \left[ I + \epsilon Q \cos(4\gamma(t) -\Phi )  + 
  \epsilon U \sin ( 4\gamma(t) -\Phi ) \right],
\label{eqn:polcal1}
\end{equation}
where the incident polarized radiation has Stokes vector $I, Q, U$,
$\epsilon$ is the \ac{PME}, $\gamma = \omega_{hwp} t $ is the instantaneous angle of the \ac{HWP}, $\omega_{hwp} = 2\pi f_{hwp}$
 is the rotation rate of the \ac{HWP}, 
and $\Phi$ is an overall offset that encodes the angle between the orientation 
that defines  $\gamma = 0$ and a frame that defines the $Q, U$ coordinate system.  
In EBEX, $\gamma$ is the angle between the double slot of the angle encoder and its LED; $\gamma = 0 $ 
is when the double-slot is illuminated. For $Q,U$ 
it is typical to use either a frame that is locked to the instrument
or one that is locked to the celestial sphere.  The EBEX polarization calibration relied only on ground-based 
measurements as there were no bright astrophysical mm-wave sources with sufficiently high polarization 
to use as calibrators during flight.  
Therefore, for the calibration we used a $Q,U$ system that is referenced to the instrument. 
The orientation $+Q$ was in the symmetry plane of the optical system, perpendicular
to the optical axis of the receiver, and aligned with the $x$ direction shown in 
Figure~\ref{fig:raydiagram}. The $+U$ direction is at $+45$~degrees relative to $Q$ (positive angles
follow the right hand rule).
Calibration of the instrumental polarimetric response consists of 
determining the offset angle $\Phi$ (see Section~\ref{sec:polarizationrotation}), 
and the \ac{PME} $\epsilon$ (see Section~\ref{sec:pme}).

Equation~\ref{eqn:polcal1} is idealized. It neglects instrumental polarization and ignores spurious input Stokes vectors. 
Adding these elements, Equation~\ref{eqn:polcal1} becomes
\begin{eqnarray}
  D(t) & =  & \frac{1}{2} [ I + \epsilon Q \cos(4\gamma(t) -\Phi )  +
  \epsilon U \sin ( 4\gamma(t) -\Phi ) ] + \nonumber \\
  &  & \sum_{j=0}^{j=\infty} A_{j} \cos( j\gamma(t) + \phi_{j} ).  
\label{eqn:polcal2}
\end{eqnarray}
We have lumped all spurious effects into the coefficients $A_{j}$, the phases $\phi_{j}$, and phenomenologically allowed them 
to be present at all harmonics of the rotation frequency. In practice the first few terms in the expansion are dominant and we find that they vary with time on time scales slow compared to the rotation rate. 

\subsubsection{Polarization Rotation} 
\label{sec:polarizationrotation}


Calibrating the polarization rotation consists of illuminating the experiment with nearly 100~\% linearly 
polarized light of known orientation, expressible as $\alpha_{in} = (1/2)\arctan(U/Q)$, and determining $\Phi$
from the \ac{TOD}.  We used a standard lock-in technique. 
Multiplying Equation~\ref{eqn:polcal2} by $\cos(4\gamma)$ 
and $\sin(4\gamma)$ and low-passing to keep only the terms at zero frequency gives
\begin{eqnarray}
    D\cos(4\gamma)  & = & \frac{\epsilon}{4} \left( \frac{Q}{\cos(2\alpha_{in})} \right) 
                  \cos(2\alpha_{in} + \Phi) + A_4 \cos(\phi_4) \\
    D\sin(4\gamma) & =  & \frac{\epsilon}{4} \left( \frac{Q}{\cos(2\alpha_{in})} \right) 
                  \sin(2\alpha_{in} + \Phi) - A_4 \sin(\phi_4) .
\end{eqnarray}
The spurious term characterized by $A_{4}$ and $ \phi_{4}$ represents, for example, incident polarization 
other than the intended source. Without it we readily have 
\begin{equation}
\Phi = \arctan \left( \frac{D \sin (4\gamma) }{ D \cos (4\gamma) } \right) -2\alpha_{in}.
\label{eqn:phi_expression1}
\end{equation}
To discriminate against spurious polarization signals we temporally modulated, a.k.a `chopped', the 
source at frequency $\omega_{c}=2 \pi f_{c}$; in reference to Equation~\ref{eqn:polcal2}, 
$I, Q$, and $U$ where chopped but the term $A_{4}$ was not.  
The \ac{HWP} was rotating continuously at frequency $\omega_{hwp}$. The combination of the temporal 
modulation of the polarized source together with the continuous rotation of the \ac{HWP} placed the 
polarization signals of interest at two sidebands of $4\omega_{hwp}$: $4\omega_{hwp} + \omega_{c}$ and
 $4\omega_{hwp} - \omega_{c}$. 
We used a {\it double} demodulation technique to reject
the spurious polarization. We describe the double demodulation 
technique in Appendix~\ref{sec:polcaleqns} and show how it gives Equation~\ref{eqn:phi_expression1}. 
The bolometer response time $\tau$ introduced a phase shift between the recorded and input orientation of a 
polarized signal. We corrected for the time constant by estimating each bolometer's time constant $\tau$ 
and deconvolving the \ac{TOD} using a single pole filter response. 
The process of time constant estimation and deconvolution is described in Appendix~\ref{sec:timeconstants}. 
Detailed time constant measurements showed that a single pole filter model is only 
approximate and introduces a bias in the determination of $\Phi$. We determined the bias $\delta \Phi$
using simulations, as described in Appendix~\ref{sec:timeconstants}. We report a measured $\Phi$ that 
already includes this bias correction. 



We measured the angle $\Phi$ using two configurations: a configuration with only the receiver (without
the warm telescope), which we call `receiver only', and a configuration with the entire instrument including the warm telescope, 
to which we refer as `entire instrument'. 
For the `receiver only' calibration, a wire grid polarizer was mounted to the vacuum window 
of the cryostat. The signal to the detectors was modulated between a room temperature chopper blade blackened 
with Eccosorb LS140 and a styrofoam bucket containing Eccosorb CV3\footnote{Emerson and Cuming, Inc.}
and liquid nitrogen that filled the light throughput entering the receiver. Measurements of the orientation 
of the wire grid transmission axis gave the orientation of the incident polarization to better than $0.1^{\circ}$. 
\begin{figure}[htp]
\begin{center}
\includegraphics[width=10cm]{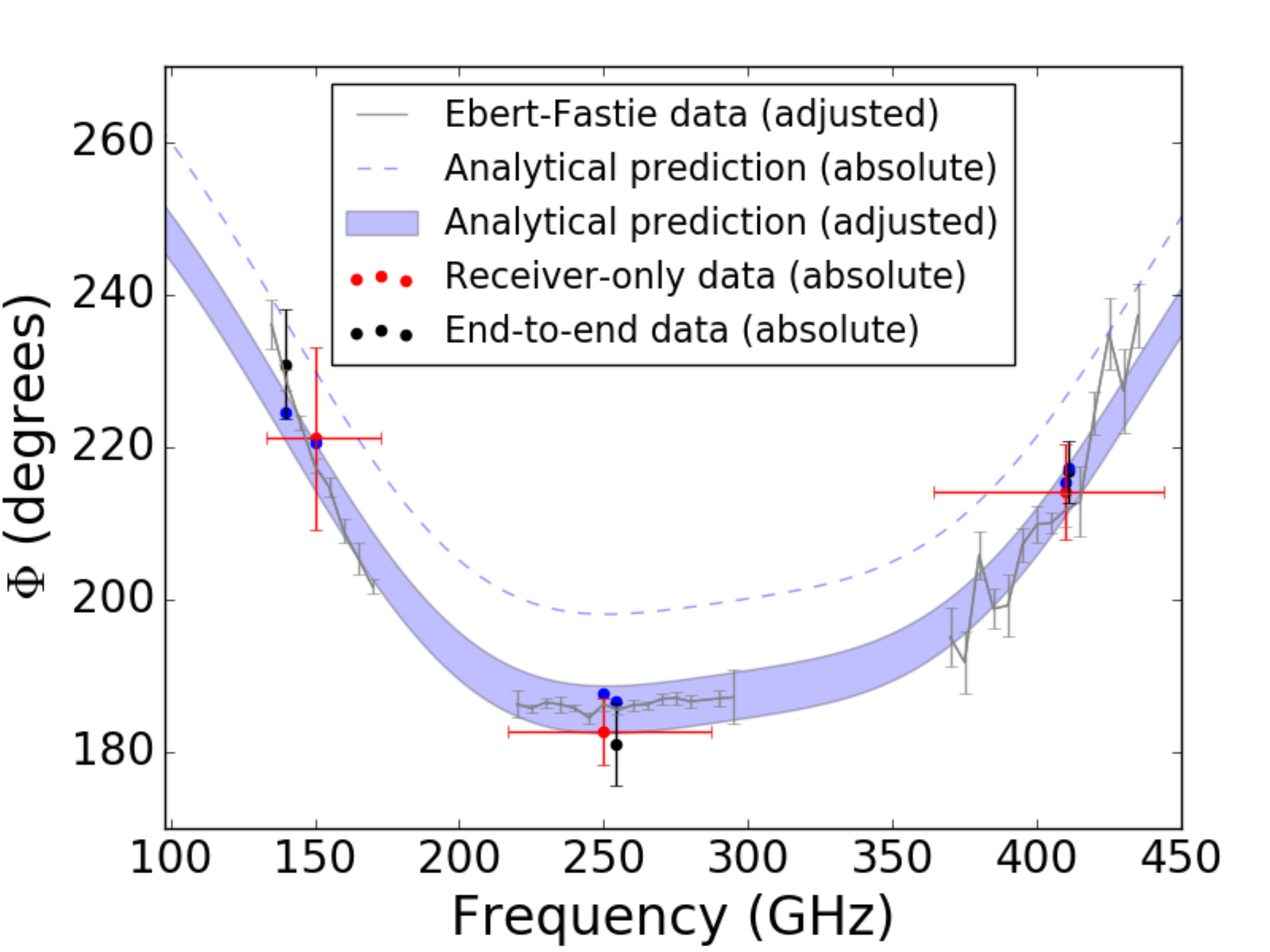} 
\caption{Results of the two polarization rotation calibration tests (black and red data points) and the 
predicted phase $\Phi$ as a function of frequency  (blue dash and band, see text). We also measured the relative 
phase within each band using an Ebert-Fastie spectrometer (grey); all the data points for a given band 
were adjusted with a single arbitrary phase offset to fit the predicted phase.
\label{fig:polcalresults} }
\end{center}
\end{figure}

For the end-to-end calibration we positioned a coherent source atop a 50~m tower that was 104~m 
away from the payload. 
The source was chopped on/off electronically, and a polarizing wire grid was placed 
at the output of the horn so as not to rely on its polarization properties.   
The entire instrument, including receiver, warm telescope, and gondola, scanned the source to construct 
antenna patterns of the telescope. 
We determined the relative orientation of the source and telescope by measuring the orientation of each relative 
to the gravitational acceleration vector using commercial tiltometers. The transmission axis of the polarizing grid 
was aligned parallel
to the symmetry plane of the optical system to within $0.1^{\circ}$, giving input polarized light aligned with
$+Q$ in our reference frame.

The measured values of $\Phi$ for the calibrations are shown in Figure~\ref{fig:polcalresults}. Since the receiver only 
calibration used a broad-band source the data shown are an average over the band. For the calibration with the entire
instrument we used a coherent source and the data are valid only at the frequencies shown. The error bars 
are the standard deviations of the values measured among different detectors of a given frequency band. 
For the receiver only (end-to-end) we measured 100, 22, and 69, (42, 56, 151) detectors for the 150, 250, 
and 410~GHz bands, respectively. 

We also used the Ebert-Fastie spectrometer described in Section~\ref{sec:filtersandbands} to ascertain that the 
change in the phase $\Phi$ {\it within} each band matched predictions. We placed a wire grid polarizer at the 
output of the spectrometer. Using the double demodulation technique described in Appendix~\ref{sec:polcaleqns}
we extracted a phase angle $\phi$ as a function of frequency $\nu$. We only measured the variation of $\phi(\nu)$ 
within a band; we did not attempt to determine an offset that would determine $\phi(\nu)$ relative to $\Phi(\nu)$.  
The data are shown in Figure~\ref{fig:polcalresults}; the data of each band was give a single arbitrary phase offset 
for the entire data of each band to match the predictions. 

\begin{figure}[ht!]
\begin{center}
\includegraphics[height=2.5in]{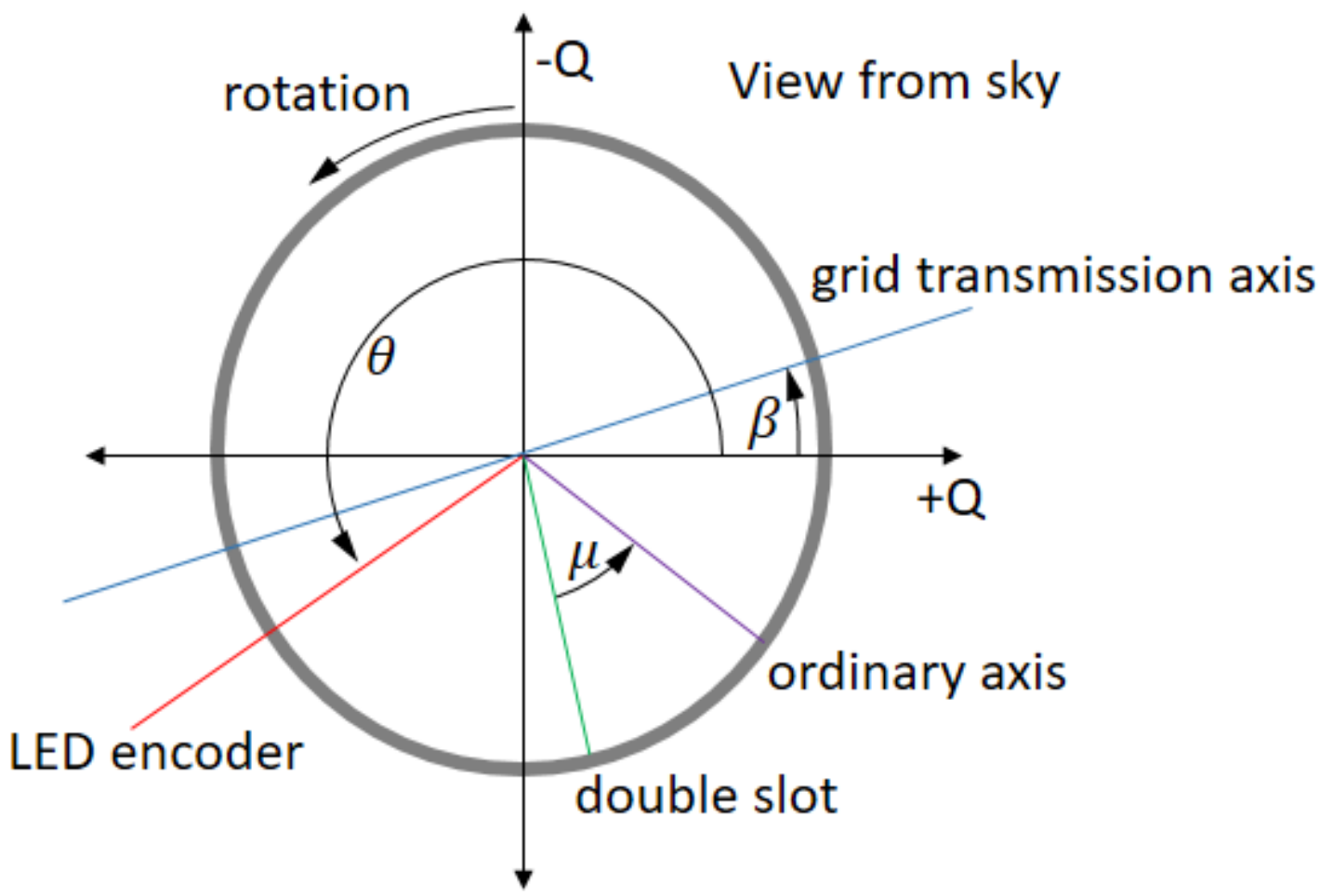}
\par\end{center}
\caption{The angles that determine the calibration of polarization rotation. 
\label{fig:polcal_angles}}
\end{figure}

We compared the measured $\Phi$ to predictions. The phase $\Phi$ is given by  
\begin{equation}
  \Phi =  - 4\theta + 2\beta - 4\mu - 4\delta_{readout} - 4\Delta_{\phi}(\nu) ,
\label{eqn:phi_expression2}
\end{equation}
where $\theta$ is the angle between the angle-encoding LED and the $+Q$ orientation, 
$\beta$ is the angle between the transmission axis of the polarizing grid and $+Q$, and $\mu$ is the 
the angle between the ordinary axis of the \ac{AHWP} and the double-slot; see Figure~\ref{fig:polcal_angles}. 
Time delays between the readout of the \ac{HWP} angle and the detector data
give rise to an offset angle encoded by $\delta_{readout}$. 
The achromatic nature of the \ac{HWP} gives a frequency dependent offset 
$\Delta_{\phi}(\nu)$ (see Section~\ref{sec:hwpandgrid}). Following the right hand rule, 
a positive angle corresponds to a counter-clockwise rotation when viewing the focal plane from above.

\begin{table}[htpb]
\begin{center}
\begin{tabular}{|c|c|c|}
\hline 
 & Receiver-only Test & End-to-end Calibration\tabularnewline
\hline
\hline 
150 GHz & $2.8^{\circ}$ & $2.6^{\circ}$\tabularnewline
\hline 
250 GHz & $10.4^{\circ}$ & $10.6^{\circ}$\tabularnewline
\hline 
410 GHz & $2.9^{\circ}$ & $3.5^{\circ}$\tabularnewline
\hline
\end{tabular}
\end{center}
\caption{Phase shift due to the achromatic nature of the \ac{AHWP}, for each
of the calibration test setups. 
\label{tab:AHWP_phase} }
\end{table}

We used a coordinate measuring machine to determine the axes defining 
the angles $\theta, \beta$, and $\mu$.  Through an analysis of the firmware we determined that $\delta_{readout}$ 
was 10.69 ms. This delay corresponded to a 
7.6$^{\circ}$ rotation of the AHWP at the 
1.98~Hz rotation frequency of the \ac{AHWP} used 
during both calibrations.  The orientation of the effective ordinary axis of the \ac{AHWP} is frequency-dependent. 
The effective ordinary axis over a band up to $\sim$600~GHz 
was determined and marked on the \ac{AHWP} using the Fourier transform spectroscopy transmission 
measurements described in Section~\ref{sec:hwpandgrid}. 
We used measured values for the index of refraction of sapphire at
cryogenic temperatures~\citep{Loewenstein73, Afsar91} to model the expected frequency
dependence of the \ac{AHWP} at its operating temperature. 
The results for the prediction of $\Phi$ are shown in Figure~\ref{fig:polcalresults} (blue dash). 
We find an offset of 12~degrees between 
the prediction and the measurements and we ascribe this difference to an offset of 
3~degrees in the determination of the axis defining $\mu=0$. This is the ordinary axis of the 
\ac{HWP} as determined during spectroscopy in the zero-path difference position. The adjusted 
prediction is shown in the figure as a blue band. 
There is an uncertainty in the determination of each of the terms in Equation~\ref{eqn:phi_expression2}. 
This uncertainty leads to an uncertainty in the predicted $\Phi$, which gives the width of the blue band. 
The sources of uncertainty are tabulated in Table~\ref{tab:Phi_uncertainties}. 
\begin{table}[htpb]
\begin{center}
\begin{tabular}{|c|c|c|}
\hline 
 Parameter & Uncertainty & Source\tabularnewline
\hline 
$\theta$  & $<0.1^{\circ}$  & Geometrical measurement error  \tabularnewline
\hline 
$\beta$ & \begin{tabular}{@{}c@{}} $<0.1^{\circ}$ \\ $0.25^{\circ}$\end{tabular} & \begin{tabular}{@{}c@{}} Geometrical measurement error\\ Uncertainty in transmission axis measurement\end{tabular}  \tabularnewline
\hline 
$\mu$  & $<0.1^{\circ}$  & Geometrical measurement error  \tabularnewline
\hline 
$\Delta_{\phi}$  & $ .7^{\circ}$ &  Uncertainty in \ac{AHWP} model \tabularnewline
\hline
$\delta_{readout}$ & $<0.1^{\circ}$ & \ac{AHWP} speed variations  \tabularnewline
\hline
\cline{1-2}
Total & $0.8^{\circ}$ \tabularnewline
\cline{1-2}
\end{tabular}
\end{center}
\caption{Uncertainties for each of the parameters in the prediction of $\Phi$. The total is a quadrature addition 
of the individual contributions. 
\label{tab:Phi_uncertainties} }
\end{table}


\subsubsection{Polarization Modulation Efficiency} 
\label{sec:pme}

\begin{figure}[ht!]
\begin{center}
\includegraphics[width=12cm]{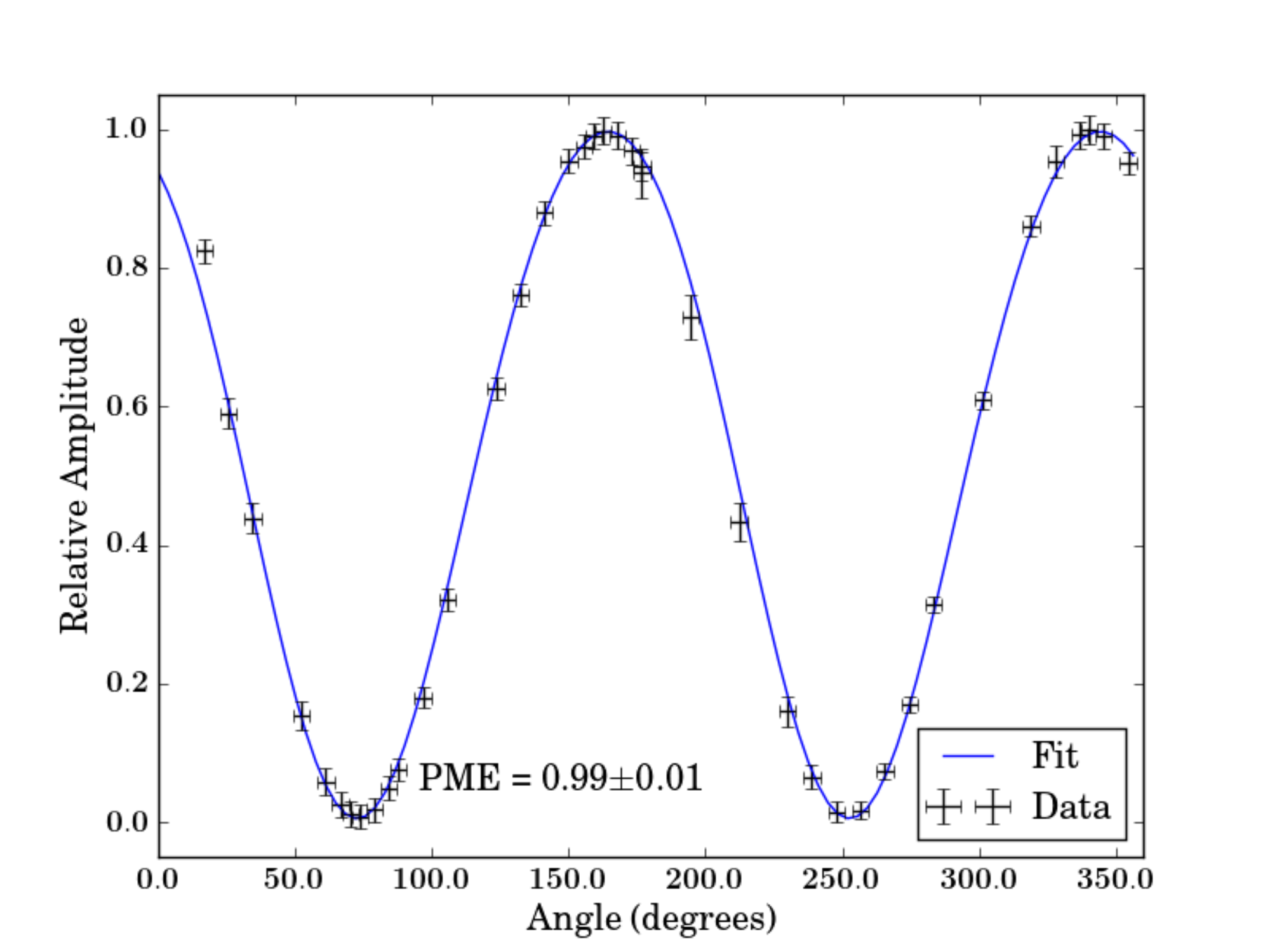} 
\caption{A measurement of the \ac{PME} with one 250~GHz detector. Data points are the 
detector response at the frequency of the temporally modulated input radiation as a function of the rotation angle 
of the grid defining the input polarization. }
\label{fig:pme}
\end{center}
\end{figure}

The \ac{PME} of a polarimeter is the ratio of the measured to input level of polarization. 
We deduce the predicted \ac{PME} of the \ac{AHWP} from Fourier transform spectroscopy transmission 
measurements as described in Section~\ref{sec:hwpandgrid}. 
The predicted \ac{PME} for the three frequency bands is given in Table~\ref{tab:pme}. 
We measured the \ac{PME} using the Ebert-Fastie
spectrometer, described in Section~\ref{sec:filtersandbands}. We replaced the diffraction 
grating with a flat aluminum panel to enable broad band measurements with the black body source. 
We placed a polarizing grid at the output of
the Ebert-Fastie so the light entering the receiver was to a good approximation fully linearly
polarized. The polarized light was temporally chopped by the chopper at the output of the spectrometer 
and coupled to one detector at a time. We measured the \ac{PME} by 
monitoring the detector response to chopped light as a function of orientation of the Ebert-Fastie polarizer.  
An example measurement of a 250~GHz detector is shown in Figure~\ref{fig:pme}.
The average \ac{PME} measured for each of the three frequency bands is 
given in Table~\ref{tab:pme}. To our knowledge the \ac{EBEX} \ac{AHWP} fractional bandwidth of 109~\% is 
the broadest band response of any \ac{HWP} reported to date.

\begin{table}[ht]
\begin{centering}
\begin{tabular}{|c|c|c|}
\hline 
 & Prediction & Measured\tabularnewline
\hline
\hline 
150 GHz & 0.99 & $0.98 \pm 0.06$\tabularnewline
\hline 
250 GHz & 1.0 & $0.98 \pm 0.02$\tabularnewline
\hline 
410 GHz & 0.92 & $0.92 \pm 0.06$\tabularnewline
\hline
\end{tabular}
\par\end{centering}
\caption{Predicted and measured \ac{PME} for each frequency band. The predictions were 
calculated using the Fourier transform spectroscopy measurements described in 
Section~\ref{sec:hwpandgrid} and assume a flat input spectrum. The measured values are 
a weighted average over 7, 4, and 9 detectors at 150, 250, and 410~GHz, respectively. 
\label{tab:pme}  }
\end{table}

\newpage

\section{Summary}
\label{sec:summary}

\ac{EBEX} was the first balloon-borne experiment to probe the CMB polarization with a kilo-pixel 
array of transition edge sensor bolometric detectors. The optical system
consisted of two ambient temperature aluminum mirrors, and four cryogenically-cooled \ac{PE} lenses to 
increase the throughput and to form a flat and telecentric focal surface for each of the two focal planes. 
The total throughput per focal plane was 115~cm$^{2}$sr. 

We used a stack of 5 sapphire half-wave plates to form an \ac{AHWP} that had a fractional bandwidth of 109~\%. 
To our knowledge, this is the broadest fractional bandwidth with which an \ac{AHWP} was used to date. 
The \ac{AHWP} was levitated by means 
of a superconducting magnetic bearing, the first such application in astrophysics. It was rotated continuously at a 
rate of 1.235~Hz with power dissipation of 15~mW on the 4~K stage. Six mechanical bearings that were implemented as 
part of the drive train executed more than 1.5 million rotations over 10 days. Rotational speed 
stability was 0.45\% over period of 10 hours when averaging angular decoding over 1.5~degrees. 
We reconstructed angle with an uncertainty of 0.01~deg, more than a factor of 
10 smaller than required for measurements of the tensor to scalar ratio of $r=0.05$.  
We calibrated the polarimetric response using double temporal modulation to eliminate sources
of systematic uncertainty.  

\ac{EBEX} had three frequency bands centered on 150, 250, and 410~GHz, all sharing the same 
refractive/reflective optical  train. 
We used broad-band anti-reflection coating 
that consisted of glued layers of porus Teflon on the lenses and a combination of TMM and Teflon on the \ac{AHWP}. 
Polyethylene and Stycast were used as glues. 
As a consequence the system has an approximately achromatic response with 60-70\% transmission 
over the entire range of frequencies. For an optical system using refractors, this is the broadest fractional 
bandwidth operated simultaneously by any CMB experiment to date. 

The receiver that was built to house the cold optics performed according to specifications, giving focal plane temperatures
near 250~mK and temperature stability with $1/f$ knee at 2~mHz. We implemented a double vacuum window 
mechanism to reduce emission from a thick vacuum window at float altitudes. 

With \ac{EBEX} we were the first to implement on a balloon platform many of the technologies characterizing 
present-day large throughput CMB polarimetric instruments. These use large throughput optical systems and 
arrays with thousands of transition-edge bolometric sensors that are massively multiplexed. For polarimetry we used a 
continuously rotating \ac{AHWP}, one of an array of techniques being explored by the CMB community in the search 
for the inflationary signal. 

Two companion papers provide additional details about the EBEX detectors, their readout, and their flight performance (\ac{EP2}), 
and about the gondola, the attitude control system, and other support systems (\ac{EP3}).

\begin{acronym}
    \acro{ACS}{attitude control system}
    \acro{ADC}{analog-to-digital converters}
    \acro{ADS}{attitude determination software}
    \acro{AHWP}{achromatic half-wave plate}
    \acro{AMC}{Advanced Motion Controls}
    \acro{ARC}{anti-reflection coating}
    \acro{ATA}{advanced technology attachment}
    \acro{BRC}{bolometer readout crates}
    \acro{BLAST}{Balloon-borne Large-Aperture Submillimeter Telescope}
    \acro{CANbus}{controller area network bus}
    \acro{CMB}{cosmic microwave background}
    \acro{CMM}{coordinate measurement machine}
    \acro{CSBF}{Columbia Scientific Balloon Facility}
    \acro{CCD}{charge coupled device}
    \acro{DAC}{digital-to-analog converters}
    \acro{DASI}{Degree~Angular~Scale~Interferometer}
    \acro{dGPS}{differential global positioning system}
    \acro{DfMUX}{digital~frequency~domain~multiplexer}
    \acro{DLFOV}{diffraction limited field of view}
    \acro{DSP}{digital signal processing}
    \acro{EBEX}{E~and~B~Experiment}
    \acro{EBEX2013}{EBEX2013}
    \acro{ELIS}{EBEX low inductance striplines}
    \acro{EP1}{EBEX Paper~1}
    \acro{EP2}{EBEX Paper~2}
    \acro{EP3}{EBEX Paper~3}
    \acro{ETC}{EBEX test cryostat}
    \acro{FDM}{frequency domain multiplexing}
    \acro{FPGA}{field programmable gate array}
    \acro{FCP}{flight control program}
    \acro{FOV}{field of view}
    \acro{FWHM}{full width half maximum}
    \acro{GPS}{global positioning system}
    \acro{HPE}{high-pass edge}
    \acro{HWP}{half-wave plate}
    \acro{IA}{integrated attitude}
    \acro{IP}{instrumental polarization} 
    \acro{JSON}{JavaScript Object Notation}
    \acro{LDB}{long duration balloon}
    \acro{LED}{light emitting diode}
    \acro{LCS}{liquid cooling system}
    \acro{LC}{inductor and capacitor}
    \acro{LPE}{low-pass edge}
    \acro{MLR}{multilayer reflective}
    \acro{MAXIMA}{Millimeter~Anisotropy~eXperiment~IMaging~Array}
    \acro{NASA}{National Aeronautics and Space Administration}
    \acro{NDF}{neutral density filter}
    \acro{PCB}{printed circuit board}
    \acro{PE}{polyethylene}
    \acro{PME}{polarization modulation efficiency}
    \acro{PSF}{point spread function}
    \acro{PV}{pressure vessel}
    \acro{PWM}{pulse width modulation}
    \acro{RMS}{root mean square}
    \acro{SLR}{single layer reflective}
    \acro{SMB}{superconducting magnetic bearing}
    \acro{SQUID}{superconducting quantum interference device}
    \acro{SQL}{structured query language}
    \acro{STARS}{star tracking attitude reconstruction software}
    \acro{TES}{transition edge sensor}
    \acro{TDRSS}{tracking and data relay satellites}
    \acro{TM}{transformation matrix}
    \acro{TOD}{time ordered data}

\end{acronym}

\acknowledgments
Support for the development and flight of the EBEX instrument was provided by 
NASA grants NNX12AD50G, NNX13AE49G, 
NNX08AG40G, and NNG05GE62G, and by NSF grants AST-0705134 and ANT-0944513.   
We acknowledge support from the Italian INFN INDARK Initiative.  
Ade and Tucker acknowledge the 
Science \& Technology Facilities Council for its continued support of the 
underpinning technology for filter and waveplate development.  
We also acknowledge support by the Canada Space Agency, the Canada 
Research Chairs Program, the Natural Sciences and 
Engineering Research Council of Canada, the Canadian Institute for Advanced Research, 
the Minnesota Supercomputing Institute, the National Energy Research 
Scientific Computing Center, the Minnesota and Rhode Island 
Space Grant Consortia, our collaborating institutions, and Sigma Xi the 
Scientific Research Society. 
Baccigalupi acknowledges support from the RADIOFOREGROUNDS 
grant of the European Union's Horizon 2020 
research and innovation program (COMPET-05-2015, grant agreement number 687312) 
and the INDARK INFN Initiative.
Didier acknowledges a NASA NESSF fellowship NNX11AL15H.  
Reichborn-Kjennerud acknowledges an NSF Post-Doctoral Fellowship AST-1102774, 
and a NASA Graduate Student Research Fellowship. Raach and Zilic 
acknowledge support by the Minnesota Space Grant Consortium.  
We very much thank Danny Ball and his colleagues at the Columbia 
Scientific Balloon Facility for their dedicated support of the EBEX program.  
We are grateful for contributions to the fabrication of optical elements by Enzo Pascale and 
Lorenzo Moncelsi. Xin Zhi Tan's and Chiou Yang Tan's help with figures is acknowledged 
and much appreciated. 


\newpage
\appendix

\section{Double Modulation/Demodulation and Extracting $\Phi$}
\label{sec:polcaleqns}

We consider the signal recorded due to a source that is chopped between two intensities, 
polarized at angle $\alpha_{in}$, and detected by a polarimeter with a \ac{HWP} and polarizing 
grid. We are only interested in linearly polarized light therefore, 
the input Stokes vector is 
\begin{equation}
    S_{in} =
    \left[    \begin{array}{c}
              I   \\
             Q  \\
             U  
        \end{array}    \right]
\end{equation}
with
\begin{equation}
X = \sum_{l} X_{l} \cos( l\omega_{c}t + \phi_{c, l} ), \,\,\, X = I, Q, U,
\label{eqn:xi}
\end{equation}
where $\omega_{c}$ is the chopping frequency.  We include an arbitrary number of $l$ harmonics
in $X$ because the chopping is not necessarily sinusoidal. Typically the $l=0,1$ components dominate
giving 
\begin{equation}
X_{01} =  X_{0} + X_{1}\cos( \omega_{c}t + \phi_{c} ), \,\,\, X = I, Q, U.   
\label{eqn:xii}
\end{equation}
The rotation of the \ac{HWP} modifies $S_{in}$ to give
\begin{equation}
    S_{in} =
    \left[  \begin{array}{c}
            I \\
            \epsilon \left[Q \cos( 4\gamma ) + U \sin( 4\gamma \right]   \\
             \epsilon \left[Q \sin( 4\gamma ) - U \cos( 4\gamma \right]  \end{array}
    \right],
\end{equation}
where $\gamma = \omega_{hwp} t$ is the rotation angle. After the analyzer grid, which is oriented at an 
angle $\beta$, we have the \ac{TOD} 
\begin{equation}
    D(t) = \frac{1}{2} \left\{ I + \epsilon \left[ Q \cos(4\gamma - 2\beta) + U \sin(4\gamma - 2\beta ) \right] \right\}. 
\label{eqn:tod}
\end{equation}
In the ideal case the \ac{HWP} is rotating at a constant rate $\dot{\gamma}= \omega_{hwp} $. In practice, 
and as we showed in Figure~\ref{fig:HWP-speed}, the rotation rate is not constant. The 
constant term is dominant but there are also sub-dominant terms at  
frequencies that are harmonics of the rotation rate. Thus 
the angle $\gamma(t)$ should more generally be written as 
\begin{equation}
    \gamma(t) = \omega_{hwp} t + \sum_{m} b_{m} \cos (m \omega_{hwp} t). 
\label{eqn:gamma}    
\end{equation}
We first highlight the most dominant properties of the \ac{TOD} by assuming a purely 
sinusoidal chop and constant rotation rate; that is using Equation~\ref{eqn:xii} and  
$\dot{\gamma}= \omega_{hwp} $. We also include the phase $\Phi$ and spurious polarization 
signals that are not chopped (see Section~\ref{sec:polarizationcalibration} )
\begin{eqnarray}
D(t) & =  & \frac{1}{2} \left(I_{0} + I_{1} \cos(\omega_{c}t + \phi_{c}) \right) + \nonumber \\
           &     &  \frac{\epsilon}{2}   \{ Q_{0}\cos( 4\gamma - \Phi)   + 
           Q_{1}\left[ \cos( 4 \omega_{hwp} t - \Phi)\cos(\omega_{c}t + \phi_{c}) \right] \} +  \nonumber \\
           &     &  \frac{\epsilon}{2} \{ U_{0}\sin( 4\gamma - \Phi) + 
            U_{1} \left[ \sin(4 \omega_{hwp} t - \Phi)\cos(\omega_{c}t + \phi_{c})\right] \}  +  \nonumber \\
           &     &     \sum_{j} A_{j} \cos( j\omega_{hwp} + \phi_{j} ) .
\label{eqn:doubledemod_rawtod}
\end{eqnarray}
The $Q_{0}, U_{0}$ signals are at frequency of $4\omega_{hwp}$ but so is 
the spurious signal term $A_{4}$. The components $Q_{1}, U_{1}$ are 
at frequencies $4\omega_{hwp} \pm \omega_{c}$, which are sidebands of $4\omega_{hwp}$. Since 
$Q_{1}, U_{1}$ are known -  specifically, the incident polarization angle is given
by $ \alpha_{in} = (1/2) \arctan \left( U_1 / Q_1 \right)$ -  
analysis of the signal at these sidebands gives $\Phi$ without contamination by the 
spurious signal terms. 

The product of the harmonic series in $X$ (Equation~\ref{eqn:xi}) 
and the multiplicity of frequencies in $\gamma$ (Equation~\ref{eqn:gamma}) generate 
additional sidebands, some of which overlap with $4\omega_{hwp} \pm \omega_{c}$. 
Band-passing the raw \ac{TOD} around these sidebands 
before the double demodulation rejects other sidebands. We estimated the contributions 
of terms that do not include $Q_{1}, U_{1}$ to the sidebands $4\omega_{hwp} \pm \omega_{c}$ 
and concluded that they would change the determination of $\Phi$ by $0.2^{\circ}$ at most.
We therefore proceeded with the analysis using Equation~\ref{eqn:doubledemod_rawtod}. 

Double demodulation consists of multiplying the bandpassed version of $D(t)$
by a reference chopper signal and a reference \ac{HWP} signal. One can bandpass around and 
use both sidebands, or bandpass around and use just one of the sidebands. Using  
both sidebands gains a factor of 2 in signal relative to a single sideband; doing the analysis 
separately on each sideband can be used as an internal cross-check, or to determine the bolometer 
time constant, as discussed in Appendix~\ref{sec:timeconstants}. Below we assume an analysis 
that uses both sidebands. We derive the reference chopper signal by 
bandpassing $D(t)$ with a $0.7~Hz$ filter centered on $\omega_{c}$. We derive
the reference \ac{HWP} signal using the angular encoding information of the \ac{HWP}. 

After bandpassing the \ac{TOD} is
\begin{eqnarray}
D'(t) & =  & \frac{\epsilon}{4}  \left\{ Q_{1} \left[ \cos(4 \omega_{hwp} t - \Phi + \omega_{c}t + \phi) + 
                      \cos(4\omega_{hwp} t -\Phi-\omega_{c}t-\phi) \right] \right\} + \nonumber \\
           &      & \frac{\epsilon}{4} \left\{ U_{1} \left[ \sin(4\omega_{hwp} t - \Phi + \omega_{c}t + \phi)  + 
                      \sin(4\omega_{hwp} t - \Phi - \omega_{c}t -\phi) \right] \right\} .
\end{eqnarray}
Multiplying by the chopper reference signal gives
\begin{eqnarray} 
 D''(t) & = & D'(t)\cos(\omega_{c}t+\phi) \\
            & = &  \frac{\epsilon}{4} Q_{1} \{ \cos(4\omega_{hwp} t - \Phi) + 
                        \frac{1}{2} [ \cos(4\omega_{hwp} t-\Phi-2\omega_{c}t-2\phi) +  \nonumber \\
            &    &   \cos(4\omega_{hwp} t-\Phi+2\omega_{c}t+2\phi) ]  \} + \nonumber \\
            &     &  \frac{\epsilon}{4} U_{1} \{ \sin(4\omega_{hwp} t - \Phi) + 
                        \frac{1}{2} [ \sin(4\omega_{hwp} t - \Phi - 2\omega_{c}t -2\phi) + \nonumber  \\
            &    &    \sin(4\omega_{hwp} t - \Phi + 2\omega_{c}t +2\phi) ] \} . 
\end{eqnarray}
Multiplying by the two \ac{HWP} reference signals $\cos4\gamma$ and
$\sin4\gamma$ and lowpassing at $0.7~Hz$ gives
\begin{eqnarray}
    D''(t)\cos 4\gamma\, _{DC} & = & \frac{\epsilon}{8}\left[Q_{1}\cos(\Phi) - U_{1}\sin(\Phi)\right] \\
    D''(t)\sin 4\gamma\, _{DC} & = & \frac{\epsilon}{8}\left[Q_{1}\sin(\Phi) + U_{1}\cos(\Phi)\right] ,
\end{eqnarray}
where the susbscript $_{DC}$ indicates that we are keeping only the zero frequency term; the low pass
filter suppresses the other terms. Because $\alpha_{in} = (1/2) \arctan  (U_{1}/Q_{1})$
we can rewrite $D''(t)$ as
\begin{eqnarray}
    D''(t) \cos 4\gamma \,_{DC} & = & \frac{\epsilon}{8} \left( \frac{Q_{1}}{\cos(2\alpha_{in})} \right) 
                 \left[ \cos(2\alpha_{in})\cos(\Phi) - \sin(2\alpha_{in})\sin(\Phi) \right]  \nonumber \\
        & =  & \frac{\epsilon}{8} \left( \frac{Q_{1}}{\cos(2\alpha_{in})} \right) \cos(2\alpha_{in}+\Phi)   \\
    D''(t) \sin 4\gamma \,_{DC} & =  & \frac{\epsilon}{8}\left( \frac{Q_{1}}{\cos(2\alpha_{in})} \right)
                \left[\cos(2\alpha_{in})\sin(\Phi) + \sin(2\alpha_{in})\cos(\Phi)\right]  \nonumber \\
        & = & \frac{\epsilon}{8} \left( \frac{Q_{1}}{\cos(2\alpha_{in})} \right) \sin(2\alpha_{in}+\Phi) ,
\end{eqnarray}
and solve for $\Phi$
\begin{equation}
    \Phi = \arctan 
    \left( \frac{ \left[D''(t)\sin(4\gamma) \right]_{DC} }{ \left[D''(t)\cos(4\gamma)\right]_{DC} } \right) 
    - 2\alpha_{in}.
\end{equation}

\section{Estimates of Bolometer Time Constants}
\label{sec:timeconstants}

\subsection{Single Pole Response} 
\label{sec:bolotausinglepole}

We assumed that the bolometer time constant follows a single pole response with critical 
frequency $\omega_{b} = 2 \pi f_{b} = 1 / \tau_{b}$ and used the polarization calibration data to estimate it. 
A single pole response gives rise to attenuation 
of signals at frequencies near and above $\omega_{b}$ and to a frequency dependent phase shift. 
We investigated the use of both effects to estimate the time constant. 

As described in Section~\ref{sec:polarizationrotation} 
and in Appendix~\ref{sec:polcaleqns}, the polarization calibration relies on 
signals that are in frequencies $4\omega_{hwp} + \omega_{c}$ and $4\omega_{hwp} - \omega_{c}$.
The `attenuation' method relies on the observation that for the relevant range of
 $\omega_{b},\, \omega_{hwp}$, and $\omega_{c}$  
there is more attenuation of power in the higher sideband than in the lower. 
We used the ratio of powers in the two sidebands to estimate $\tau_{b}$.
 The `phase' method relies on the fact that the 
signal at the higher frequency sideband undergoes a larger phase shift compared to the 
signal at the lower frequency. Thus an analysis extracting the calibration angle $\Phi$ from each of these 
signals {\it without} first deconvolving the bolometer temporal response gives 
a difference in angle $\Phi$ with a magnitude that depends on $\tau_{b}$. 
We used the phase method because simulations showed that attenuation method 
was prone to larger bias at high noise levels. Here 
we describe the phase method; \citet{klein_thesis} gives more details on both 
approaches. 

\begin{figure}[h]
\begin{center}
\includegraphics[height=2.4in]{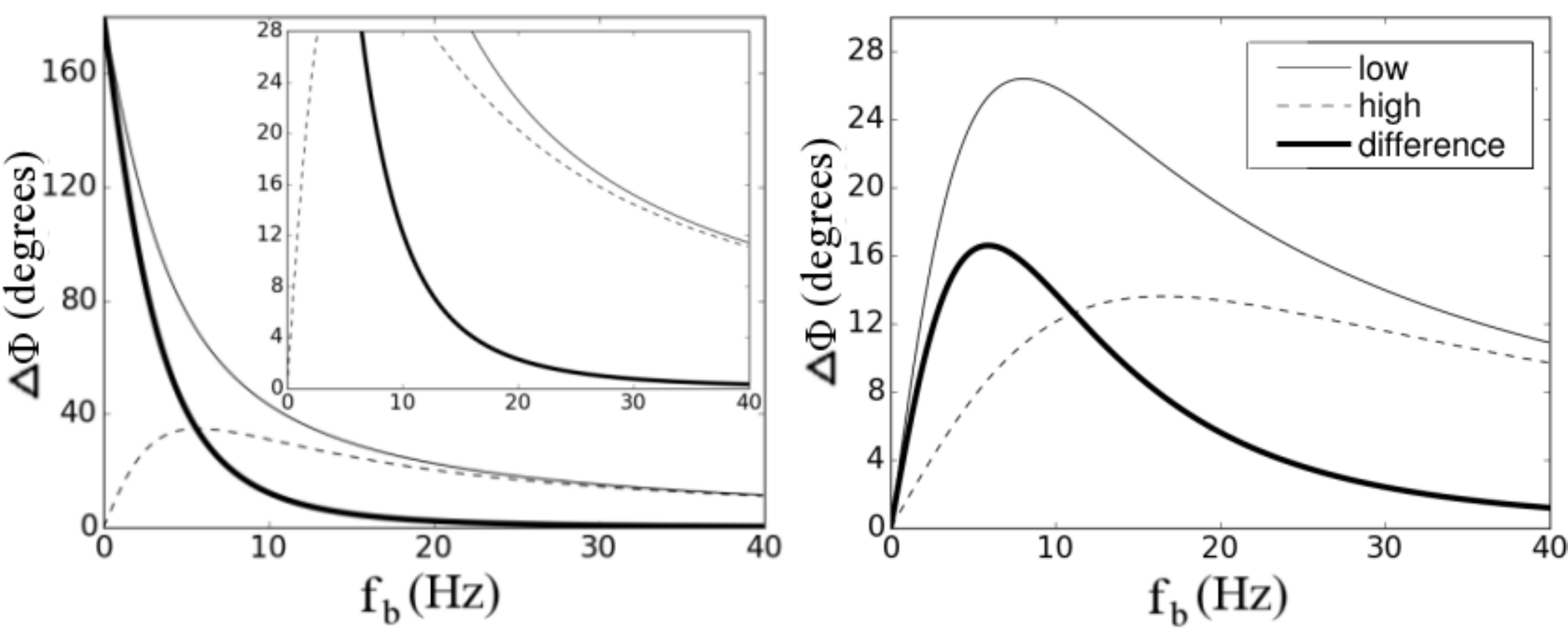}
\par\end{center}
\caption{The value of $\Delta \Phi_{obs}$ (dark) as a function of $f_{b}$
for $f_{hwp}=2$~Hz and $f_{c}=3$~Hz that were used for the receiver only polarization 
calibration (left), and for $f_{hwp}=2$~Hz and $f_{c}=13$~Hz that were used for the entire-instrument 
polarization calibration (right). The panels also show the phases $\delta_{low}$ and $\delta_{high}$ (light grey).  
For $f_{c} < 4f_{hwp}$ we show $\Delta \Phi_{obs}$ over all possible phase differences and in the 
inset over the range that matches the panel for for $f_{c} > 4f_{hwp}$.  
\label{fig:tc_diff} }
\end{figure}

Relative to an infinitely fast detector, each frequency component of the signal undergoes
a phase shift 
\begin{equation}
\delta(\omega) = -\arctan(\omega \tau_{b}).
\end{equation}
Analyzing the raw data for $\Phi$ as described in Appendix~\ref{sec:polcaleqns}, that is, 
without deconvolving a bolometer response function, and using each 
sideband separately we find 
\begin{equation}
 \Phi_{obs, low} = \Phi_{0} - \frac{1}{2}(\delta_{low} + \delta_{c}), 
\end{equation}
and
\begin{equation}
 \Phi_{obs, high} = \Phi_{0} - \delta_{high} - \delta_{c},
\end{equation}
where $\Phi_{obs, low}$ and $\Phi_{obs, high}$ are the extracted $\Phi$ values
from the low and high sidebands, respectively. $\Phi_{0}$ is the nominal value, i.e.
the value we would extract with an infinitely fast time constant (or with a perfect deconvolution). 
The quantities 
$\delta_{low}$, $\delta_{high}$, and $\delta_{c}$ are the phase shifts due to the bolometer 
time constant at the low and high sidebands, and the chop frequency, respectively. The 
phase difference $\Delta \Phi_{obs} \equiv \Phi_{obs, low} - \Phi_{obs, high}$ is given by:
\begin{eqnarray}
\Delta \Phi_{obs} & = & \delta_{high} - \delta_{low} - 2\delta_{c}  \nonumber \\
                 &  = & \arctan((4\omega_{hwp} - \omega_{c})\tau) - 
                               \arctan((4\omega_{hwp} + \omega_{c})\tau) \nonumber \\ 
                  &    &          + 2\arctan(\omega_{c}\tau) .
\end{eqnarray}
Figure~\ref{fig:tc_diff} shows $\Delta \Phi_{obs}$ as a function of $\omega_{b}$. 
When $\omega_{c} < 4\omega_{hwp}$ a value of $\Delta \Phi_{obs}$ uniquely determines $\omega_{b}$. 
When $\omega_{c} > 4\omega_{hwp}$ a value of $\Delta \Phi_{obs}$ gives two 
possible values of $\omega_{b}$. 

In the receiver-only polarization calibration $f_c$ and $4f_{hwp}$ were 3, and 8~Hz, 
respectively. In the entire-instrument calibration $f_{c}$ and $4f_{hwp}$ were 13, and 8~Hz, 
respectively. For this test we used the attenuation method in combination with the phase method
to break the degeneracy between the two possible values of $f_{b}$. Typical values for $f_{b}$ were close
to 5~Hz.

\subsection{Deviations from Single Pole Response}
\label{sec:bolotaudeviation}

\begin{figure}[h]
\begin{center}
\includegraphics[height=3in]{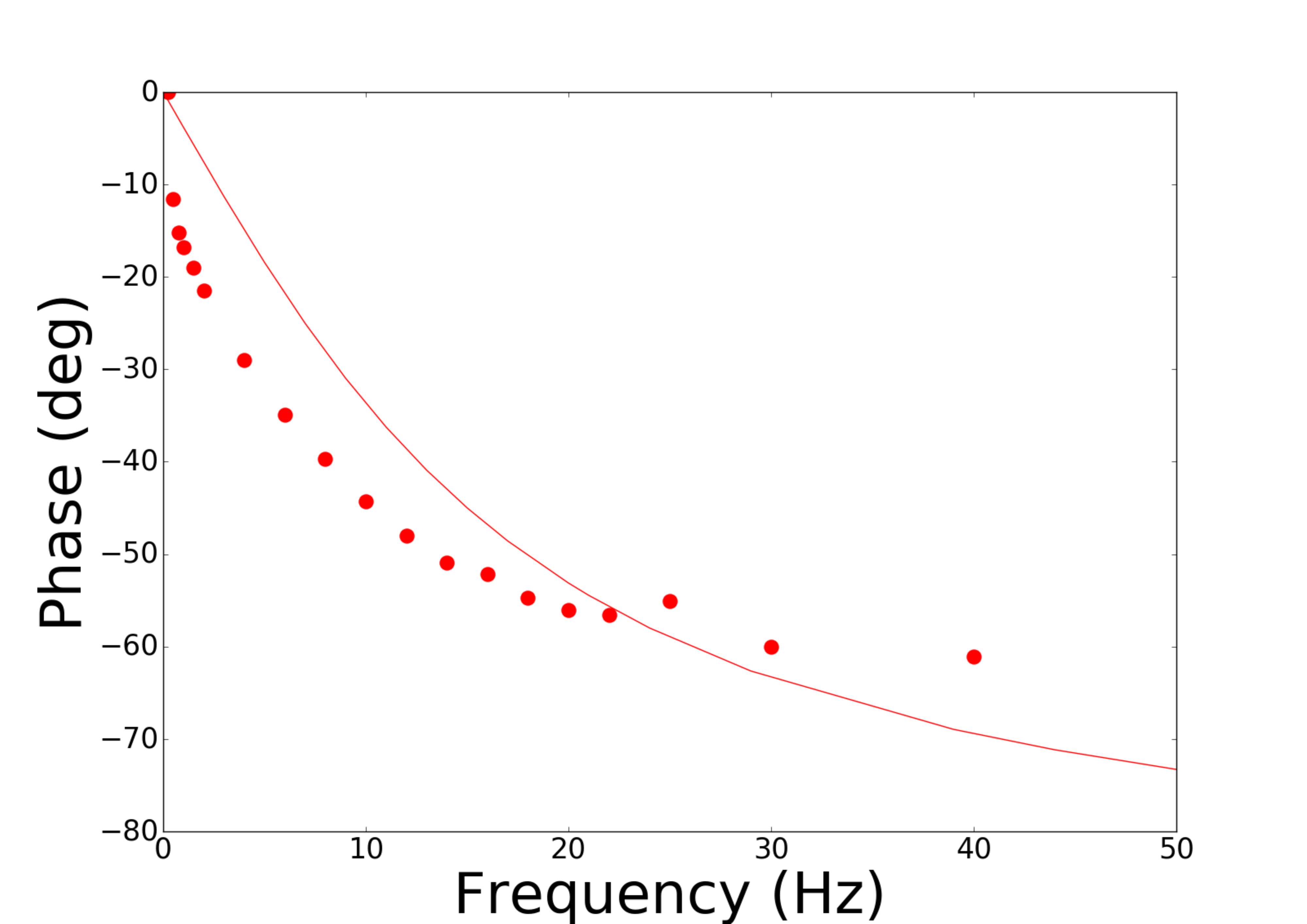}
\par\end{center}
\caption{Relative phase between an input and measured chopper signal as a function 
of the frequency of the chopper (dots) for one detector. The relative phase is due to the
bolometer time constant, which does not follow a pure single pole model (solid).      
\label{fig:measurement_tau_phase} }
\end{figure}

We measured the frequency response due to the bolometer time constant by correlating the 
chopper signal as recorded by the detectors and the input chopper signal for chopper 
frequencies between near 0 and 40~Hz. The difference in phase gave the frequency response
of the bolometer. For the 350 bolometers 
measured, we found the bolometer time constant deviated from a single pole response. An example 
is shown in Figure~\ref{fig:measurement_tau_phase}. The deviation
entails a necessary correction to the extraction of the phase $\Phi$. Because we did not
measure the individual frequency response of all the bolometers, we 
applied an overall correction based on the ensemble of frequency responses measured. 
We now describe this correction. 

We generated simulated 
\ac{TOD} for each of the 350 measured frequency responses with 
a known input $\Phi$. The simulated data 
were subject to the polarization calibration data analysis, still under the assumption that the 
detectors were modeled by a single-pole filter. The difference between the input and the mean 
extracted $\Phi$ quantified the bias introduced by the assumption of a single pole response. 
Table~\ref{tab:Bias-due-to-TC} 
gives the phase shifts $\Phi$ measured for each of the two polarization rotation
calibrations and for each frequency band.


\begin{table}[htpb]
\begin{center}
\begin{tabular}{|c|c|c|}
\hline 
 & Receiver-only Test & End-to-end Calibration\tabularnewline
\hline
\hline 
150 GHz & $4.1^{\circ}$ & $0.0^{\circ}$\tabularnewline
\hline 
250 GHz & $1.5^{\circ}$ & $3.5^{\circ}$\tabularnewline
\hline 
410 GHz & $1.9^{\circ}$ & $2.1^{\circ}$\tabularnewline
\hline
\end{tabular}
\end{center}
\caption{Polarization rotation correction due to the assumption of single pole response
\label{tab:Bias-due-to-TC} }
\end{table}

\clearpage

\bibliographystyle{aasjournal}
\bibliography{EBEXPaper1}

\end{document}